\documentclass[preprint,prd,aps,showpacs,preprintnumbers,amsmath,amssymb]{revtex4-1}
\usepackage{graphics,epsfig,subfigure}
\usepackage{diagbox}
\usepackage[usenames]{color}
\usepackage{color}
\usepackage{graphicx}
\usepackage{amsfonts}
\usepackage{txfonts}
\usepackage{indentfirst}
\usepackage{booktabs}
\usepackage{hyperref}
\hypersetup{hypertex=ture,backref=true,colorlinks=ture,linkcolor=blue,anchorcolor=blue,citecolor=blue}
\usepackage{float}
\usepackage{multirow}
\usepackage[section]{placeins}
\usepackage{array}
\usepackage{amsmath}

\begin{document}

\title{\Large \bf Proca stars in AdS Ellis wormholes}
\author{Gen Li}
\author{Chen-Hao Hao}
\author{Xin Su}
\author{Yong-Qiang Wang\footnote{yqwang@lzu.edu.cn, corresponding author}}
\affiliation{ $^{1}$Lanzhou Center for Theoretical Physics, Key Laboratory of Theoretical Physics of Gansu Province,
	School of Physical Science and Technology, Lanzhou University, Lanzhou 730000, China\\
	$^{2}$Institute of Theoretical Physics $\&$ Research Center of Gravitation, Lanzhou University, Lanzhou 730000, China}

\begin{abstract}

In this paper, we  study Proca stars in asymptotically anti-de Sitter (AdS) Ellis wormholes. This study distinguishes itself from the analysis of the Proca stars in asymptotically flat spacetimes. In the AdS Ellis wormhole background, the mass of the wormhole solutions vanishes. Consequently, we employ numerical techniques to investigate in detail the impact of the cosmological constant on both the matter field and the wormhole geometry, while categorizing the solutions in accordance with the symmetries of the Proca field. The results show that when the cosmological constant decreases, not only does the characteristic spiral behavior of the Proca star solutions, namely the charge-frequency relation $Q$-$\omega$, gradually disappear, but the throat or both sides of the wormhole may also develop a ``horizon", presenting a ``black bounce" characteristic.

\end{abstract}

\maketitle

\section{INTRODUCTION}\label{Sec1}

Proca stars constitute a family of vector Boson stars arising from spin-1 massive vector fields, a notion originating from Wheeler's 1955 conception of ``geons" \cite{Wheeler:1955zz,Power:1957zz}. ``Geons" are unstable soliton solutions formed by classical electromagnetic waves in a gravitational field. Subsequent advancements emerged in 1968 when Kaup et al. \cite{Kaup:1968zz} reformulated geon theory by replacing massless vector fields with massive complex scalar fields, thereby pioneering the Klein-Gordon geon — Boson stars formed through gravitational coupling of spin-0 fields. The recognition that spin-1 Proca fields analogously admit solitonic gravitational bound states, now termed Proca stars, was established much later by Brito et al. in 2015 \cite{Brito:2015pxa}. Subsequent investigations have significantly expanded this framework, encompassing charged Proca configurations \cite{SalazarLandea:2016bys}, structures in anti-de Sitter (AdS) spacetime \cite{Duarte:2016lig}, flat-spacetime Proca Q-balls \cite{Loginov:2015rya,Brihaye:2017inn,Heeck:2021bce,Dzhunushaliev:2021tlm,Herdeiro:2023lze}, and diverse extensions \cite{Herdeiro:2019mbz,Aoki:2022woy,Aoki:2022mdn,Ma:2023bhb,Zhang:2023rwc,Lazarte:2024jyr,Minamitsuji:2018kof,Babichev:2017rti}. Due to the Proca field forming stable structures, Proca stars have garnered considerable interest in cosmological and astrophysical contexts, serving as viable dark matter candidates \cite{Jockel:2023rrm,GRAVITY:2023azi} or potential explanations for anomalous signatures in the GW190521 gravitational wave event \cite{Romero-Shaw:2020thy,CalderonBustillo:2020fyi}.

Wormholes represent intriguing hypothetical astrophysical constructs linking disparate spacetime regions via traversable throats, with speculations even suggesting their utility as time machines \cite{Morris:1988tu}.  The concept of a wormhole was first proposed by Flamm in 1916, around the same time the Schwarzschild black hole solution was introduced. In 1935, to exclude singularities of the field, Einstein and Rosen constructed a spacetime structure that connects two identical spatial sheet via a ``bridge" which later came to be known as the Einstein-Rosen bridge \cite{Einstein:1935tc}. In 1957, Wheeler named this structure a wormhole \cite{Misner:1957mt}, pointing out that when matter or information attempts to pass through the throat of such a wormhole, the throat would quickly collapse into a singularity, thereby preventing the transfer of matter or information \cite{Kruskal:1959vx,Fuller:1962zza}. Ellis and Bronnikov achieved a critical breakthrough in 1973 with their traversable wormhole solutions \cite{Ellis:1973yv,Ellis:1979bh,Bronnikov:1973fh,Kodama:1978dw}. Subsequently, systematic analyses by Morris and Thorne demonstrated that in order for the wormhole throat to be traversable, it is necessary to have exotic matter that violates the null energy condition \cite{Morris:1988cz}. While quantum phenomena like the Casimir effect provide microphysical mechanisms for negative energy densities \cite{Garattini:2019ivd,Mehdizadeh:2024oyd}, ongoing efforts seek to minimize exotic matter requirements through strategies such as thin-shell wormhole \cite{Poisson:1995sv} or Einstein-Maxwell-Dirac (EMD) field couplings \cite{Blazquez-Salcedo:2020czn}. Alternative approaches employ modified gravity frameworks to circumvent energy condition violations \cite{Lobo:2009ip,Banerjee:2021mqk}.

Among wormhole models, the static asymptotically flat Ellis wormhole represents a foundational class of traversable wormhole, sustained through phantom fields \cite{Lobo:2005us,Sushkov:2005kj,Lobo:2005yv,Bronnikov:2012ch,Kleihaus:2014dla} with inverted kinetic terms that violate energy conditions. Its geometry features dual asymptotically flat regions connected by a throat. Extensions include rotating solutions \cite{Kleihaus:2014dla}, higher-dimensional analogs \cite{Dzhunushaliev:2014bya}, and couplings with diverse matter \cite{Hoffmann:2017jfs,Yue:2023ela,Ding:2023syj,Hao:2023igi,Su:2023zhh,Nozawa:2020gzz}. Astrophysical signals have been analyzed to detect potential evidence for Ellis wormholes, such as gravitational lensing effects, wormhole shadows \cite{Ohgami:2015nra, Abe:2010ap}, and gravitational wave signals during the ringdown phase \cite{Konoplya:2016hmd}, which could provide indirect observational evidence.  

Most studies of wormholes have focused on asymptotically flat spacetimes. In contrast, asymptotically AdS spacetimes exhibit unique gravitational properties. The AdS/CFT duality suggests that gravitational theory in $d+1$-dimensional AdS spacetime can be described by a conformal field theory (CFT) on its $d$-dimensional boundary \cite{Maldacena:1997re}, and through boundary interactions, eternally traversable wormholes can be constructed \cite{Maldacena:2018lmt, Freivogel:2019lej}. Additionally, the ER=EPR conjecture suggests that AdS black holes may be connected by traversable wormholes \cite{Gao:2016bin, Maldacena:2013xja}, making wormholes in AdS spacetimes an important area of research. Current studies focus on constructing wormholes, analyzing their stability, and investigating their information transmission capabilities \cite{Maldacena:2017axo, Dai:2020ffw, Kain:2023ore, Lemos:2003jb, Lemos:2004vs, Maeda:2008nz, Zhang:2024kqp, Maeda:2012fr, Wu:2022gpm, Anabalon:2018rzq, Lobo:2004uq}. Kunz et al\cite{Blazquez-Salcedo:2020nsa} extended the asymptotically flat Ellis wormhole solution to the AdS spacetime. In this paper \cite{Hao:2024hba}, the authors attempt to introduce a boson star into the AdS Ellis wormhole. Previously, studies \cite{Su:2023zhh} have investigated solutions of the asymptotically flat Ellis wormhole coupled with the Proca field. Therefore, we further extends this Proca field solution to the AdS spacetime, systematically studying its geometric structure and physical properties. Unlike the Proca star in the asymptotically flat Ellis wormhole spacetime, this solution possesses only Noether charge. The obtained solution has a finite Noether charge, which varies with frequency, the cosmological constant, and the throat size. As the cosmological constant decreases, a near ``horizon'' may emerge, potentially forming an ``horizon"-like structure akin to a ``black bounce" \cite{Simpson:2018tsi}.

The paper is organized as follows: Section~\ref{sec2} introduces the Proca star in an asymptotically AdS Ellis spacetime, the Einstein action, and the wormhole metric. Section~\ref{sec3} analyzes the boundary conditions and conducts a series expansion of the metric and Proca field to examine their asymptotic behavior. In Section~\ref{sec4}, we present numerical results of the solutions. Section~\ref{sec5} concludes the study and suggests future research directions.

\section{THE MODEL}\label{sec2}

\subsection{Action}

We consider an Einstein-Hilbert action that includes a Proca field and a phantom field, which is given by:
\begin{equation}\label{action}
S=\int\sqrt{-g} d^4x\left(\frac{\left( R - 2\Lambda \right)}{2\kappa}+\mathcal{L}_{p}+\mathcal{L}_{m}\right),
\end{equation}
where $R$ is the Ricci scalar, and $\kappa$ is the coupling constant.

The Lagrangian densities $\mathcal{L}_p$ and $\mathcal{L}_m$ are defined as:
\begin{eqnarray}
\mathcal{L}_{m} &=& -\frac{1}{4} \mathcal{F}_{\alpha\beta} \overline{\mathcal{F}}^{\alpha\beta}
- \frac{1}{2} \mu^2 A_\alpha \overline{A}^\alpha, \nonumber \\
\mathcal{L}_p &=& \nabla_\alpha \Phi \nabla^\alpha \Phi.
\end{eqnarray}
We are interested in a complex Proca field with mass
$\mu$, described by the potential 1-form $\mathcal{A}$  and the field
strength $\mathcal{F} = d\mathcal{A}$. $\Phi$ denotes the phantom field, while $\overline{A}^\alpha$ represents the conjugate of the Proca field's vector potential $A^\alpha$. 
By varying the action \eqref{action}, we obtain the Einstein equation:
\begin{equation}
\label{eq:EKG1}
R_{\mu\nu}-\frac{1}{2}g_{\mu\nu}R-\kappa T_{\mu\nu}+ \Lambda g_{\mu\nu}=0 ,
\end{equation}
where the stress-energy tensor is given by:
\begin{equation}
T_{\mu\nu} = g_{\mu\nu}({{\cal L}}_m+{{\cal L}}_p)
-2 \frac{\partial ({{\cal L}}_m+{{\cal L}}_p)}{\partial g^{\mu\nu}} .
\end{equation}

Additionally, by varying the Lagrangians of the Proca field and the phantom field, we obtain the following equations:
\begin{equation}
\label{eq:EKG2}
\nabla_\alpha \mathcal{F}^{\alpha \beta} - \mu^2 A^\beta = 0 ,
\end{equation}
and
\begin{equation}
\label{eq:EKG3}
\Box\Phi=0.
\end{equation}

It is straightforward to verify that the Proca field equation (5) directly satisfies the Lorenz condition \(\nabla_\alpha A^\alpha = 0\).

\subsection{Ansatze}

We choose the static spherically symmetric wormhole solution and adopt the following form for the metric, which can be found in references \cite{Blazquez-Salcedo:2020nsa,Hoffmann:2017jfs}:
\begin{equation}  \label{line_element1}
ds^2 = -e^{B(r)}N(r)dt^2 + \frac{p(r)}{e^{B(r)}} \left[ \frac{1}{N(r)}dr^2 + h(r) \left(d\theta^2 + \sin^2\theta \, d\phi^2\right) \right] ,
\end{equation}
where $N(r) = 1 - \frac{\Lambda r^2}{3}$, $h(r)=r^2+r_0^2$, and $r_0$ represents the radius of the wormhole throat. The radial coordinate $r$ ranges from $(-\infty, +\infty)$, corresponding to two asymptotically ads spacetime regions. In the case of pure Einstein gravity ($\Lambda=0$), the above metric simplifies to describe a static spherically symmetric Ellis wormhole with a Proca field \cite{Su:2023zhh}.

For further study, we assume that the Proca field and phantom scalar field take the following forms:
\begin{eqnarray}  \label{an2}
\mathcal{A} = \left[ H(r)dt + i G(r)dr \right] e^{-i \omega t} , \quad
\Phi = \phi(r).
\end{eqnarray}

Here, $F(r)$ and $G(r)$ are functions that depend only on the radial coordinate $r$, while $\omega$ is the frequency of the field. Additionally, the phantom field $\Phi$ is only a function of the radial coordinate and is independent of time.

By varying the action for the Proca field and the phantom field, we obtain the following system of equations:
\begin{equation}
\begin{split}
\left(e^B \mu^2 \left( -3 + r^2 \Lambda \right) + 3\omega^2 \right) G - 3\omega F^{\prime}=0,
\end{split}
\end{equation}

\begin{align}
&\frac{4r \left(-3 + 2r^2 \Lambda + r_0^2 \Lambda \right)p}{3\left(r^2 + r_0^2\right)} 
+ \frac{6 e^{-2B} \omega F p^2}{(-3 + r^2 \Lambda) G} 
+ \frac{2\left(-3 + r^2 \Lambda\right) p G'}{3 G} + \frac{1}{3}\left(-3 + r^2 \Lambda \right)p' = 0,
\end{align}

\begin{equation}
(\phi' h N \sqrt{p})' = 0 \ .
\end{equation}

Substituting the above metric into the Einstein equations and eliminating $\phi'$ gives the system of equations for $B''(r)$ and $p''(r)$:

\begin{equation}
\begin{aligned}
&\frac{2 \left( 3 r^2 + r_0^2 \right) \Lambda}{\left( r^2 + r_0^2 \right) \left( -3 + r^2 \Lambda \right)} 
- \frac{6 e^{-2 B} \left( e^{B} \Lambda \left( -3 + r^2 \Lambda \right) + 3 \mu^2 \kappa F^2 \right) p}{\left( -3 + r^2 \Lambda \right)^2} + B'' + \frac{B' p'}{2 p} \\
& \quad\quad+ \frac{2 r \left( -3 + 2 r^2 \Lambda + r_0^2 \Lambda \right) B'}{\left( r^2 + r_0^2 \right) \left( -3 + r^2 \Lambda \right)} 
+ \frac{3 e^{-B} \kappa \left( -\omega G + F' \right)^2}{-3 + r^2 \Lambda} + \frac{r \Lambda p'}{\left( -3 + r^2 \Lambda \right) p} 
 = 0,
\end{aligned}
\end{equation}

\vspace{1cm}

\begin{equation}
\begin{aligned}
&\frac{2 \left( 6 r^2 + r_0^2 \right) \Lambda p}{(r^2 + r_0^2)(-3 + r^2 \Lambda)}
- \frac{6 e^{-2B} \left(2 e^{B} \Lambda \left(-3 + r^2 \Lambda\right) + 3 \mu^2 \kappa F^2\right) p^2}{(-3 + r^2 \Lambda)^2}+ p'' - \frac{p'^2}{2 p} \\
&\quad \quad\quad \quad\quad\quad\quad\quad\quad\quad\quad\quad\quad\quad\quad\quad\quad\quad +  \frac{3 r \left(-3 + 2 r^2 \Lambda + r_0^2 \Lambda\right) p'}{(r^2 + r_0^2)(-3 + r^2 \Lambda)}
 = 0.
\end{aligned}
\end{equation}

By numerically solving equations (9), (10), (12), and (13), we can obtain the explicit solutions for $B(r)$, $p(r)$, $F(r)$, and $G(r)$. Equation (14) is commonly used to check the accuracy of the numerical calculations. Since Eq. (11) results in zero, the equation can be further simplified as:
\begin{equation}
\phi' = \frac{\sqrt{\cal D}}{h N \sqrt{p}}\ ,
\end{equation}
where $\cal D$ represents the scalar charge of the phantom scalar field, and is also used to check the precision of the calculations. Its value, as a function of frequency $\omega$, should remain consistent across different regions when the throat radius $r_0$ and cosmological constant $\Lambda$ are fixed.

The expression for the scalar charge $\cal D$ is obtained by substituting the above equation (14) into the Einstein field equations:

\begin{eqnarray}
\begin{split}
{\cal D} &= \frac{\left(r^2 + r_0^2\right)^2 \left(-3 + r^2 \Lambda\right)^2 p}{18\kappa} \Bigg[
\frac{-4 \left(3 r^4 \Lambda + r_0^2 \left(3+2r^2 \Lambda\right)\right)}{\left(r^2 + r_0^2\right)^2 \left(-3 + r^2 \Lambda\right)} 
+ 2 \mu^2 \kappa G^2 
+ \frac{18 e^{-2B} \mu^2 \kappa F^2 p}{\left(-3 + r^2 \Lambda\right)^2} \\
&\quad+ \frac{4 r \Lambda B'}{-3 + r^2 \Lambda} 
+ B'^2 
+ \frac{6 e^{-B} \left(2 \Lambda p + \kappa \left(-\omega G + F'\right)^2\right)}{-3 + r^2 \Lambda} - \frac{4 r \left(-3 + 2 r^2 \Lambda + r_0^2 \Lambda\right) p'}{\left(r^2 + r_0^2\right) \left(-3 + r^2 \Lambda\right)p} 
- \frac{p'^2}{p^2} 
\Bigg].
\end{split}
\end{eqnarray}

\section{BOUNDARY CONDITIONS}\label{sec3}

Before numerically solving the differential equations, we need to clarify the asymptotic behavior of the functions $F$, $G$, $B$, and $p$, which is equivalent to providing the necessary boundary conditions.

To study the asymptotic behavior of the metric functions as $r \rightarrow \infty$, we perform a series expansion of equations (12) and (13), yielding the following asymptotic expansions:

\begin{equation}
e^B(r)=f(r) = f_{\infty} + f_{\infty } \frac{r_0^{2}}{3 r^{2}} - \frac{f_{\infty }(12 r _0^{2}- r _0^{4} \Lambda)}{15 \Lambda r^{4}} + o\left(r^{-6}\right),
\end{equation}

\begin{equation}
p(r) = p_{\infty} - p_{\infty } \frac{r_0^{2}}{3 r^{2}} + \frac{p_{\infty }( 27 r _0^{2} + 14 r _0^{4} \Lambda)}{45 \Lambda r^{4}} + o\left(r^{-6}\right).
\end{equation}

Clearly, the odd terms are always zero, so we can obtain the asymptotic expansion of $g_{tt}$:

\begin{equation}
\begin{split}
-\left.g_{tt}\right|_{r \rightarrow \infty} = -\frac{\Lambda f_{\infty} r^2}{3} + f_{\infty} \left(1 - \frac{r_0^2 \Lambda}{9}\right) + \frac{f_{\infty} r_0^2}{15 r^2} \left( 1 + \frac{r_0^2 \Lambda}{3}\right)+o\left(r^{-4}\right).
\end{split}
\end{equation}

Since the odd terms vanish, the metric functions exhibit the same asymptotic behavior as \(r \rightarrow -\infty\) and \(r \rightarrow \infty\). We then provide the appropriate boundary conditions for the metric functions and the Proca field at infinity:

\begin{equation}
B( \pm \infty) = 0,\quad
p( \pm \infty) = 1,
\quad F( \pm \infty) = G( \pm \infty) = 0.
\end{equation}

Moreover, from the ADM expression, it is evident that the absence of the \(1/r\) term implies that the mass of these symmetric wormholes is zero. This remains true even when a matter field is added. Interestingly, when the cosmological constant \(\Lambda = 0\), this solution reduces to the Ellis wormhole with a Proca field and has a finite ADM mass. This is reflected in the asymptotic expansion of the metric components (the odd terms reappear):

\begin{equation}
g_{tt} = -1 + \frac{2M}{r} + \cdots .
\end{equation}

On the other hand, the Proca field's action exhibits a global \(U(1)\) symmetry under the transformation \(A \to e^{i\alpha} A\), where \(\alpha\) is a constant. This implies the existence of a conserved current:

\begin{equation}
j^\alpha = \frac{i}{2} \left[ \overline{\mathcal{F}}^{\alpha\beta} A_\beta - \mathcal{F}^{\alpha\beta} \overline{A}_\beta \right], \quad \nabla_\alpha j^\alpha = 0.
\end{equation}

The time component of the conserved current on a spacelike hypersurface \(\mathcal{S}\) can be integrated to obtain the Noether charge in the symmetric case:

\begin{equation}
Q = - \int j^t \sqrt{|g|} \, dr \, d\theta \, d\phi.
\end{equation}

\section{NUMERICAL RESULTS}\label{sec4}

In this study, all quantities are presented in dimensionless form, which are defined as follows:

\begin{equation}
r \to r\mu, \quad \phi \to \phi \kappa^{-1/2}, \quad \omega \to \omega/\mu.
\end{equation}

To simplify the calculations, we choose the fixed parameters $\mu_0 = 1$ and $\kappa = 2$, which do not affect the generality of the results.

For convenience in handling the radial coordinate with an infinite range, we map it to a finite interval through the following transformation:

\begin{equation}\label{transform}
x = \frac{2}{\pi}\arctan(r).
\end{equation}

This transformation compresses the radial coordinate $r \in (-\infty, +\infty)$ to $x \in (-1, 1)$. This approach allows the ordinary differential equations to be approximated as algebraic equations, thus facilitating numerical solutions. The integration region is primarily covered by 2000 grid points; under some parameters, 10,000 grid points may be required, and the relative error in the calculations is controlled within 
$10^{-4}$.

In the numerical computations, we mainly analyze two adjustable parameters: the cosmological constant $\Lambda$ and the throat radius $r_0$. To study their effects on the solution, we typically fix one parameter while varying the other. Additionally, due to the symmetry of the wormhole solution, the throat is always located at the central position $x = 0$. In the results of this study, we primarily present cases where the cosmological constant $\Lambda$ ranges from 0 to -10. Although $\Lambda$ can take any negative value, this range is sufficient to reveal the main effects of the parameters on the solution.

\subsection{Symmetric Case I}

\begin{figure}[H] 
\begin{center} 
\subfigure{\includegraphics[width=0.45\textwidth]{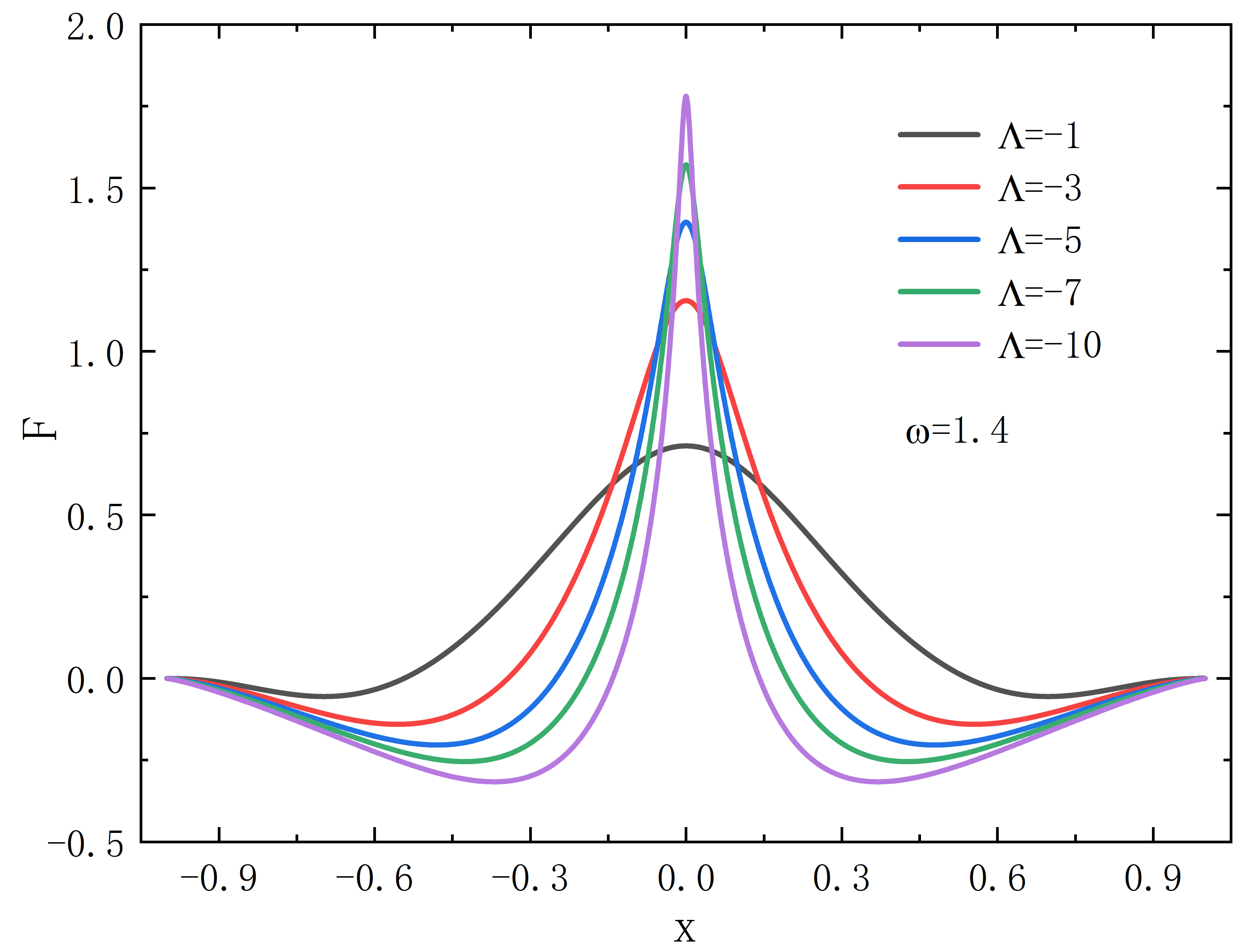}} 
\subfigure{\includegraphics[width=0.45\textwidth]{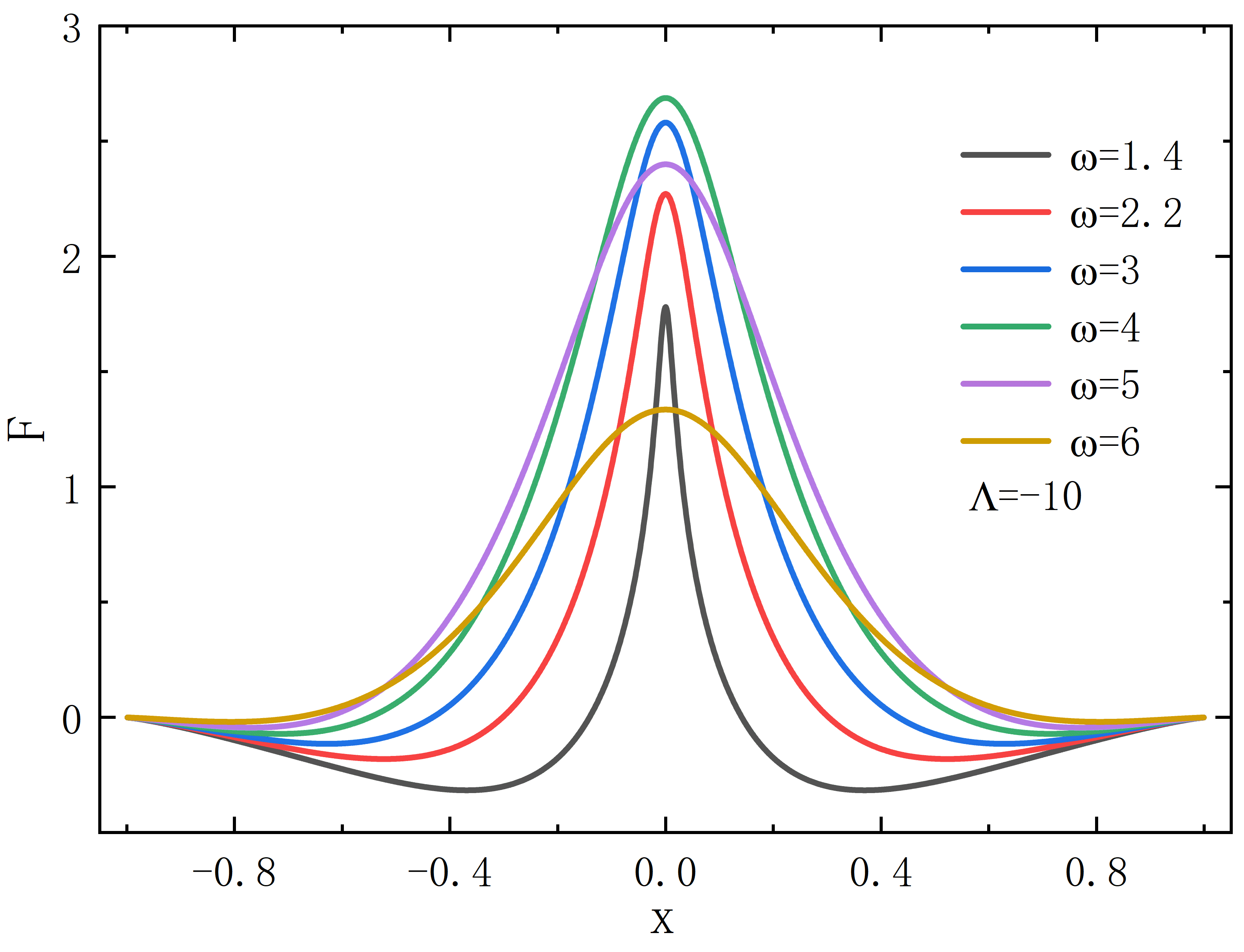}} 
\subfigure{\includegraphics[width=0.45\textwidth]{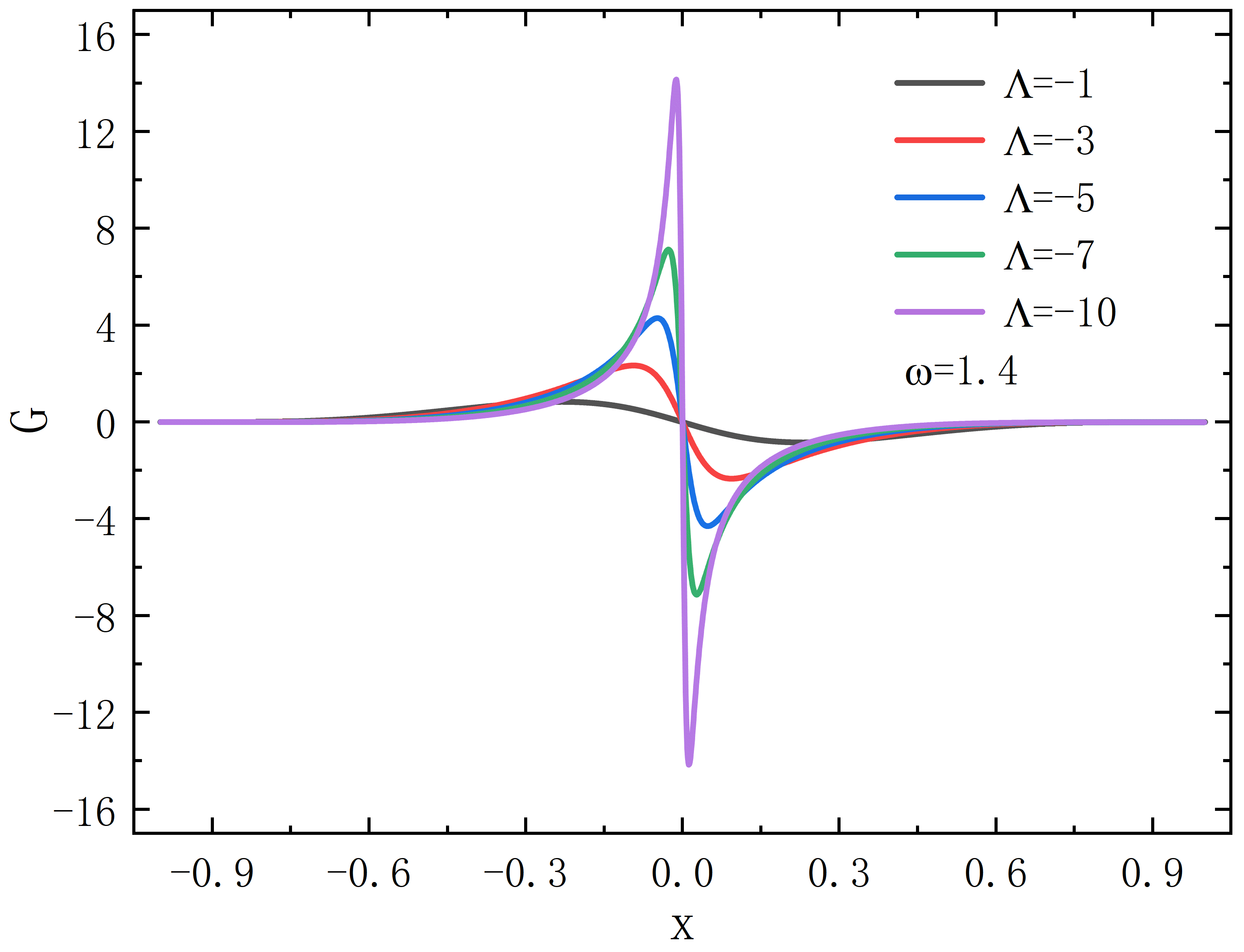}} 
\subfigure{\includegraphics[width=0.45\textwidth]{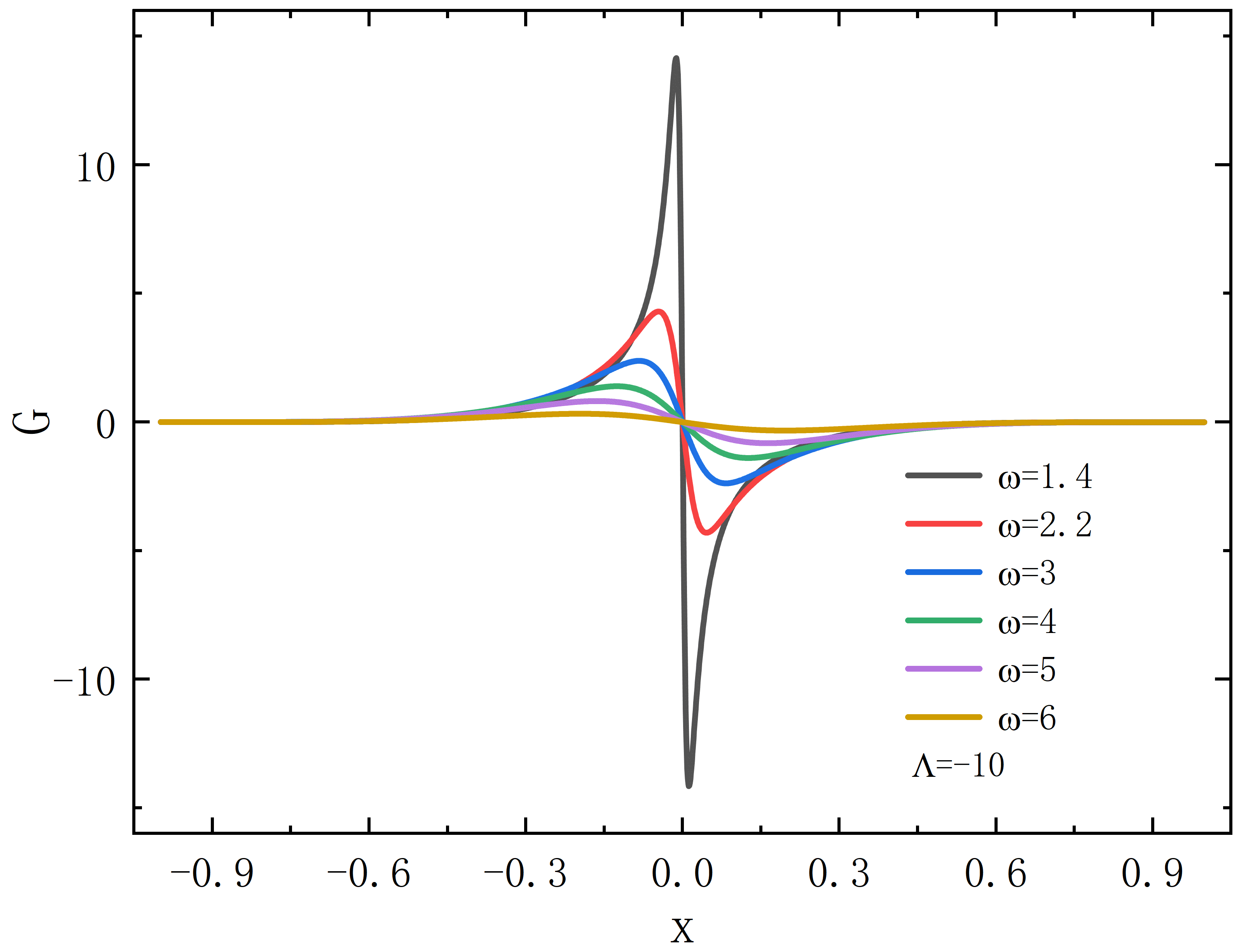}} 
\end{center} 
\caption{The matter fields $F(r)$ and $G(r)$ as functions of the radial coordinate $x$. The left panels fix $\omega$, while the right panels fix $\Lambda$. The throat radius $r_0 = 1$.} 
\label{phaseI3} 
\end{figure}

For the first class of symmetric solutions, by fixing the frequency $\omega = 1.4$ or the cosmological constant $\Lambda = -10$, we analyze the effects of $\Lambda$ and $\omega$ on the distribution of the matter field, as shown in Figure \ref{phaseI3}. The findings indicate that the field function $F(x)$ exhibits symmetry about the point $x = 0$, achieving its maximum at the throat and displaying symmetric minima on either side. A decrease in $\Lambda$ correlates with a gradual increase in the maximum value of $F$, while the minima concurrently decrease and converge towards the throat, resulting in a steeper curve. In contrast, the field function $G(x)$ demonstrates an antisymmetric distribution, with a significant increase in its slope at $x = 0$ as $\Lambda$ is reduced. Furthermore, when examining the influence of $\omega$ while maintaining a constant $\Lambda$, the maximum value of $F$ initially rises before subsequently declining as $\omega$ decreases, leading to a sharper curve. Concurrently, the slope of $G$ progressively increases as $\omega$ diminishes.

In the background of a negative cosmological constant,  the odd terms in the expansion of the metric function $g_{tt}$ vanish, implying that the ADM mass of this solution is zero. To further probe the solution's characteristics, we explore the relationship between the Noether charge $Q$ and the frequency $\omega$ for varying values of $\Lambda$, considering three sets of throat radii $r_0$ (small to large).

\begin{figure}[H] 
\begin{center} 
\subfigure{\includegraphics[width=0.45\textwidth]{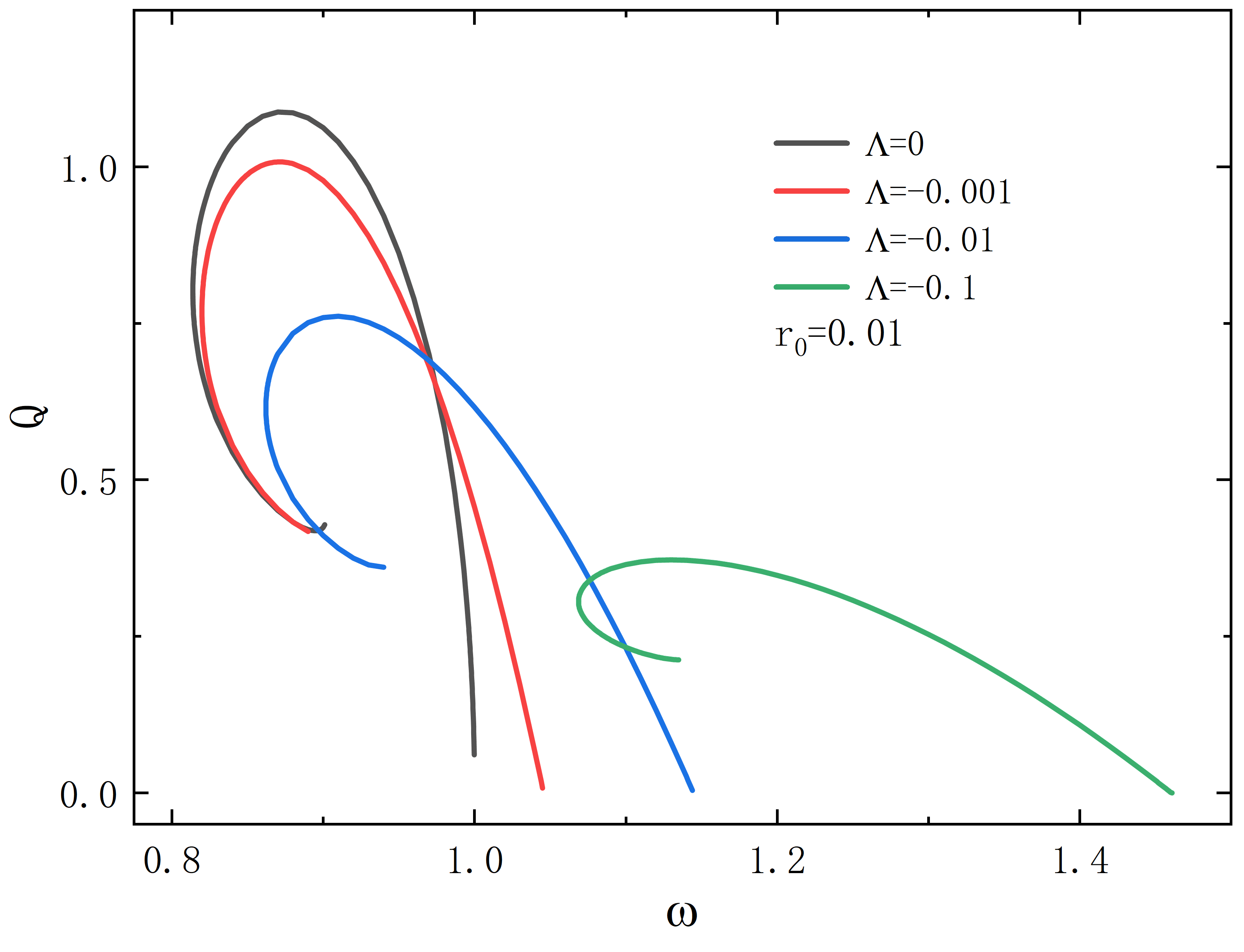}} 
\subfigure{\includegraphics[width=0.45\textwidth]{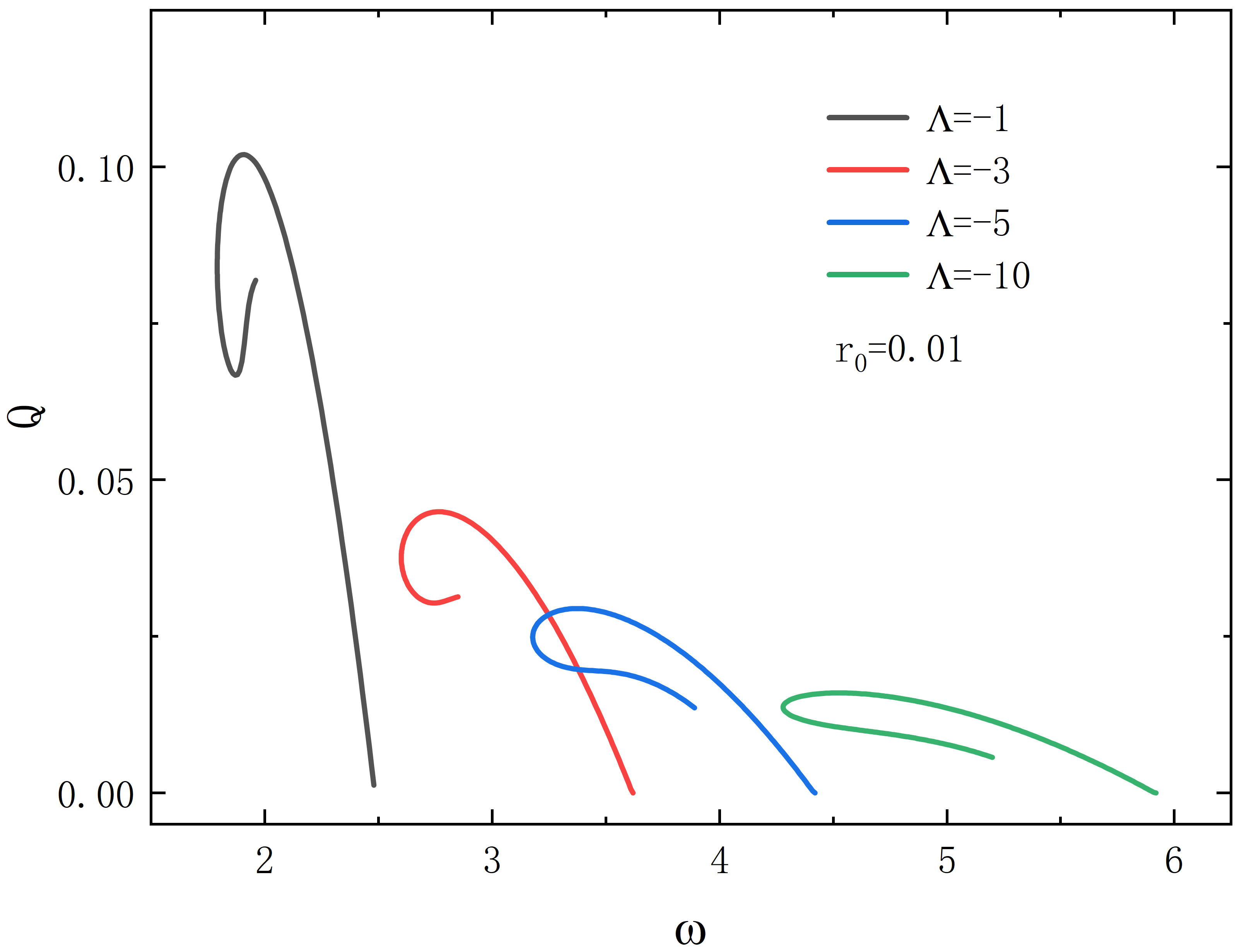}} 
\subfigure{\includegraphics[width=0.45\textwidth]{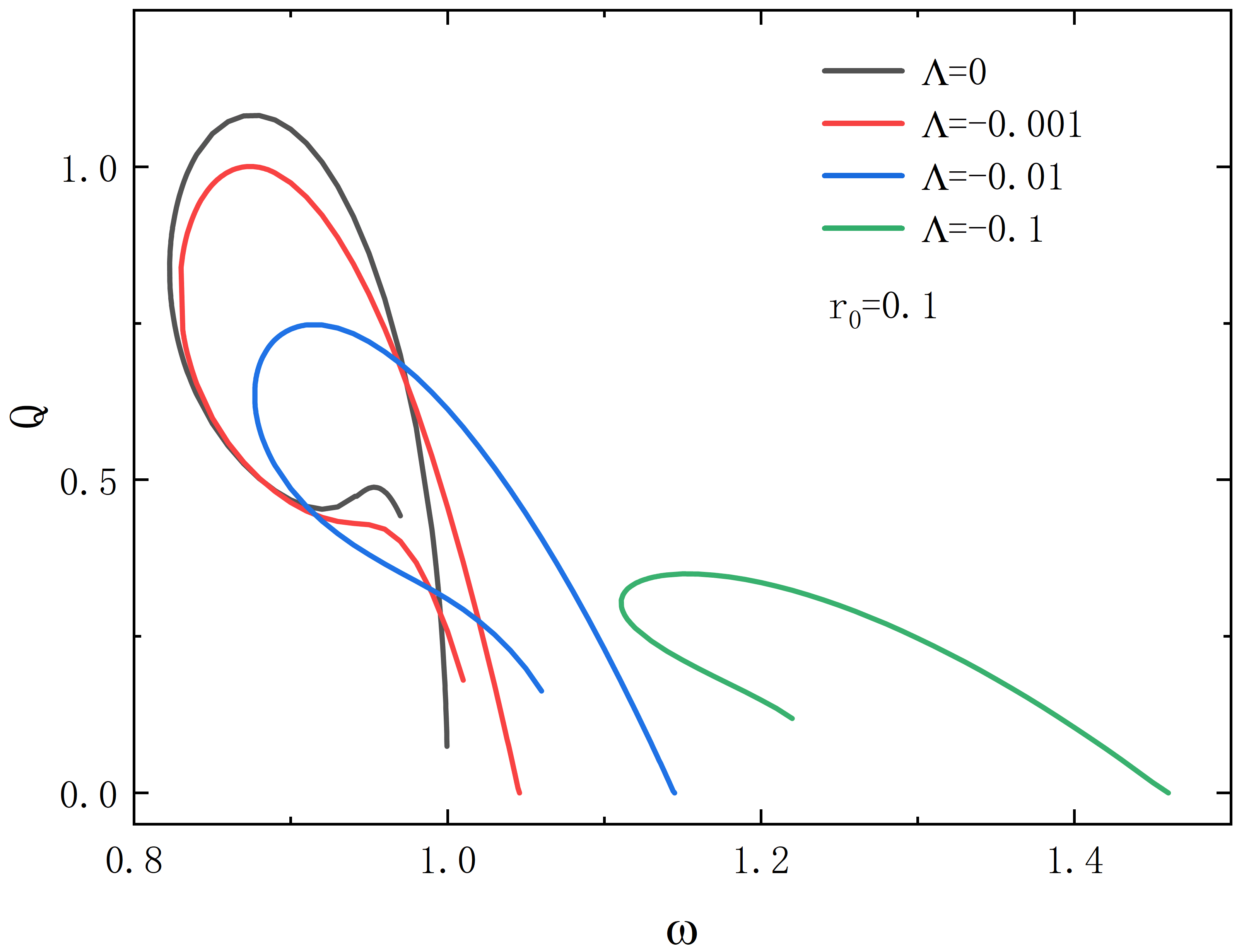}} 
\subfigure{\includegraphics[width=0.45\textwidth]{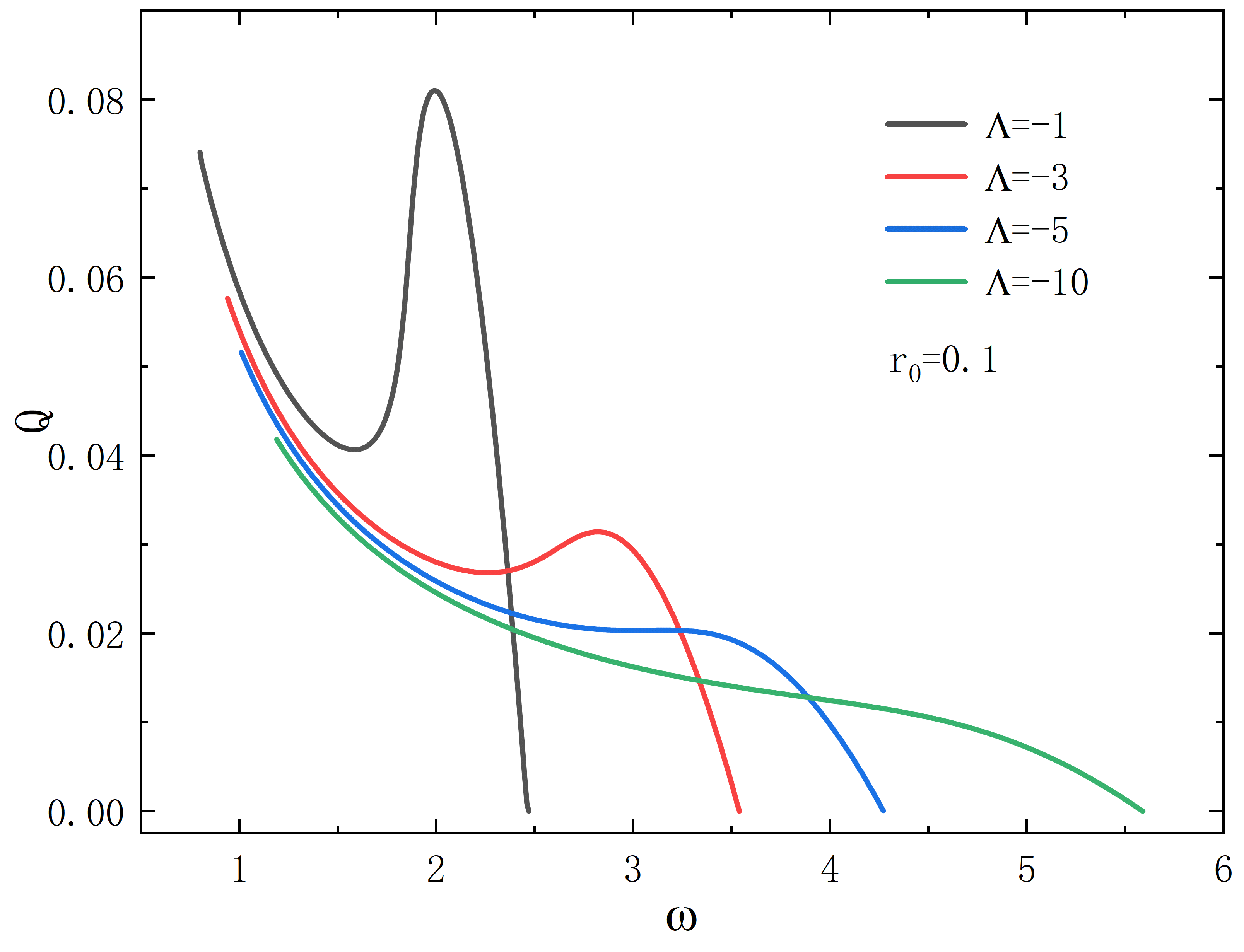}} 
\subfigure{\includegraphics[width=0.45\textwidth]{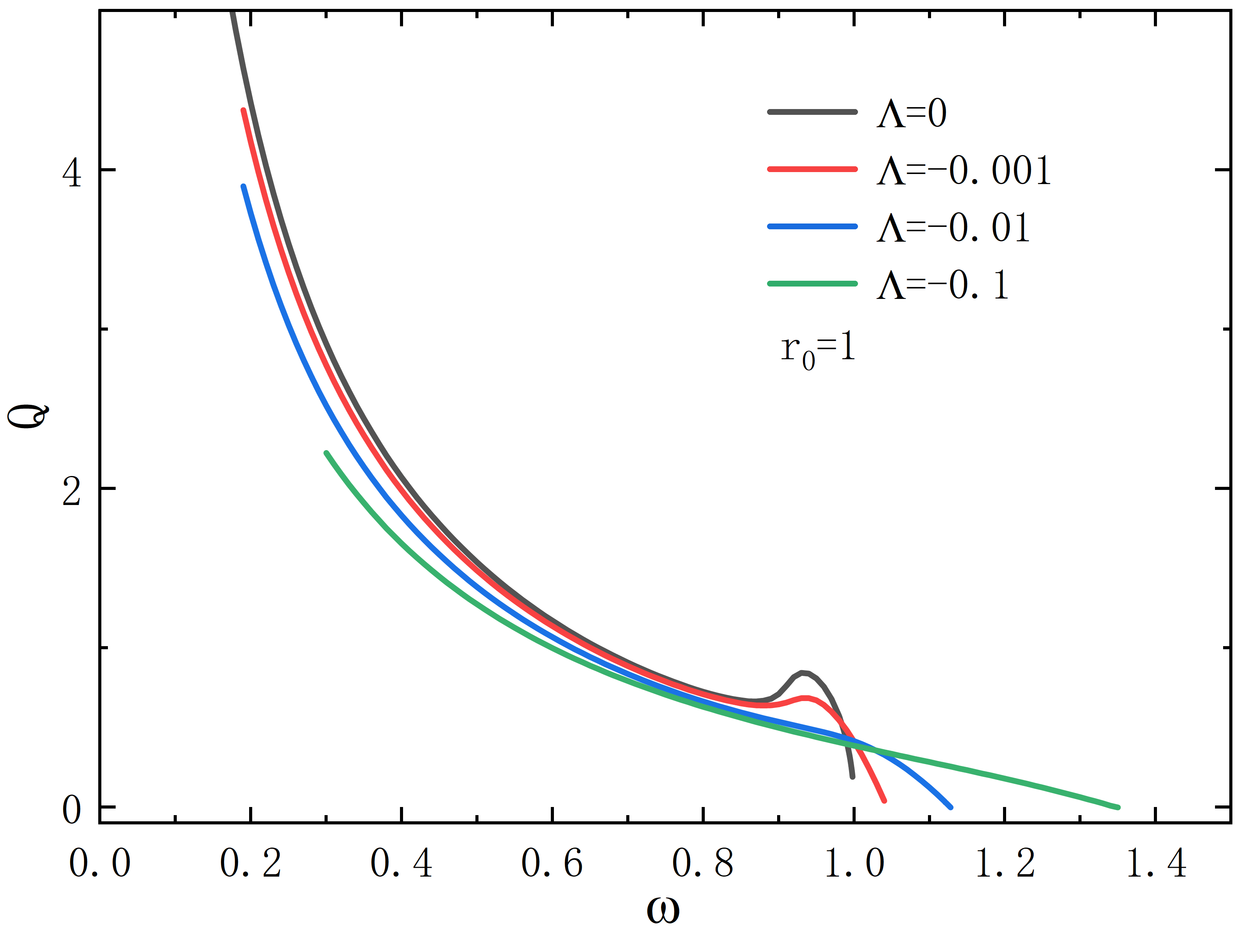}} 
\subfigure{\includegraphics[width=0.45\textwidth]{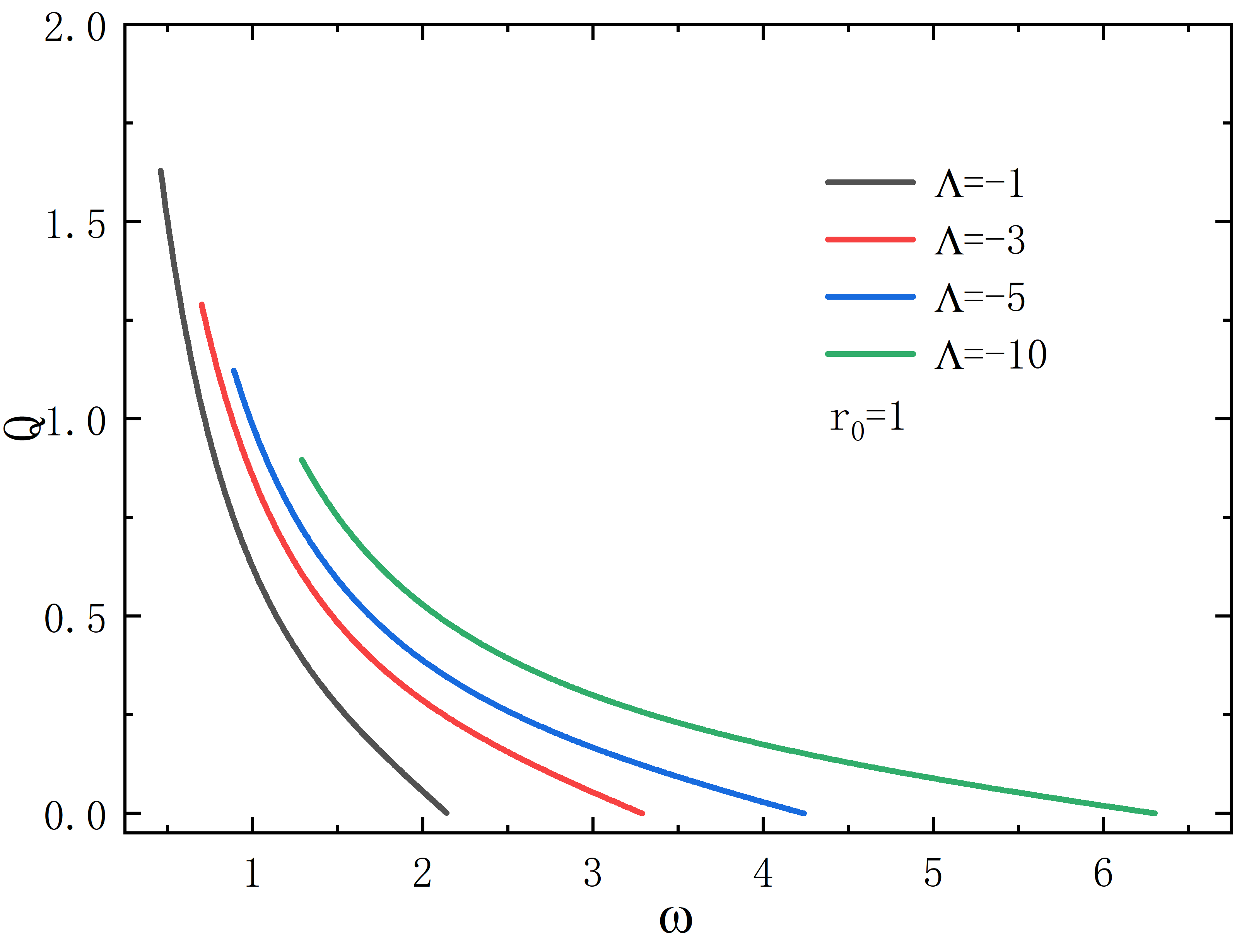}} 
\end{center} 
\caption{The Noether charge $Q$ as a function of frequency $\omega$, showing the effects of $\Lambda$ on the solution for different throat radii ($r_0 = 0.01, 0.1, 1$).} 
\label{phaseI1} 
\end{figure}

As shown in Figure \ref{phaseI1}, when $\Lambda=0$, the solution degenerates into a static, symmetric Ellis wormhole with a Proca field, where the numerical results are consistent with \cite{Su:2023zhh}. When $r_0$ is very small, the solution is constrained within a narrow range, and the curve takes the shape of the well-known spiral structure of a Proca star. However, as $r_0$ increases, the spiral structure gradually disappears. Our main focus is on the significant differences in the behavior of the Noether charge as the frequency varies for different values of $\Lambda$. For a smaller throat radius $r_0=0.01$, reducing $\Lambda$ decreases the number of branches in the curve, and also causing the spiral structure to gradually unfold, while the Noether charge decreases. A similar trend is observed for $r_0 = 0.1$. However, for a larger throat radius ($r_0 = 1$), the branching behavior disappears. Moreover, an interesting observation is that as $\Lambda$ decreases, the frequency range continuously expands and shifts to the right overall.

\begin{figure}
\begin{center} 
\subfigure{\includegraphics[width=0.45\textwidth]{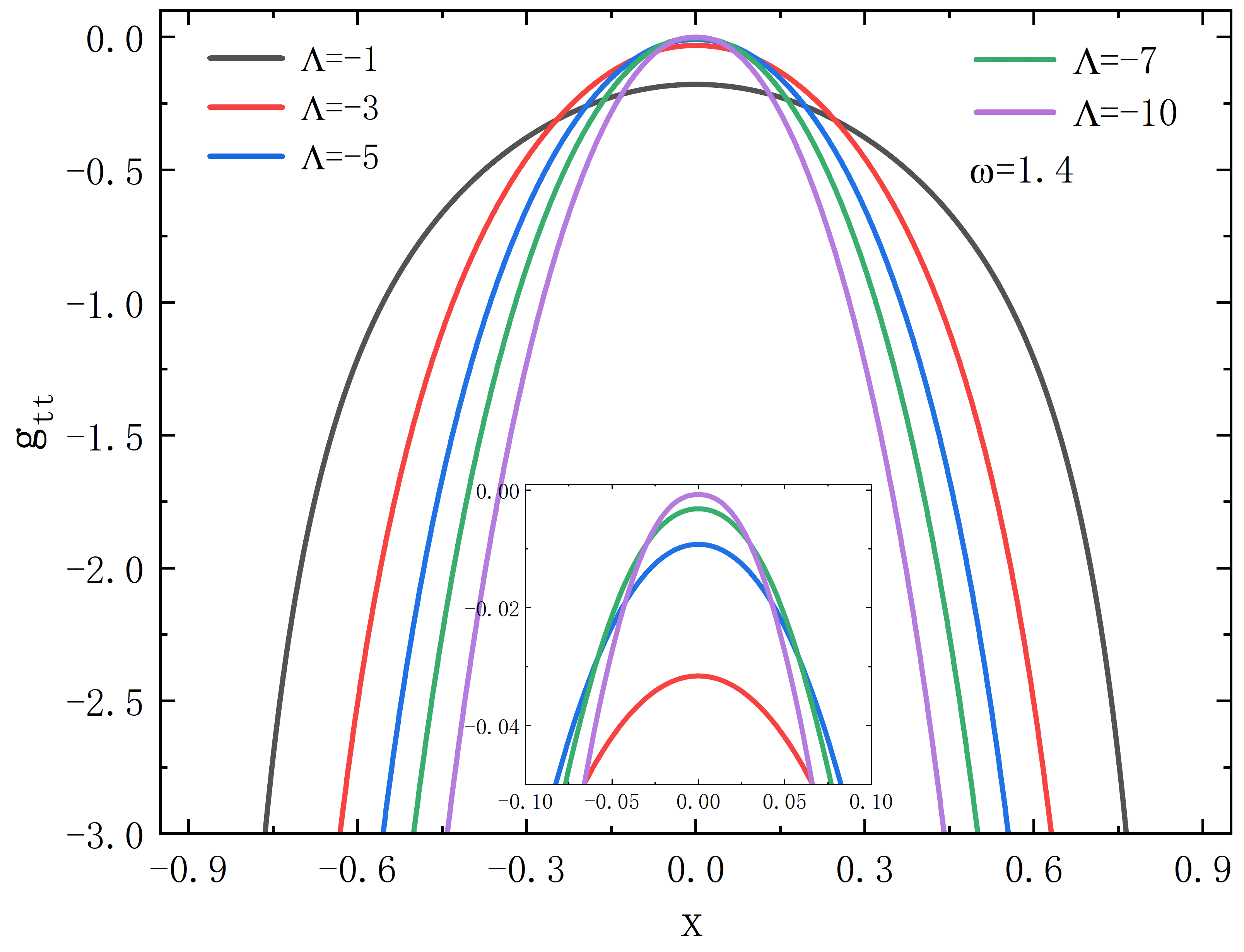}} 
\subfigure{\includegraphics[width=0.45\textwidth]{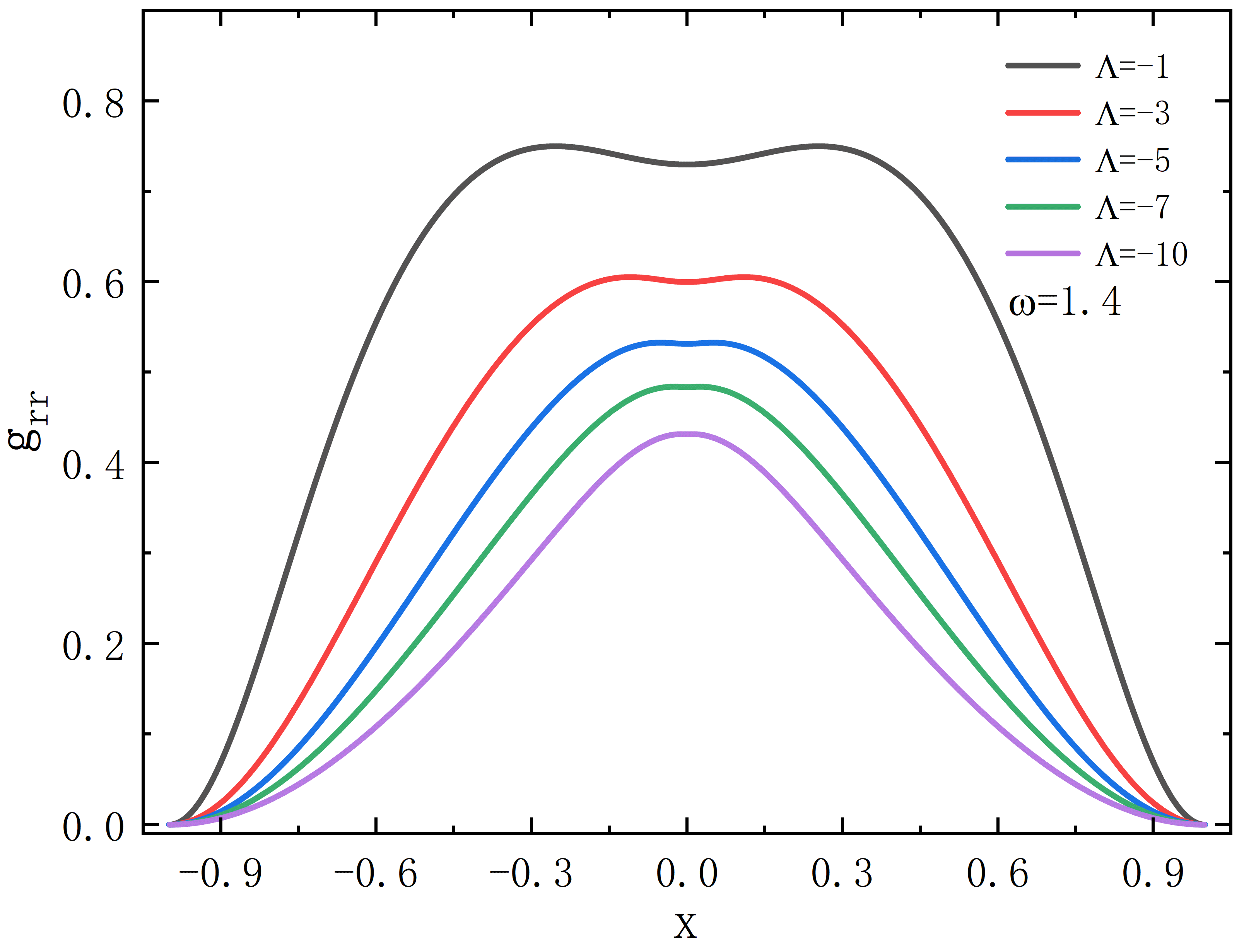}} 
\subfigure{\includegraphics[width=0.45\textwidth]{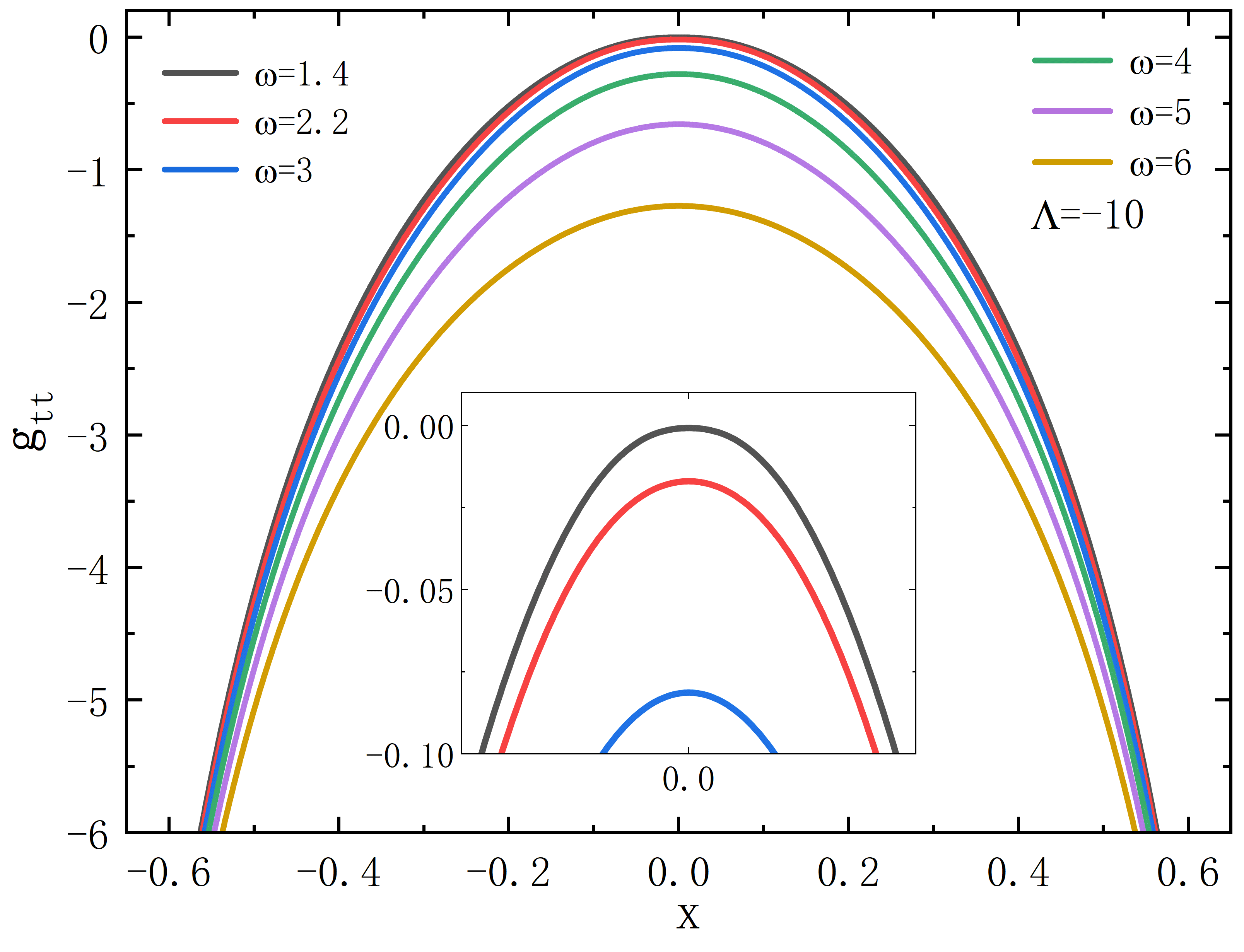}} 
\subfigure{\includegraphics[width=0.45\textwidth]{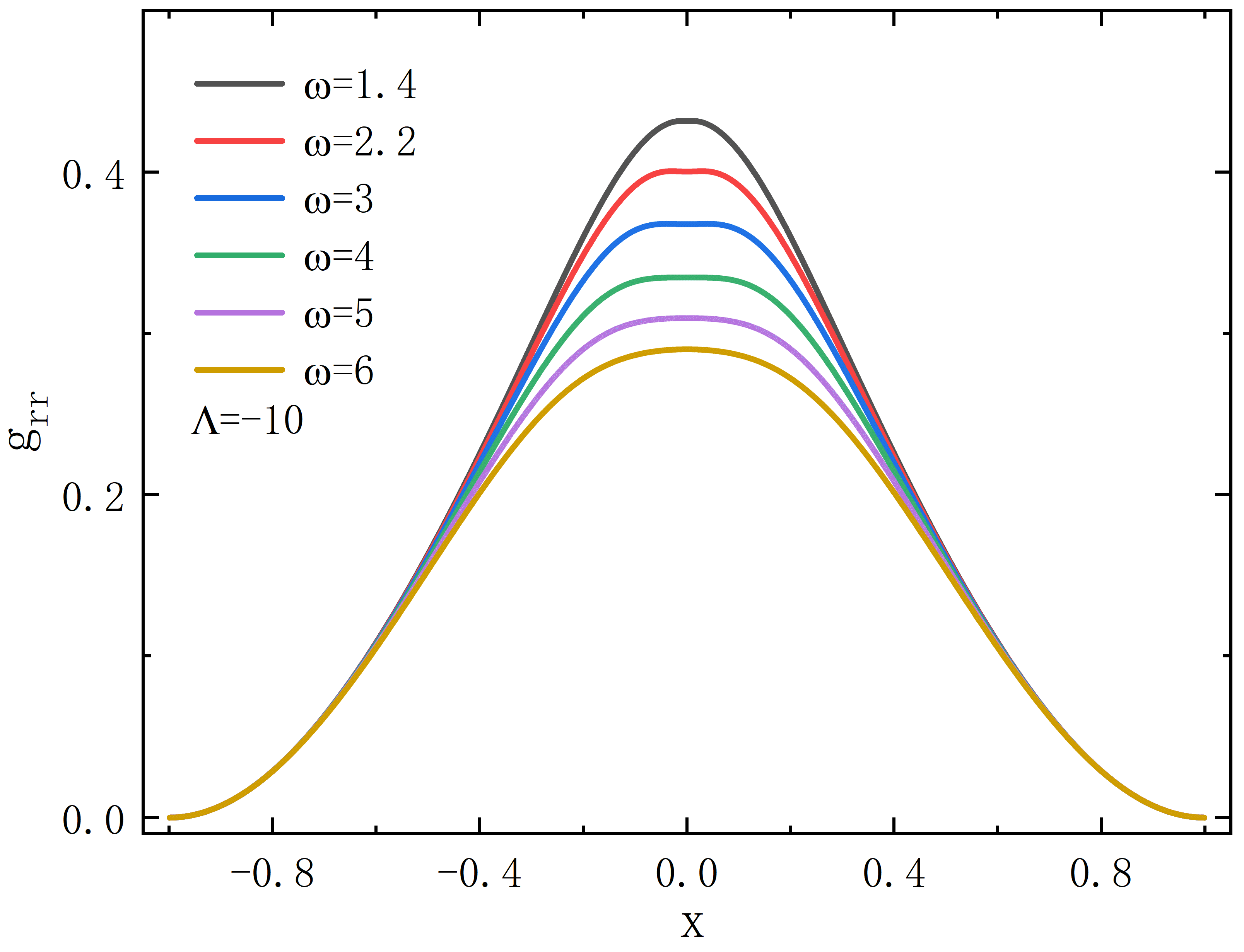}} 
\end{center} 
\caption{The metric functions $g_{tt}$ and $g_{rr}$ as functions of the radial coordinate $x$. The first row fixes $\omega = 1.4$ for different $\Lambda$ values, while the second row fixes $\Lambda = -10$ for different $\omega$. The throat radius $r_0 = 1$.} 
\label{phaseI2} 
\end{figure}

Further investigation of the variations of the metric functions $g_{tt}$ and $g_{rr}$ with the radial coordinate is shown in Figure \ref{phaseI2}. The first row shows the variations of the metric functions for different values of $\Lambda$ when $\omega = 1.4$ is fixed. As $\Lambda$ decreases, the value of $g_{tt}$ at $x=0$ approaches zero, suggesting that an approximate ``horizon" begins to form. At the same time, the peak of $g_{rr}$ moves closer to the throat on both sides and decreases. The second row shows the effect of varying $\omega$ when $\Lambda = -10$ is fixed. When $\omega$ decreases (approaching the left limit of the solution), the value of $g_{tt}$ at $x=0$ again approaches zero, suggesting that an horizon may have formed in the throat region.

\begin{figure}[H] 
\subfigure{\includegraphics[width=0.45\textwidth]{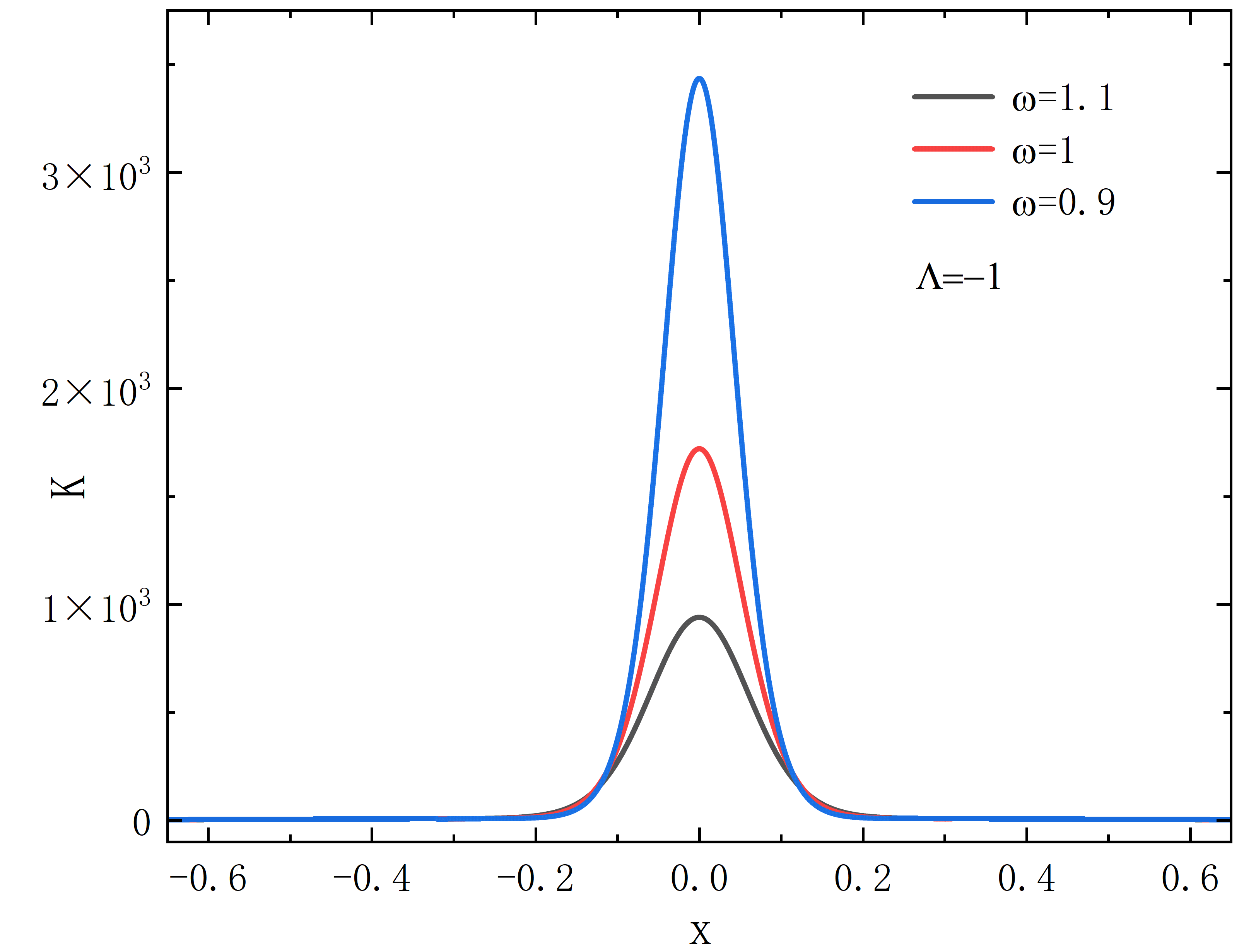}} 
\subfigure{\includegraphics[width=0.45\textwidth]{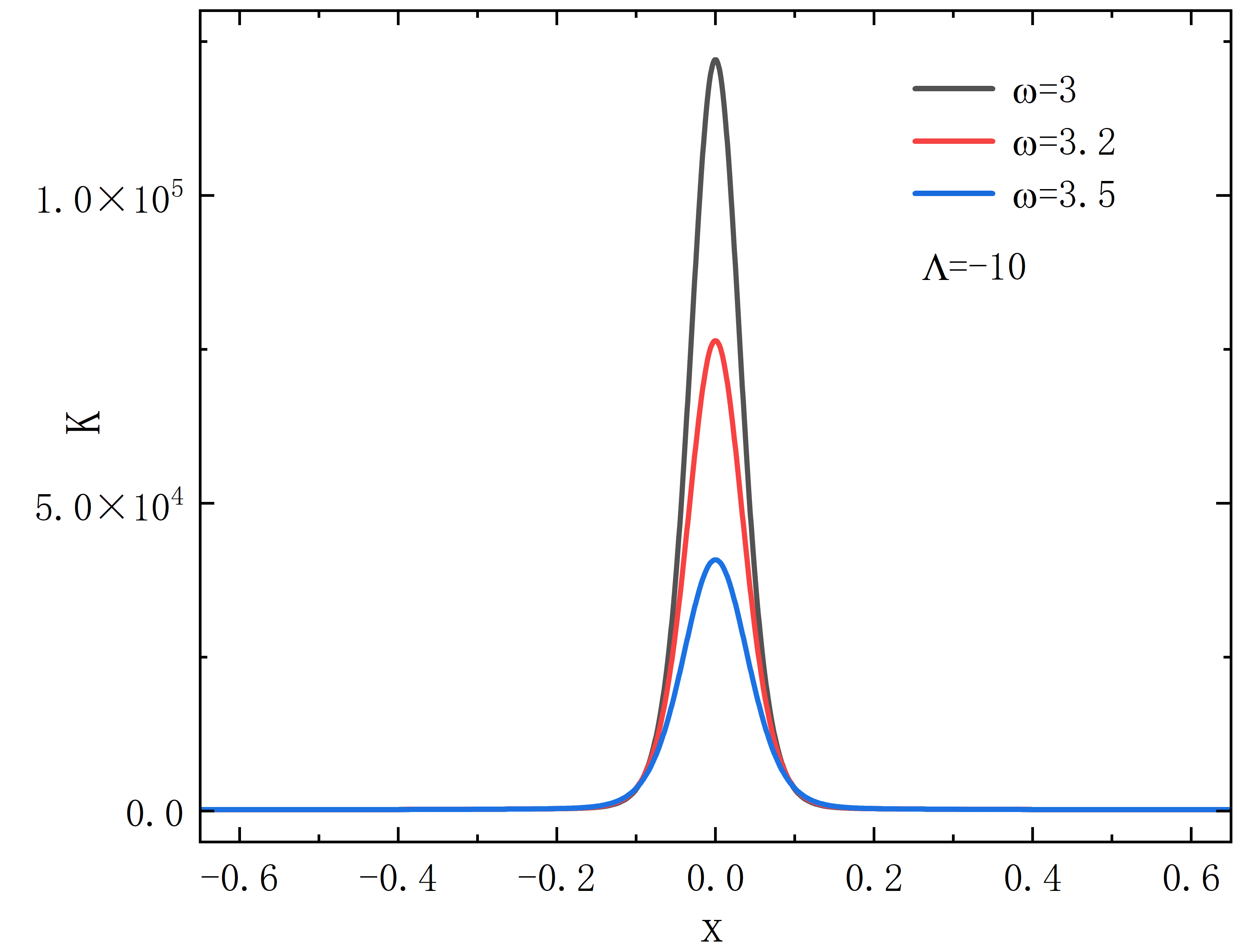}} 
\caption{The distribution of the Kretschmann scalar as a function of the radial coordinate $x$. The left image fixes $\Lambda = -1$, and the right image fixes $\Lambda = -10$. The throat radius is $r_0 = 1$.} 
\label{phaseI4}
\end{figure}

To better describe the behavior of $g_{tt}$ near the throat, Tables \ref{tab:tI1} and \ref{tab:tI2} show the minimum values of $g_{tt}$ at $x = 0$ for different conditions of $\Lambda$ and $\omega$. When either $\Lambda$ or $\omega$ becomes sufficiently small, the minimum value of $g_{tt}$ approaches $10^{-4}$, signaling the formation of an horizon. To further explore the spacetime geometry, we calculate the Kretschmann scalar $K = R_{\mu\nu\alpha\beta} R^{\mu\nu\alpha\beta}$. As shown in Figure \ref{phaseI4}, the Kretschmann scalar increases dramatically as $\omega$ decreases. The peak value of the Kretschmann scalar indicates the presence of a curvature singularity.

\begin{table}[H]
\centering
\begin{tabular}{|c||c|c|c|}
\hline
$\Lambda$ & $-3 \, (\omega=1.4)$ & $-5 \, (\omega=1.4)$ & $-10 \, (\omega=1.4)$ \\
\hline
$g_{tt}(0)$ & $-0.032$ & $-0.0091$ & $-0.00073$ \\
\hline
\end{tabular}
\caption{The minimum value of $g_{tt}$ for $\omega = 1.4$ at different values of $\Lambda = -3, -5, -10$. The throat radius is $r_0 = 1$.}
\label{tab:tI1}
\end{table}

\begin{table}[H]
\centering
\begin{tabular}{|c||c|c|c|}
\hline
$\omega$ & $1.6 \, (\Lambda=-10)$ & $1.5 \, (\Lambda=-10)$ & $1.4 \, (\Lambda=-10)$ \\
\hline
$g_{tt}(0)$ & $-0.0022$ & $-0.0013$ & $-0.00073$ \\
\hline
\end{tabular}
\caption{The minimum value of $g_{tt}$ for $\Lambda = -10$ at different values of $\omega = 1.6, 1.5, 1.4$. The throat radius is $r_0 = 1$.}
\label{tab:tI2}
\end{table}

\begin{figure}[H]
\subfigure{\includegraphics[width=0.45\textwidth]{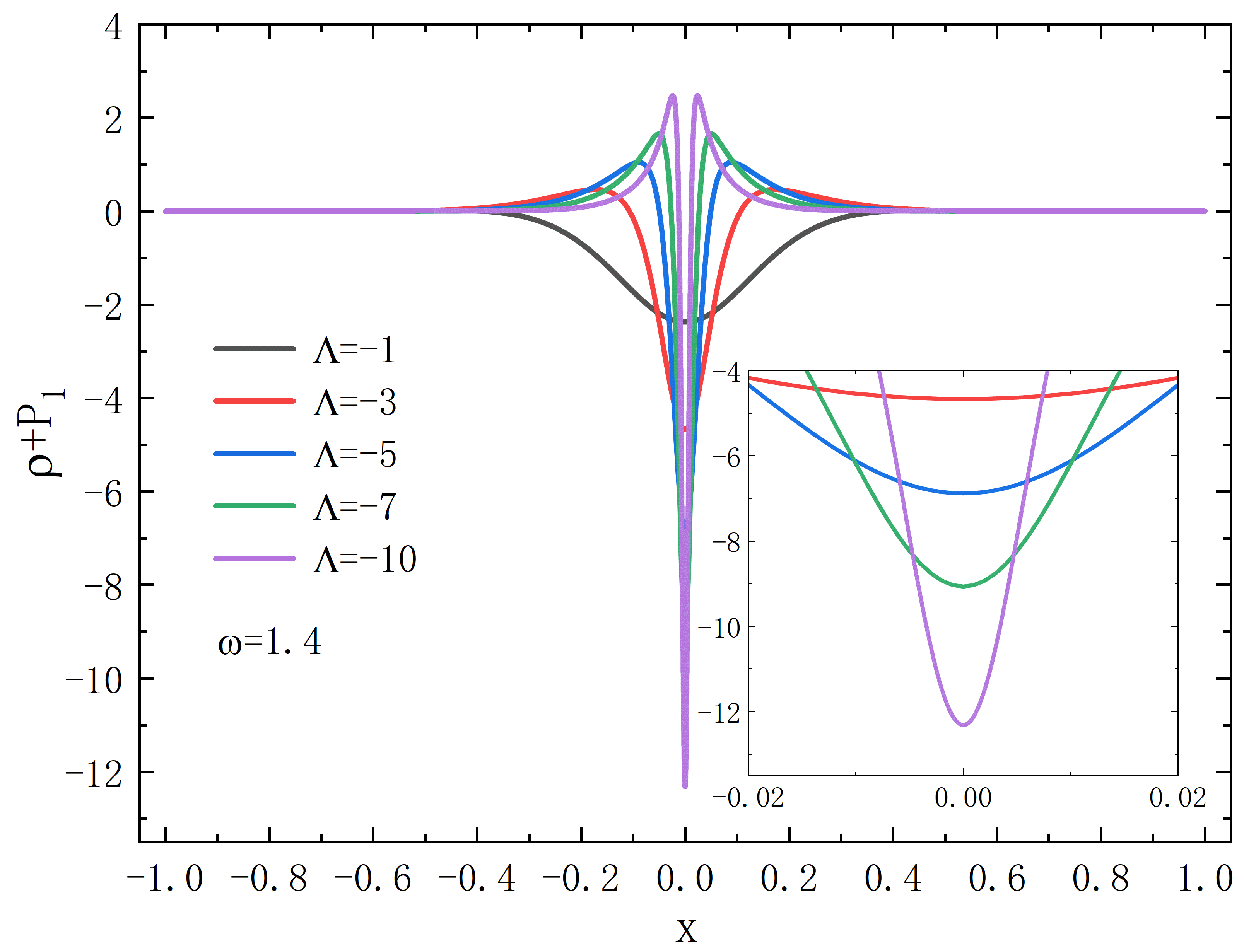}} \subfigure{\includegraphics[width=0.45\textwidth]{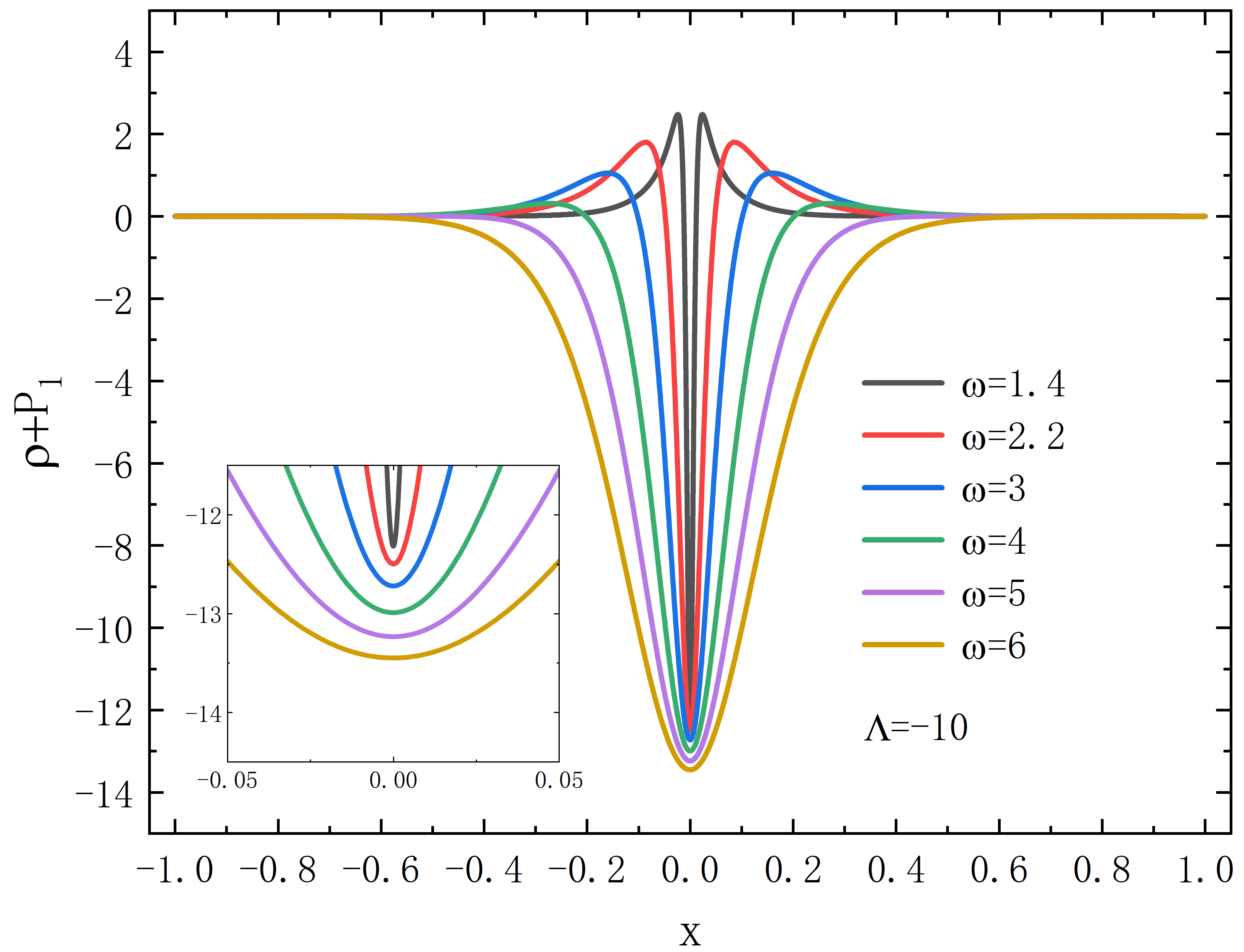}} 
\caption{The violation of NEC under different conditions of $\Lambda$ or $\omega$. The throat radius is $r_0 = 1$.} 
\label{phaseI5} 
\end{figure}

The introduction of a phantom field in the Ellis wormhole violates the null energy condition (NEC). To investigate the effects of coupling with the Proca field, we analyze the variation of the energy density $\rho$ and radial pressure $p_1$. As shown in Figure \ref{phaseI5}, we vary $\Lambda$ while fixing $\omega$, and $\omega$ while fixing $\Lambda$. The left panel shows that as $\Lambda$ decreases, the NEC violation in the throat region becomes more pronounced, and the curve sharpens. The right panel reveals that as $\omega$ decreases, the flat region of the curve shrinks, while the minimum value at the throat increases, though the change is not significant. However, in the local regions on either side of the throat, $\rho + p_1$ becomes positive, indicating that NEC is not violated in these regions, likely due to the positive energy density contribution from the Proca field.

\subsection{Symmetric Case II}

\begin{figure} 
\begin{center} 
\subfigure{\includegraphics[width=0.45\textwidth]{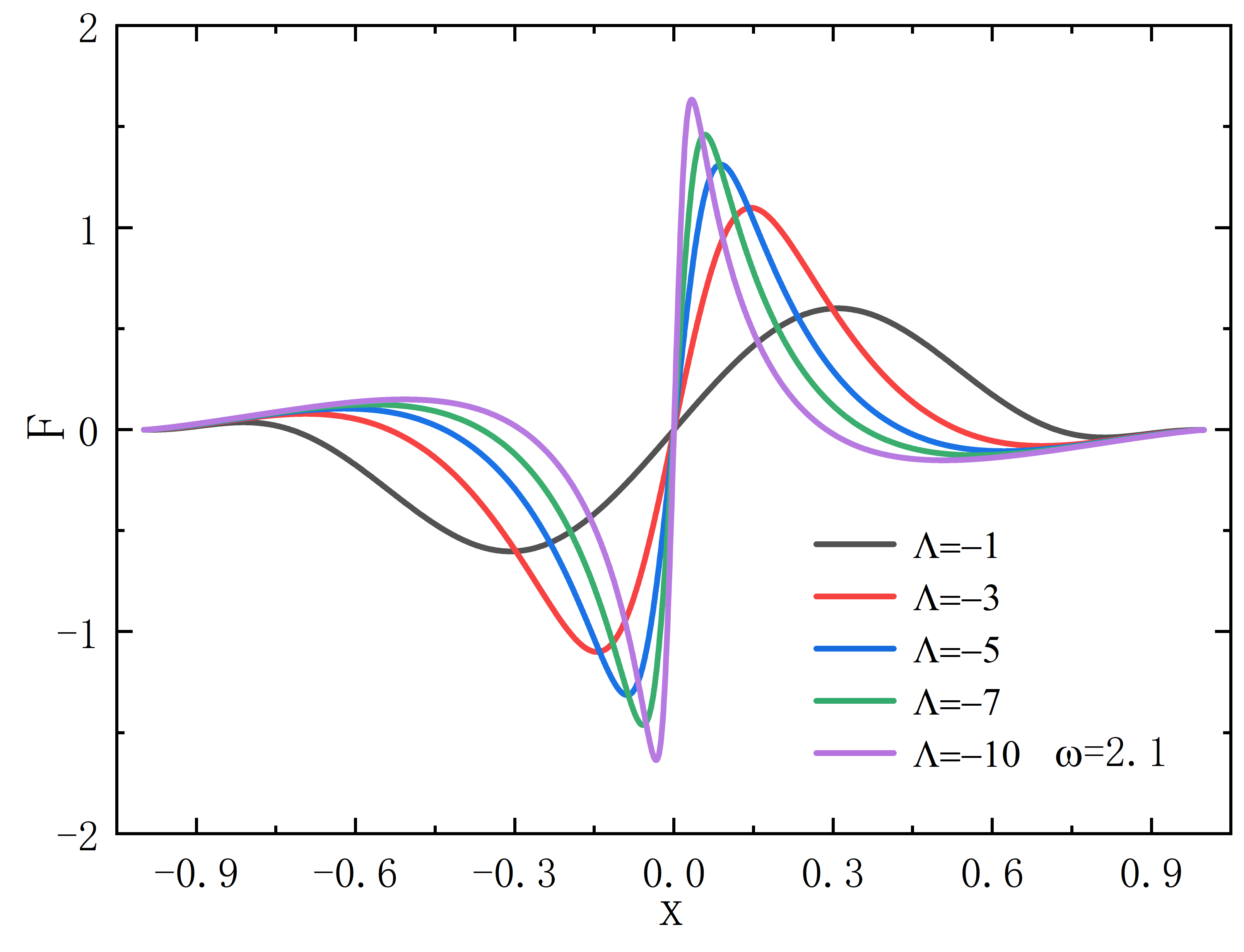}} 
\subfigure{\includegraphics[width=0.45\textwidth]{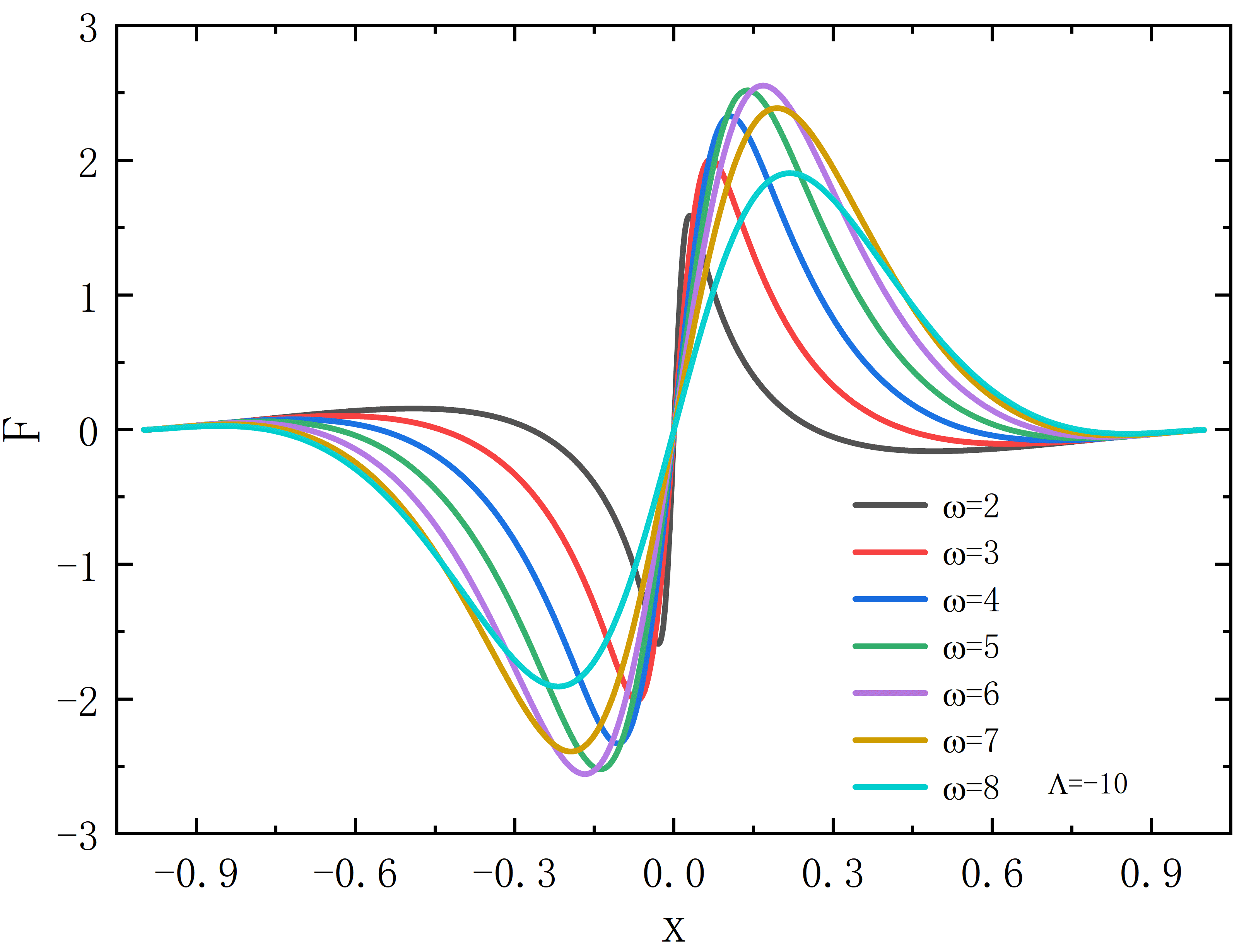}} 
\subfigure{\includegraphics[width=0.45\textwidth]{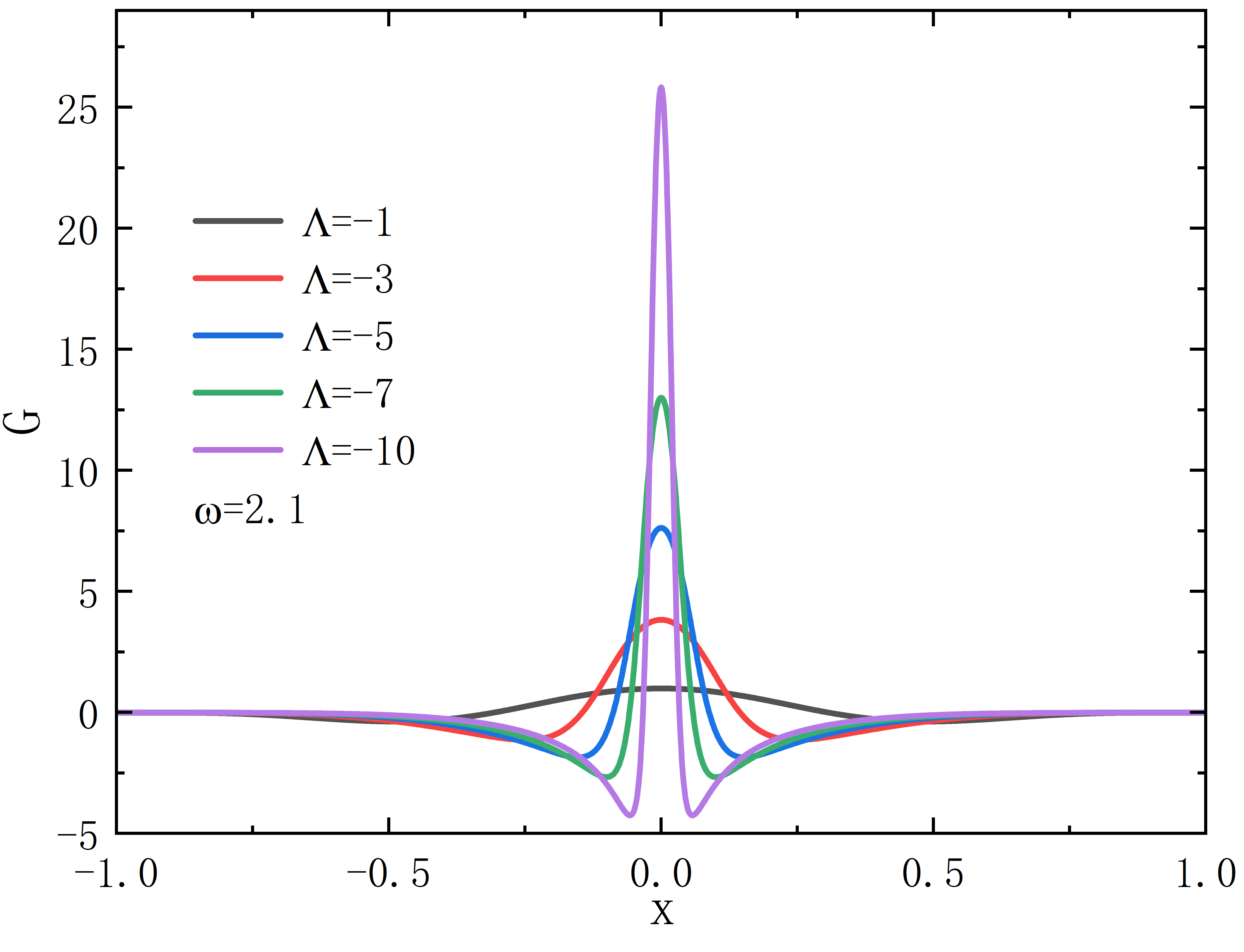}} 
\subfigure{\includegraphics[width=0.45\textwidth]{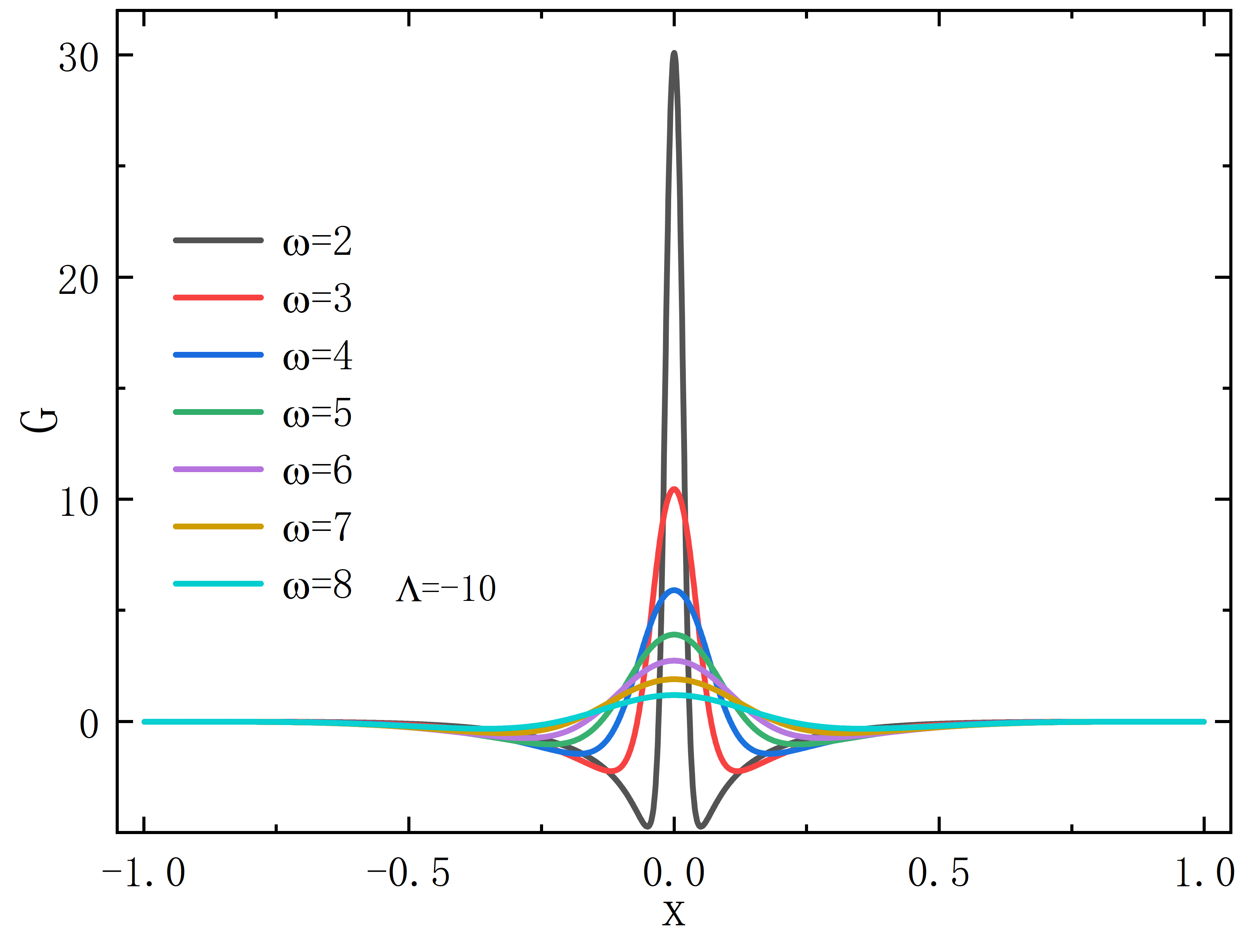}} 
\end{center} 
\caption{The matter fields $F(r)$ and $G(r)$ as functions of the radial coordinate $x$. The left panels fix $\omega$, while the right panels fix $\Lambda$. The throat radius $r_0 = 1$.} 
\label{phaseII3} 
\end{figure}

For the second class of symmetric solutions, we separately fixed the frequency $\omega = 2.1$ and the cosmological constant $\Lambda = -10$ to explore the effects of $\Lambda$ or $\omega$ on the matter fields, as shown in Figure \ref{phaseII3}. When $\omega$ is fixed, the field function $F$ is antisymmetric about $x = 0$, and as $\Lambda$ decreases, its maximum value gradually increases, with the extremum point moving closer to the origin. The field function $G$, on the other hand, is symmetric about $x = 0$, reaches its maximum at the origin, and the maximum value of $G$ increases as $\Lambda$ decreases. When $\Lambda$ is fixed, as $\omega$ decreases, the slope of the field function $F$ at $x = 0$ gradually increases, its maximum value first increases and then decreases, and the extremum point continuously moves closer to the origin. Meanwhile, the maximum value of $G$ continues to increase as $\omega$ decreases.

\begin{figure}[H]
\begin{center} 
\subfigure{\includegraphics[width=0.45\textwidth]{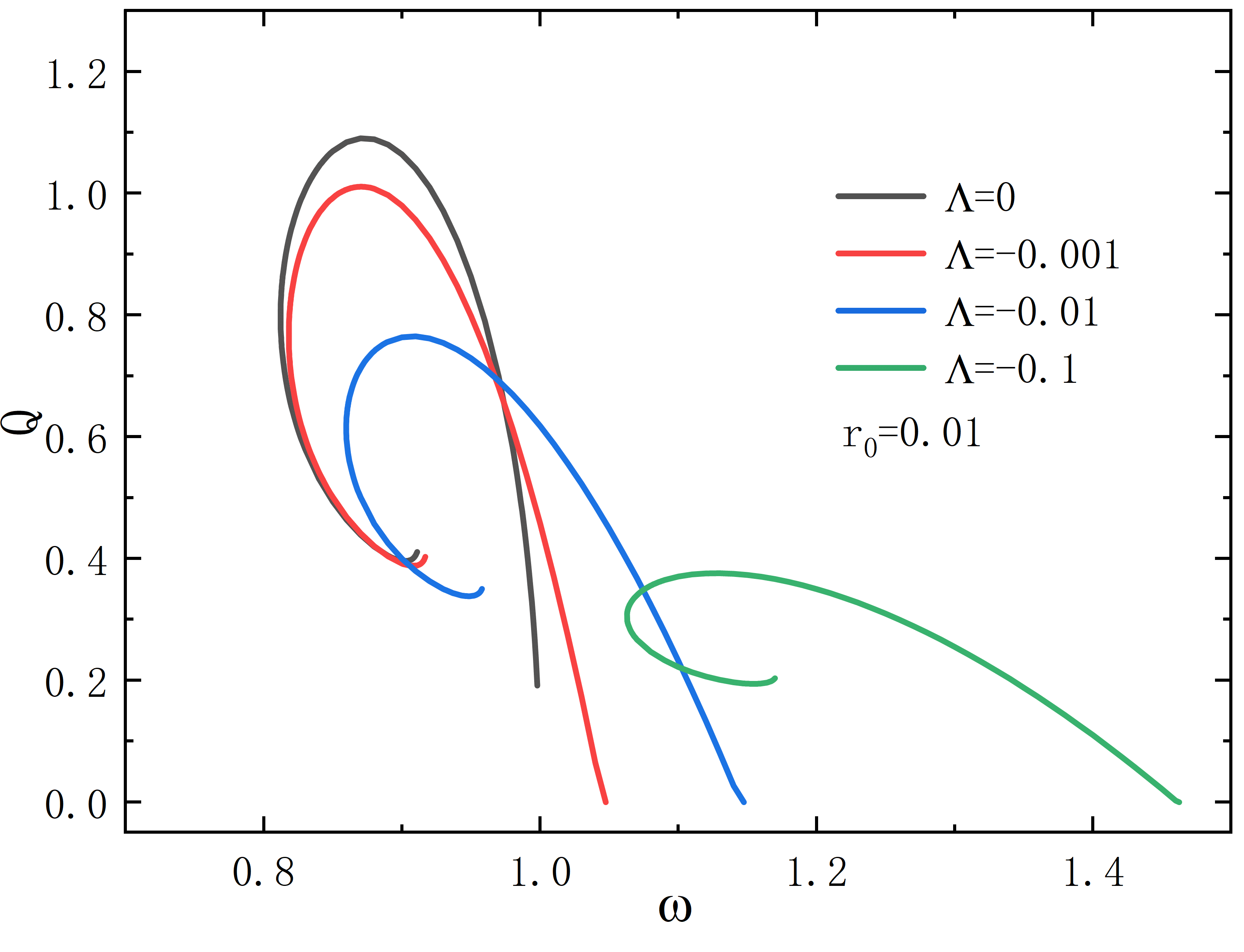}} 
\subfigure{\includegraphics[width=0.45\textwidth]{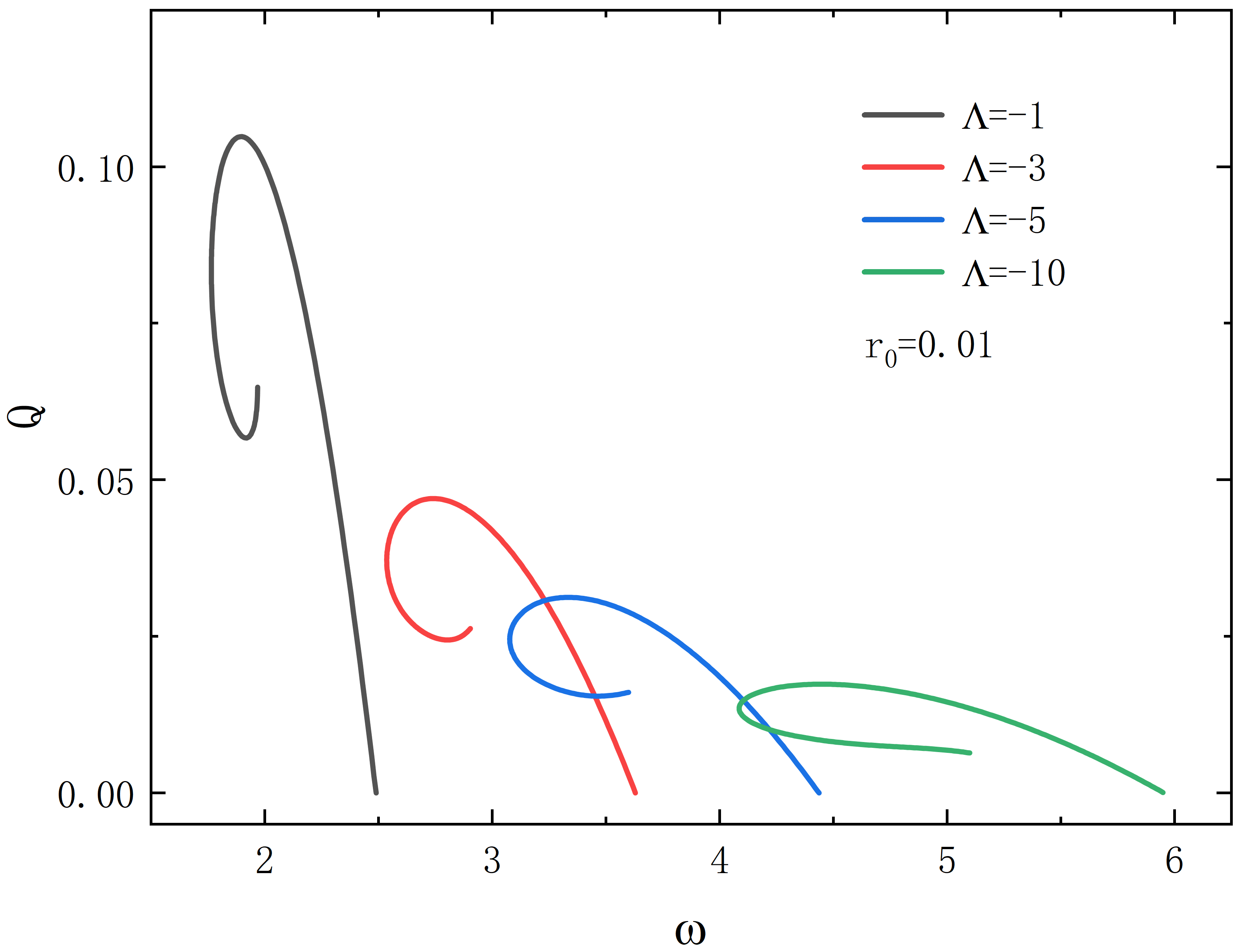}} 
\subfigure{\includegraphics[width=0.45\textwidth]{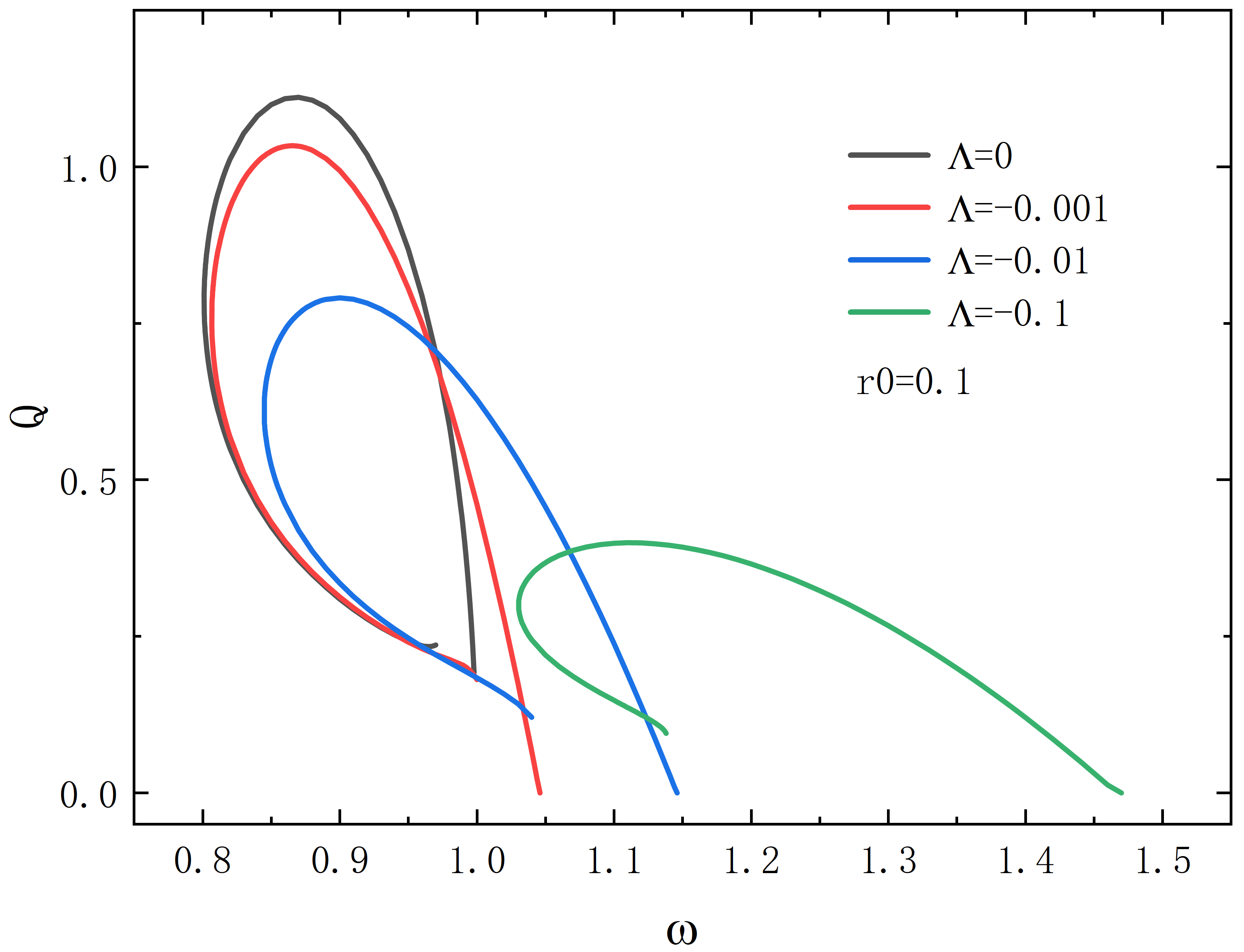}} 
\subfigure{\includegraphics[width=0.45\textwidth]{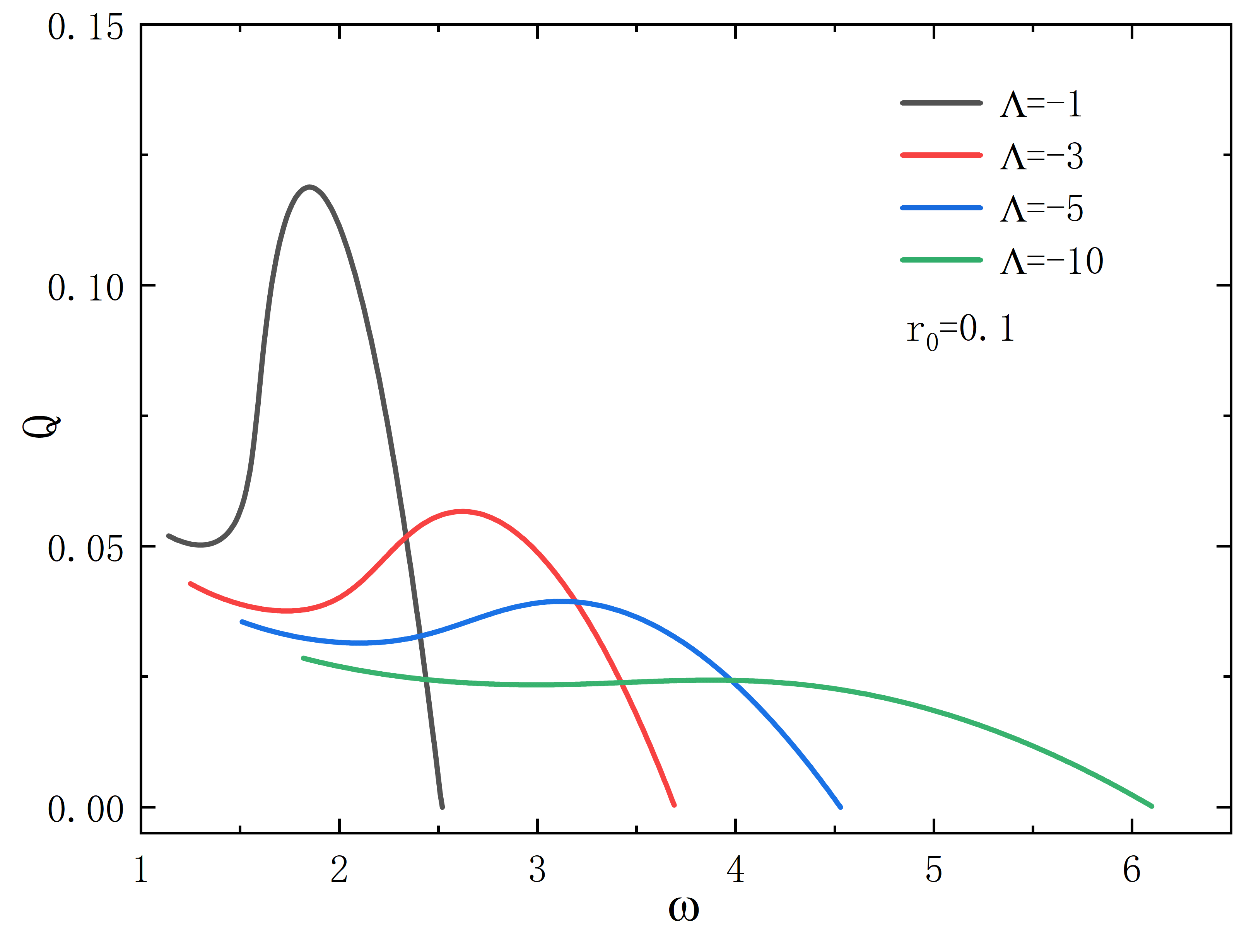}} 
\subfigure{\includegraphics[width=0.45\textwidth]{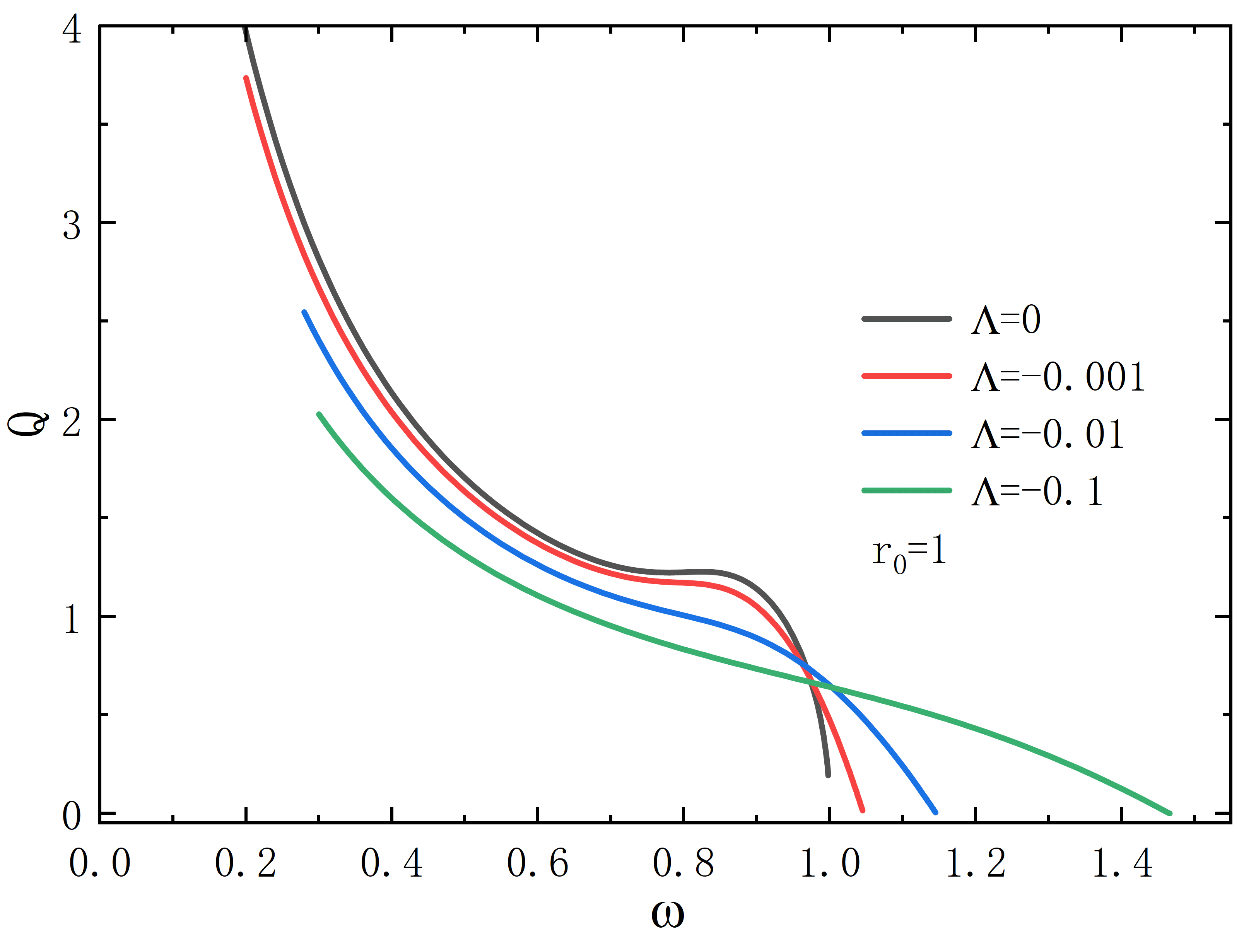}} 
\subfigure{\includegraphics[width=0.45\textwidth]{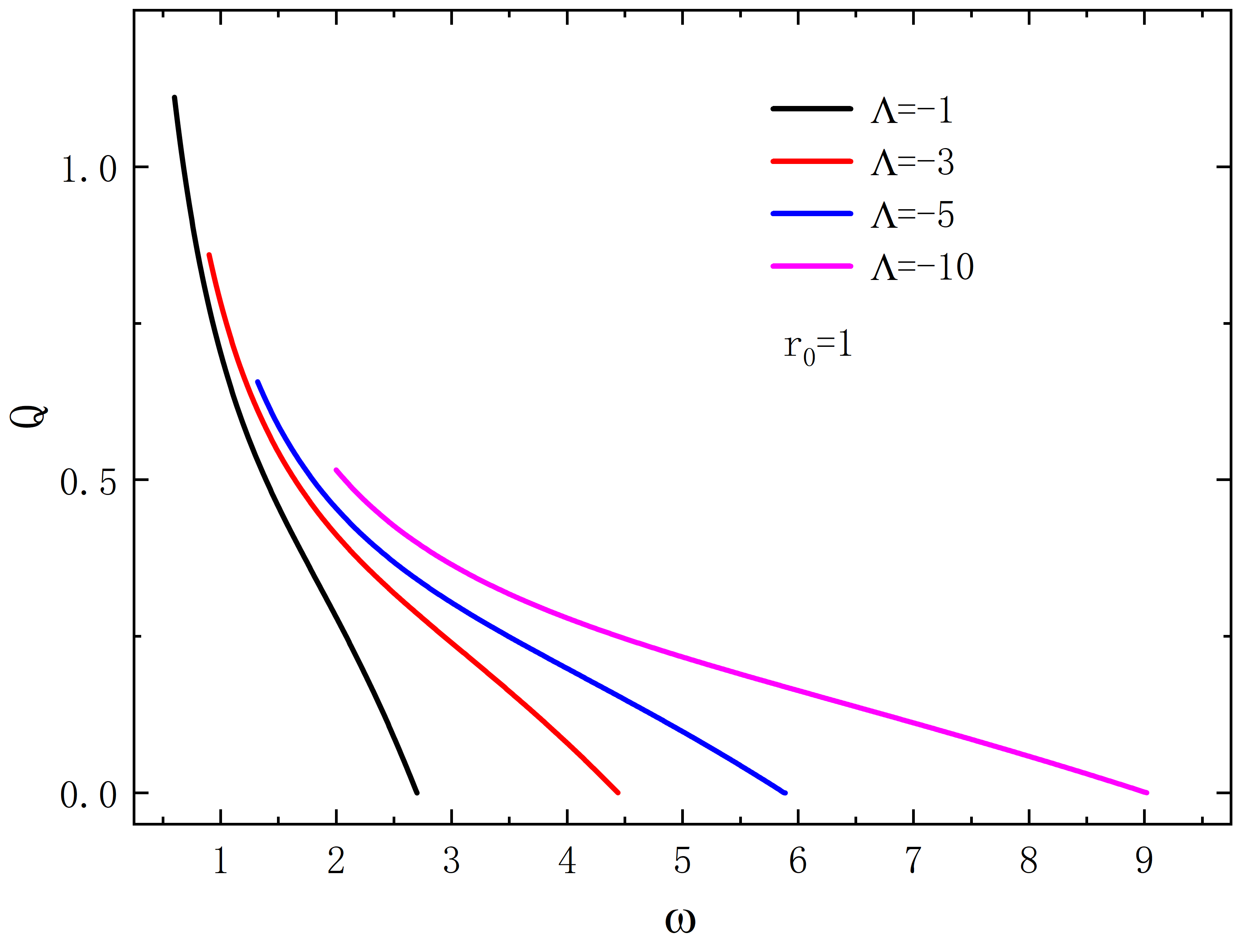}} 
\end{center} 
\caption{The Noether charge $Q$ as a function of frequency $\omega$, showing the effects of $\Lambda$ on the solution for different throat radii ($r_0 = 0.01, 0.1, 1$).} 
\label{phaseII1} 
\end{figure}

Next, we continued to investigate the relationship between the Noether charge $Q$ and the frequency $\omega$ under different cosmological constants $\Lambda$, dividing the throat radius $r_0$ into three groups (from small to large) for analysis. Figure \ref{phaseII1} shows the trend of Noether charge $Q$ as a function of frequency $\omega$ for a fixed throat radius. When $\Lambda$ is small (e.g., $\Lambda = -0.001$), the relationship curve between Noether charge $Q$ and frequency $\omega$ is largely consistent with the behavior in asymptotically flat spacetime ($\Lambda = 0$). However, as $\Lambda$ decreases further, the value of Noether charge $Q$ gradually decreases, and the spiral structure of the curve gradually unfolds. This characteristic is more pronounced for smaller throat radii (such as $r_0 = 0.1$). Additionally, as $\Lambda$ decreases, the frequency range further expands, and the entire curve shifts to the right.

\begin{figure}
\begin{center} 
\subfigure{\includegraphics[width=0.45\textwidth]{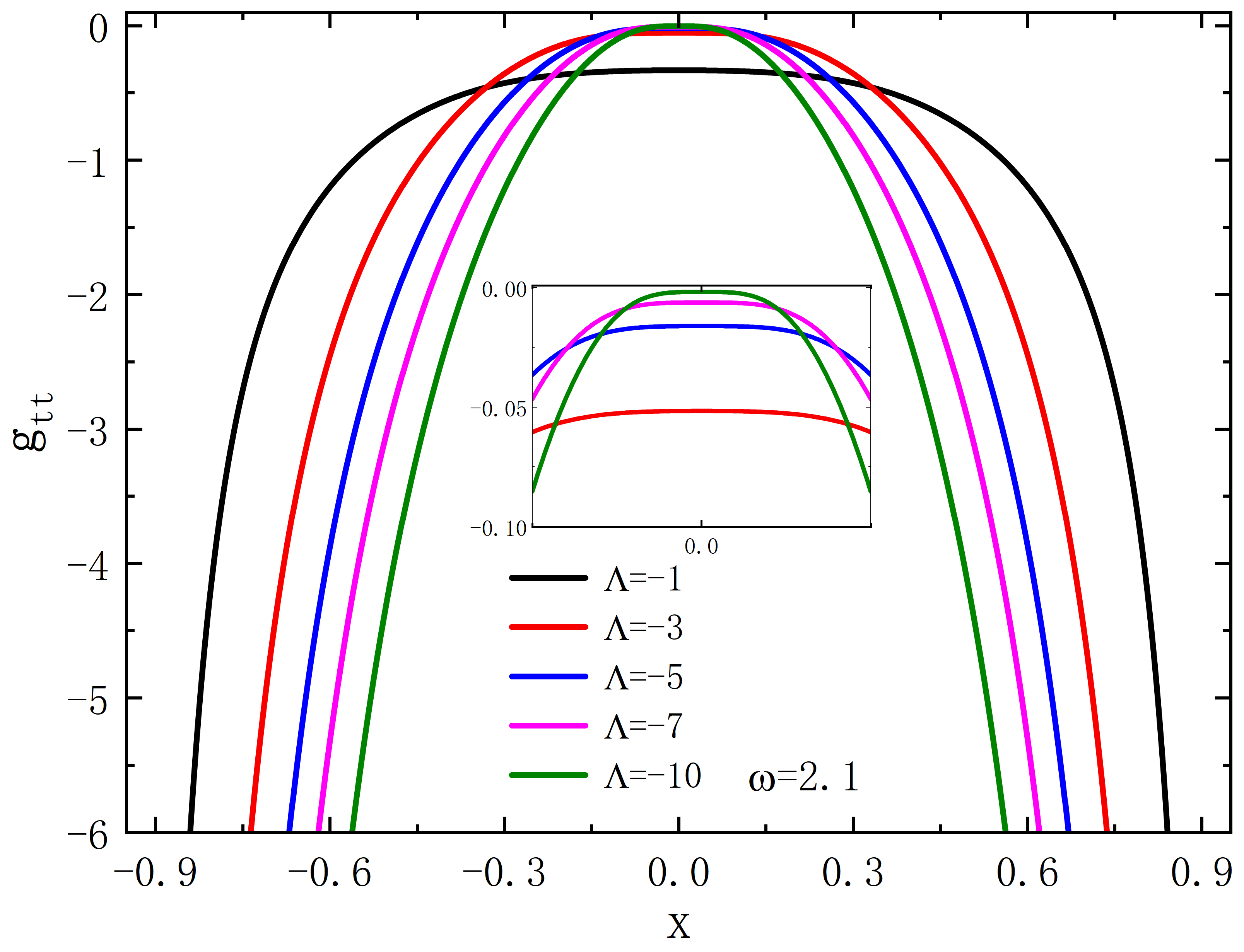}} 
\subfigure{\includegraphics[width=0.45\textwidth]{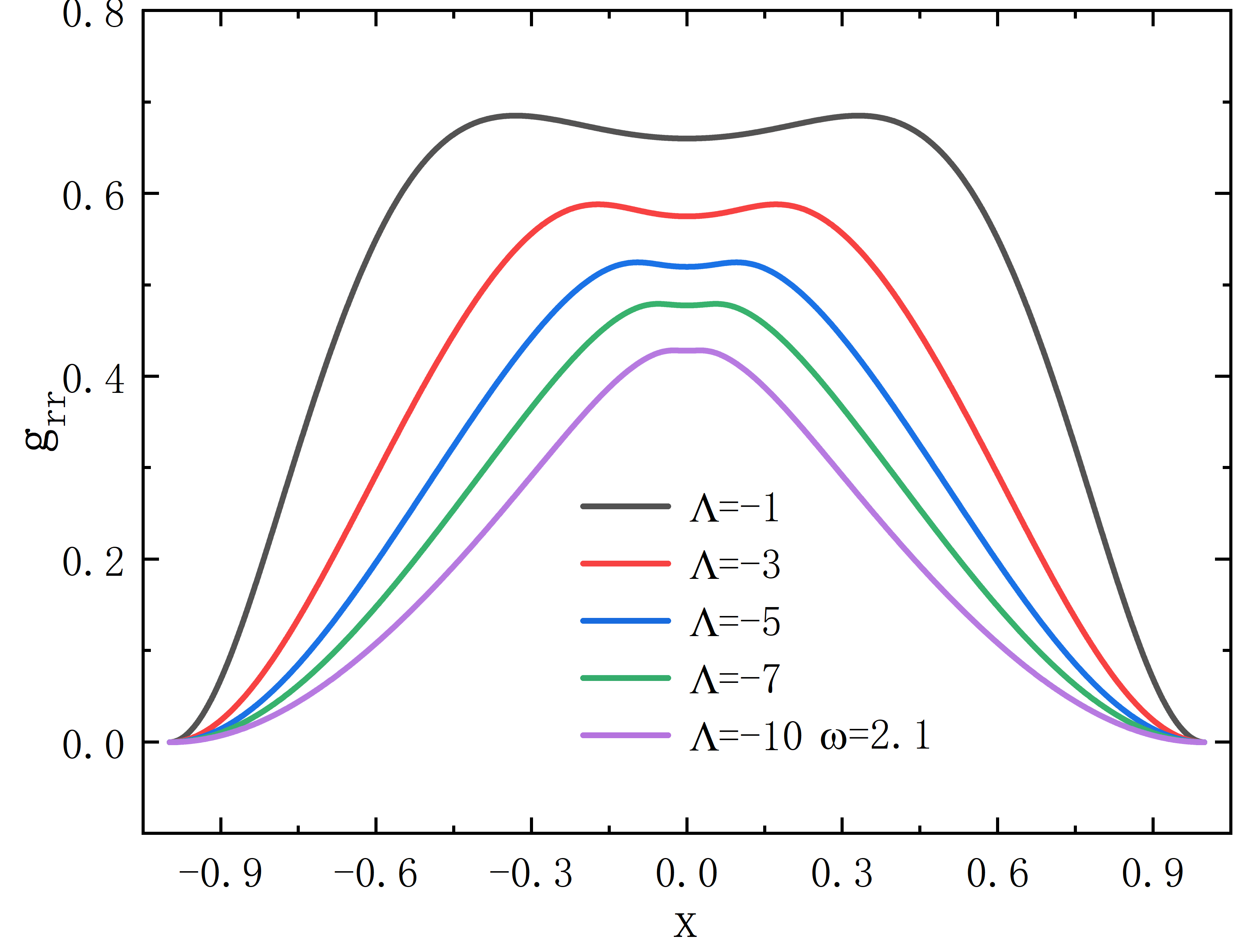}} 
\subfigure{\includegraphics[width=0.45\textwidth]{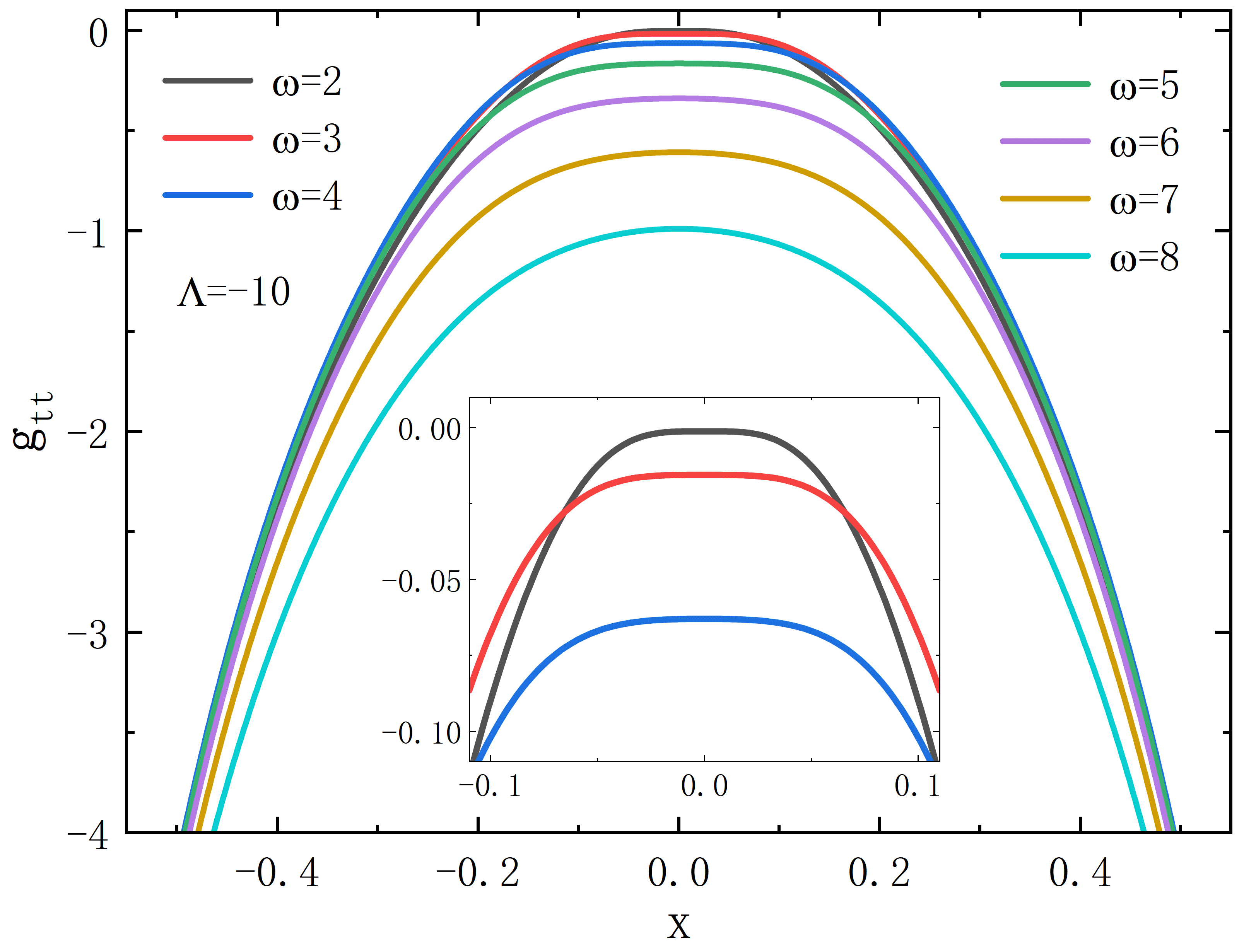}} 
\subfigure{\includegraphics[width=0.45\textwidth]{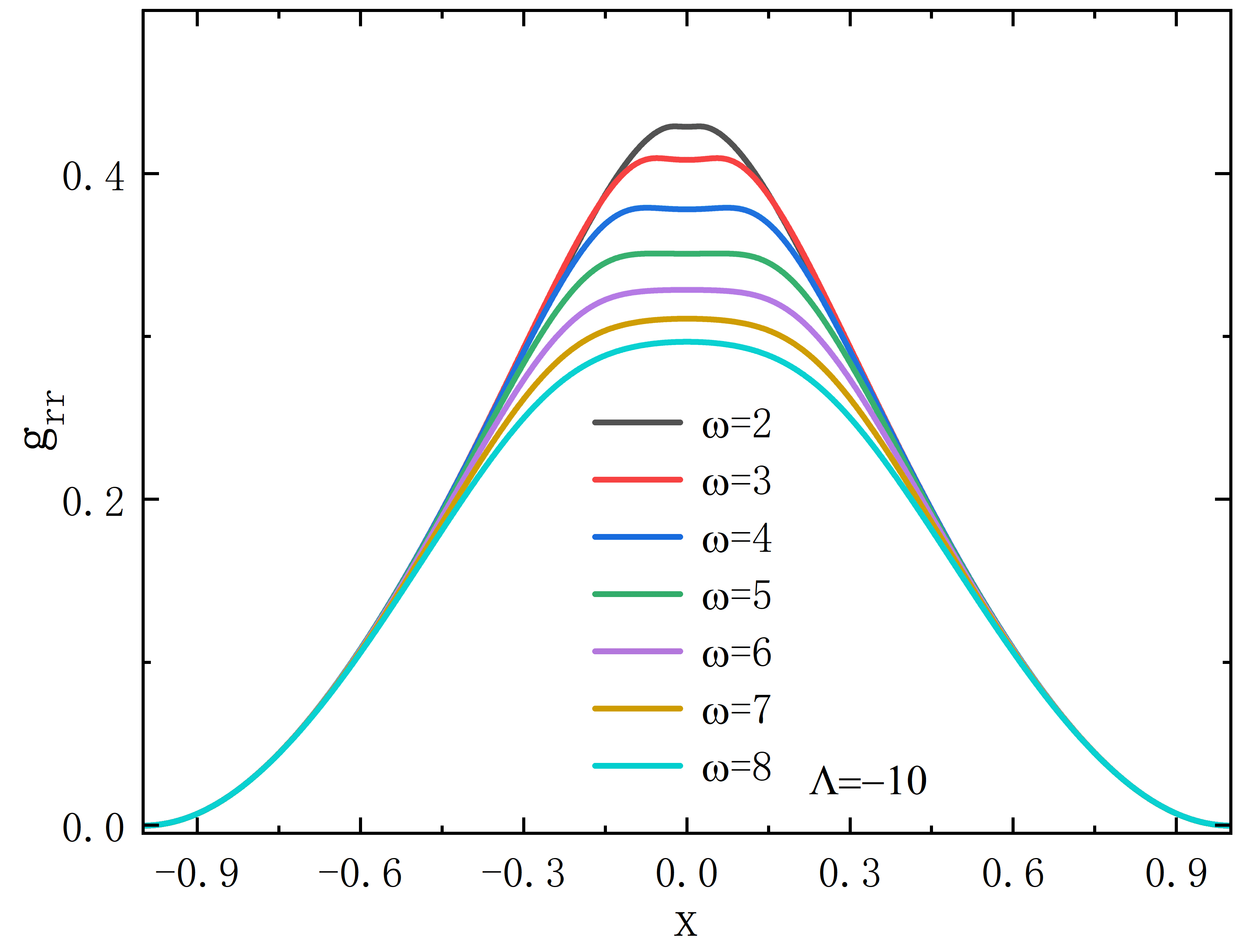}} 
\end{center} 
\caption{The metric functions $g_{tt}$ and $g_{rr}$ as functions of the radial coordinate $x$. The first row fixes $\omega = 2.1$ for different $\Lambda$ values, while the second row fixes $\Lambda = -10$ for different $\omega$ values. The throat radius $r_0 = 1$.} 
\label{phaseII2} 
\end{figure}

We then analyzed the distribution characteristics of the metric functions. Figure \ref{phaseII2} shows the behavior of the metric functions as $\Lambda$ changes when the frequency $\omega = 2.1$ is fixed. As $\Lambda$ decreases, the value of $g_{tt}$ near the throat approaches zero, while the peak of $g_{rr}$ moves closer to the throat, and the changes on both sides of the throat gradually flatten. When $\Lambda = -10$ is fixed, as $\omega$ decreases, the value of $g_{tt}$ near the throat again approaches zero, showing behavior similar to that of an horizon. Additionally, compared with symmetric case I, the flat region of $g_{tt}$ is larger, but as $\Lambda$ decreases, the flat region gradually shrinks.

To quantify the geometric behavior near the throat, Tables \ref{tab:tII1} and \ref{tab:tII2} list the minimum values of $g_{tt}$ and their distribution ranges for different values of $\Lambda$ and $\omega$. Clearly, when either $\Lambda$ or $\omega$ is sufficiently small, the minimum value of $g_{tt}$ approaches $10^{-4}$, showing characteristics similar to an horizon.

\begin{figure}[H]
\subfigure{\includegraphics[width=0.45\textwidth]{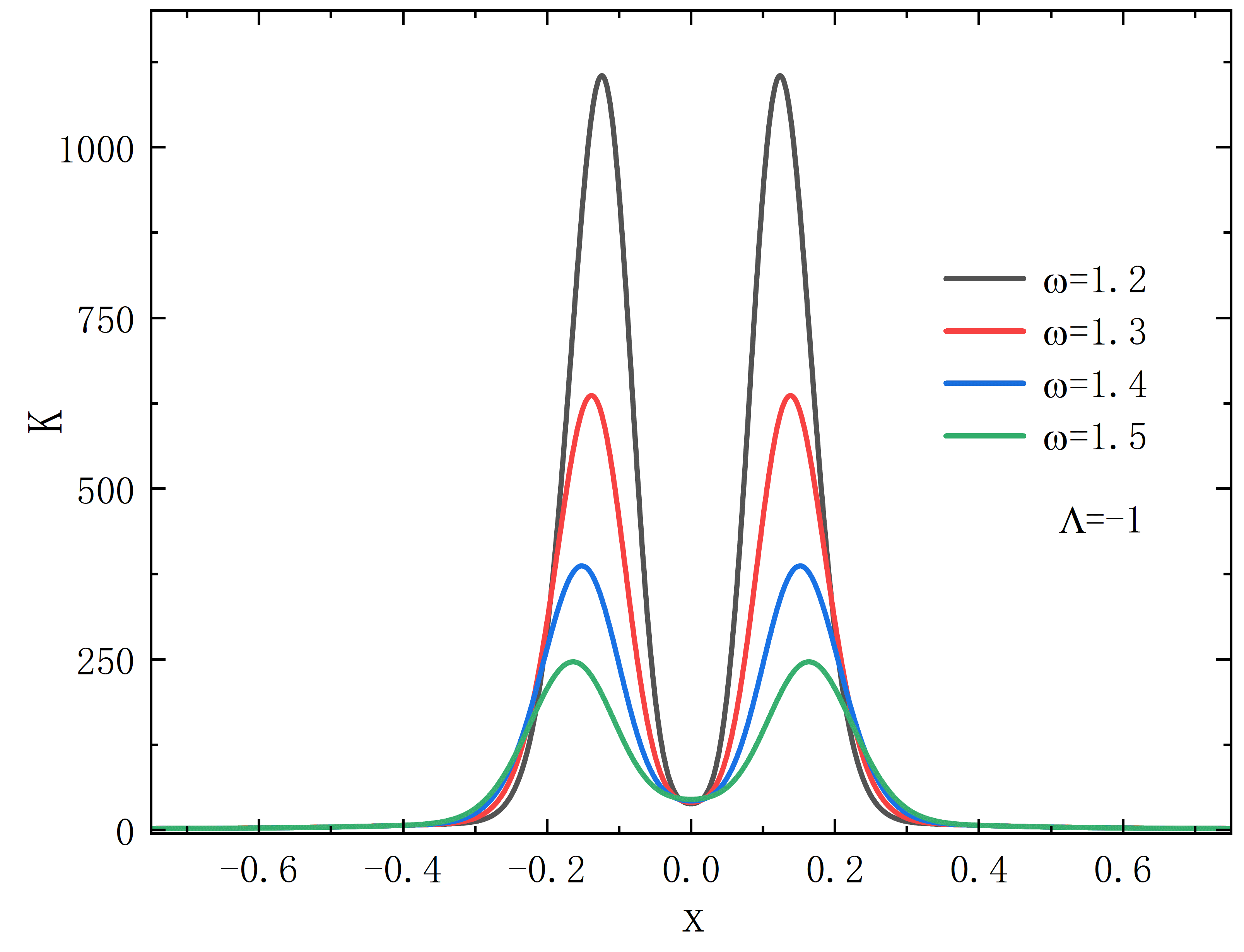}} 
\subfigure{\includegraphics[width=0.45\textwidth]{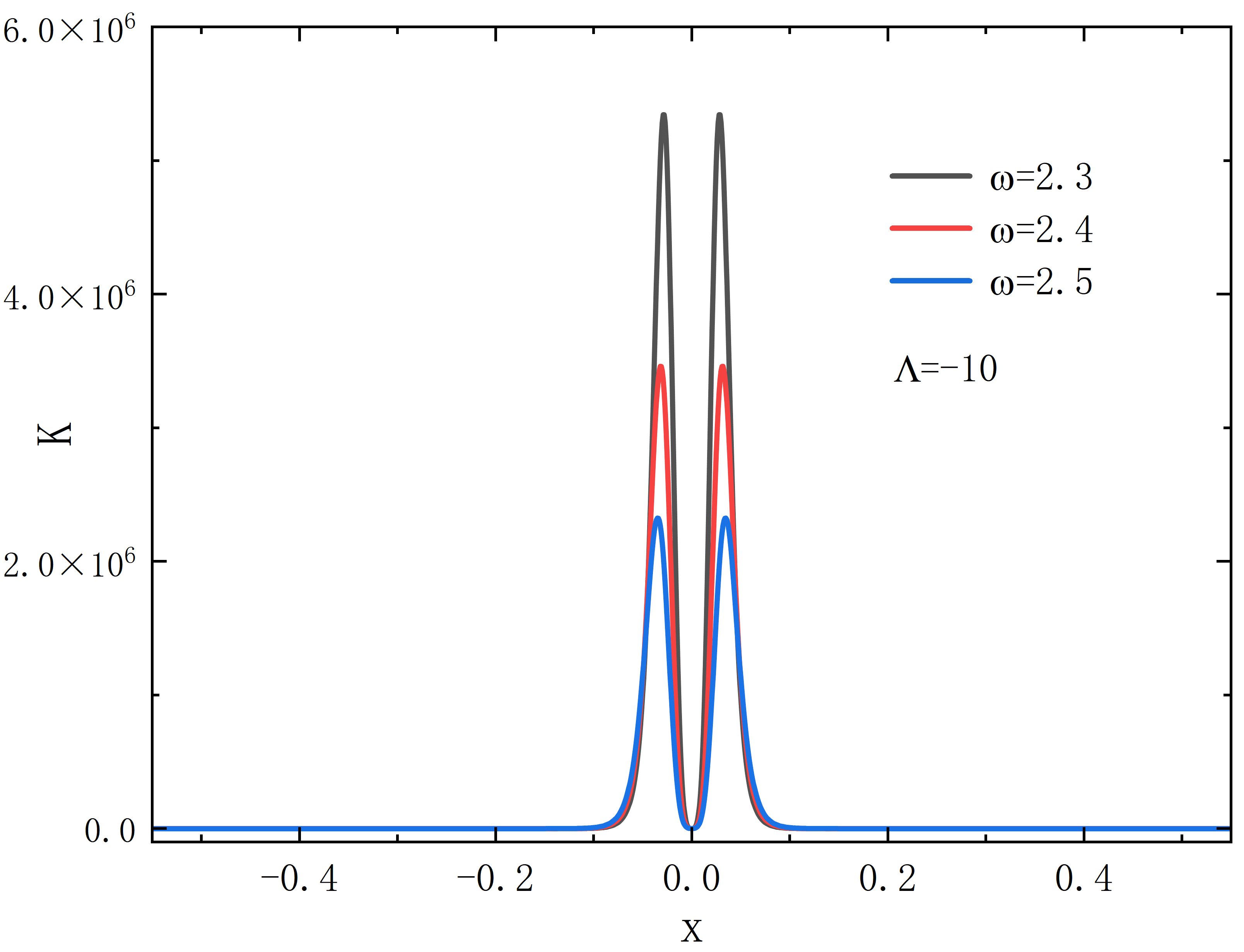}}
\caption{The distribution of the Kretschmann scalar as a function of the radial coordinate $x$. The left image fixes $\Lambda = -1$, while  the right image fixes $\Lambda = -10$. The throat radius is $r_0 = 1$.} 
\label{phaseII4}
\end{figure}

Furthermore, we considered the distribution of the Kretschmann scalar shown in Figure \ref{phaseII4}. We found that the Kretschmann curve exhibits a significant double peak structure near the throat. In contrast to symmetric case I, Figure \ref{phaseII4} shows the distribution of the Kretschmann scalar when $\Lambda$ is fixed and $\omega$ varies, which reflects the singularity in the spacetime on both sides of $x = 0$. When $\Lambda$ is fixed, as $\omega$ decreases, the value of the Kretschmann scalar increases. According to our criterion, two horizons begin to appear on either side of the throat.

\begin{table}[H]
\centering
\begin{tabular}{|c|c|c|}
\hline
$\Lambda$  ($\omega = -2$ ) & $x$ & $g_{tt}(0)$   \\ \hline
-3    & $-0.04 \leq x \leq 0.04$ & $4.1 \times 10^{-2}$ \\ \hline
-5    & $-0.015 \leq x \leq 0.015$ & $1.2 \times 10^{-2}$ \\ \hline
-10   & $-0.004 \leq x \leq 0.004$ & $1.2 \times 10^{-4}$ \\ \hline
\end{tabular}
\caption{The minimum value of $g_{tt}$ for $\omega = -2$ at different values of $\Lambda$. The throat radius is $r_0 = 1$.}
\label{tab:tII1}
\end{table}

\begin{table}[H]
\centering
\begin{tabular}{|c|c|c|}
\hline
$\omega$ ($\Lambda=-10$) & $x$ & $g_{tt}(0)$   \\ \hline
1.9  & $-0.002 \leq x \leq 0.002$ & $8 \times 10^{-4}$ \\ \hline
2.1  & $-0.005 \leq x \leq 0.005$ & $1.7 \times 10^{-3}$ \\ \hline
2.5   & $-0.01 \leq x \leq 0.01$ & $5.5 \times 10^{-3}$ \\ \hline
\end{tabular}
\caption{The minimum value of $g_{tt}$ for $\Lambda = -10$ at different values of $\omega$. The throat radius is $r_0 = 1$.}
\label{tab:tII2}
\end{table}

\begin{figure}[H]
\subfigure{\includegraphics[width=0.45\textwidth]{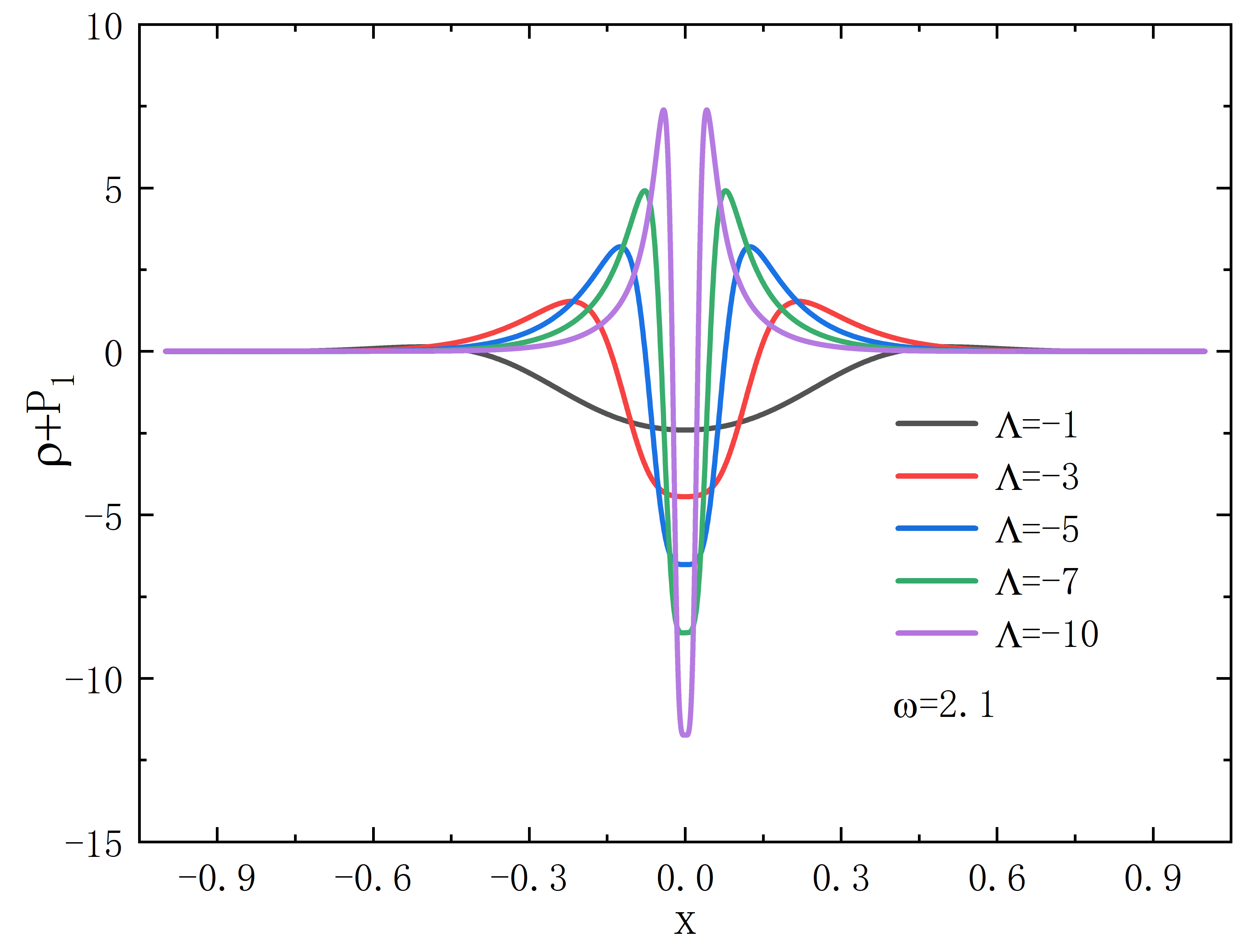}} 
\subfigure{\includegraphics[width=0.45\textwidth]{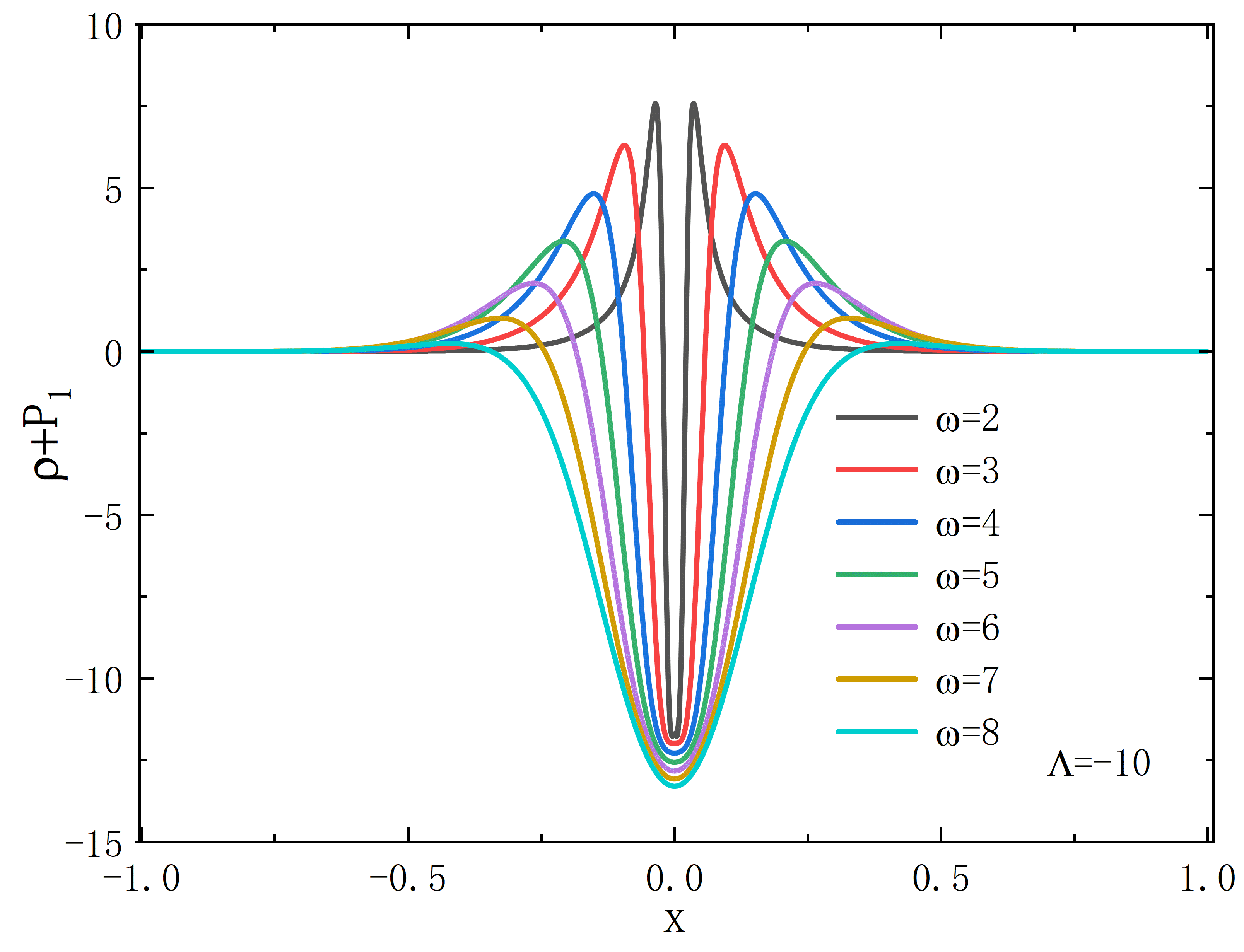}} 
\caption{The variation of NEC violation with the radial coordinate $x$, comparing different values of $\Lambda$ or $\omega$. The throat radius is $r_0 = 1$.}
\label{phaseII5}
\end{figure}

Finally, we also studied the variation of energy density in symmetric case II, as shown in Figure \ref{phaseII5}. The left panel shows that as $\Lambda$ decreases, the violation of NEC in the throat region becomes significantly stronger, and the curve becomes sharper. The right panel shows that as the frequency $\omega$ decreases, the width of the curve contracts, and although the minimum value of the throat increases slightly, the change is not significant. However, compared to symmetric case I, the variation of NEC in symmetric case II is smoother near the throat.

\subsection{The D charge}

\begin{figure}[H]
\begin{center}
\subfigure{\includegraphics[width=0.45\textwidth]{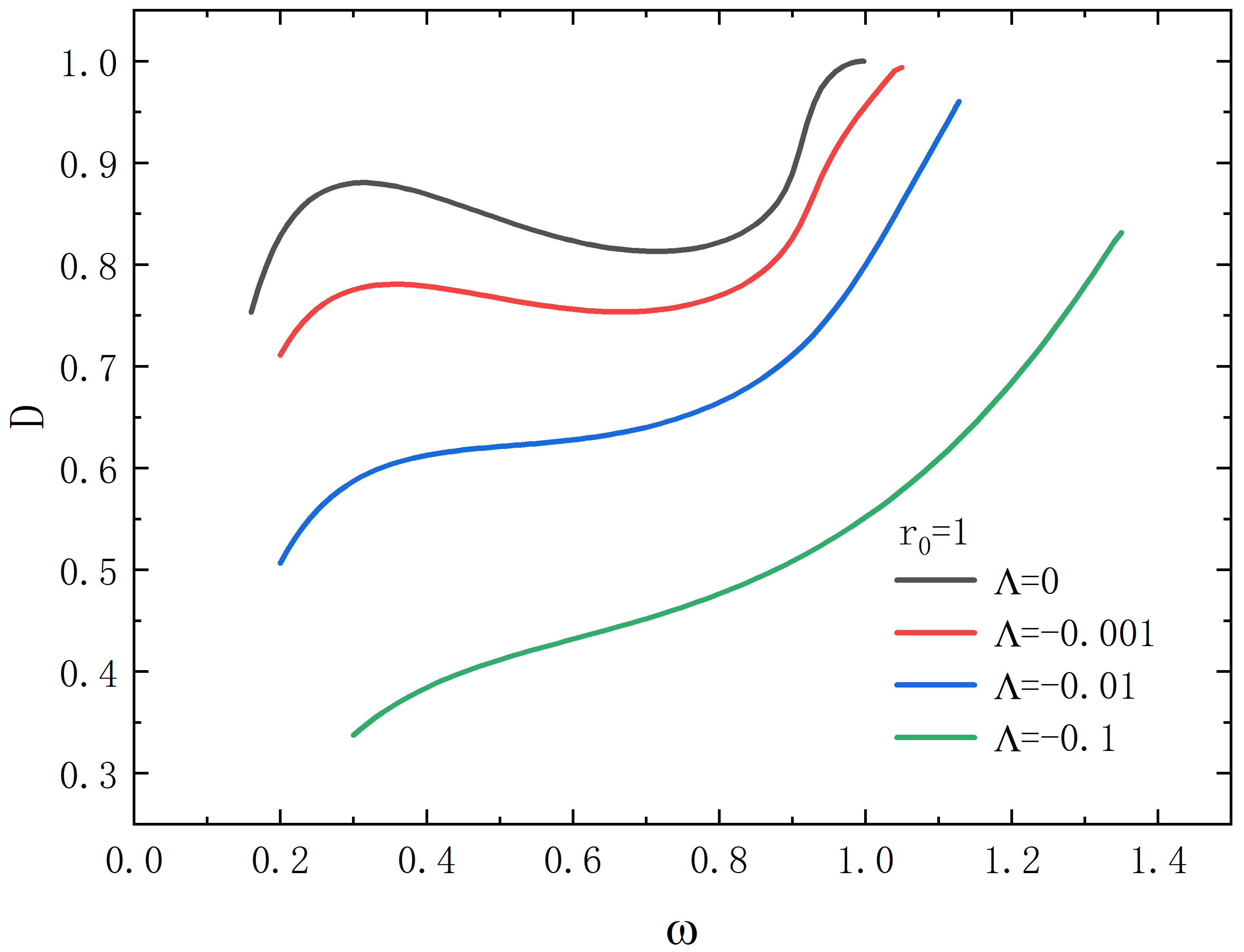}}
\subfigure{\includegraphics[width=0.45\textwidth]{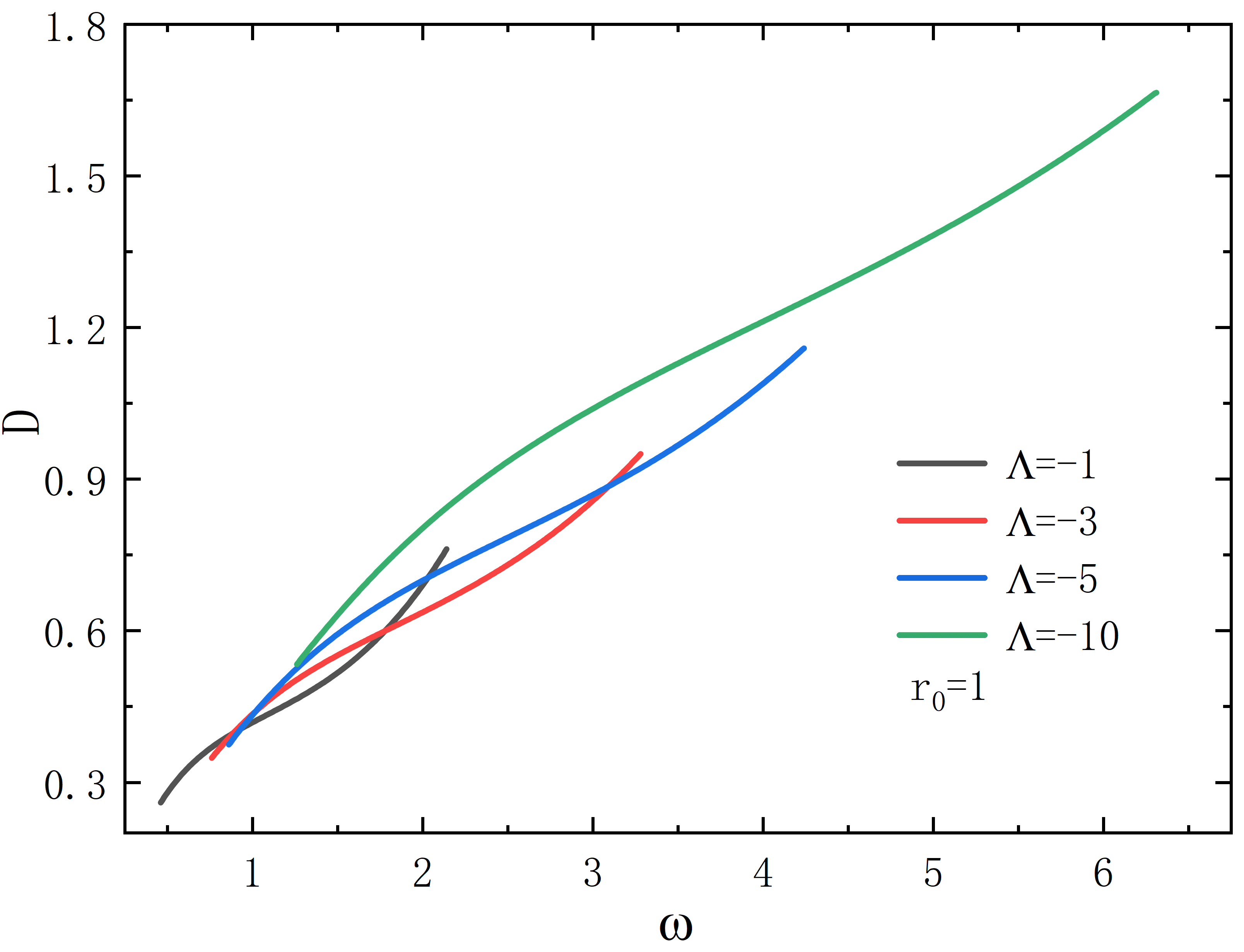}}
\subfigure{\includegraphics[width=0.45\textwidth]{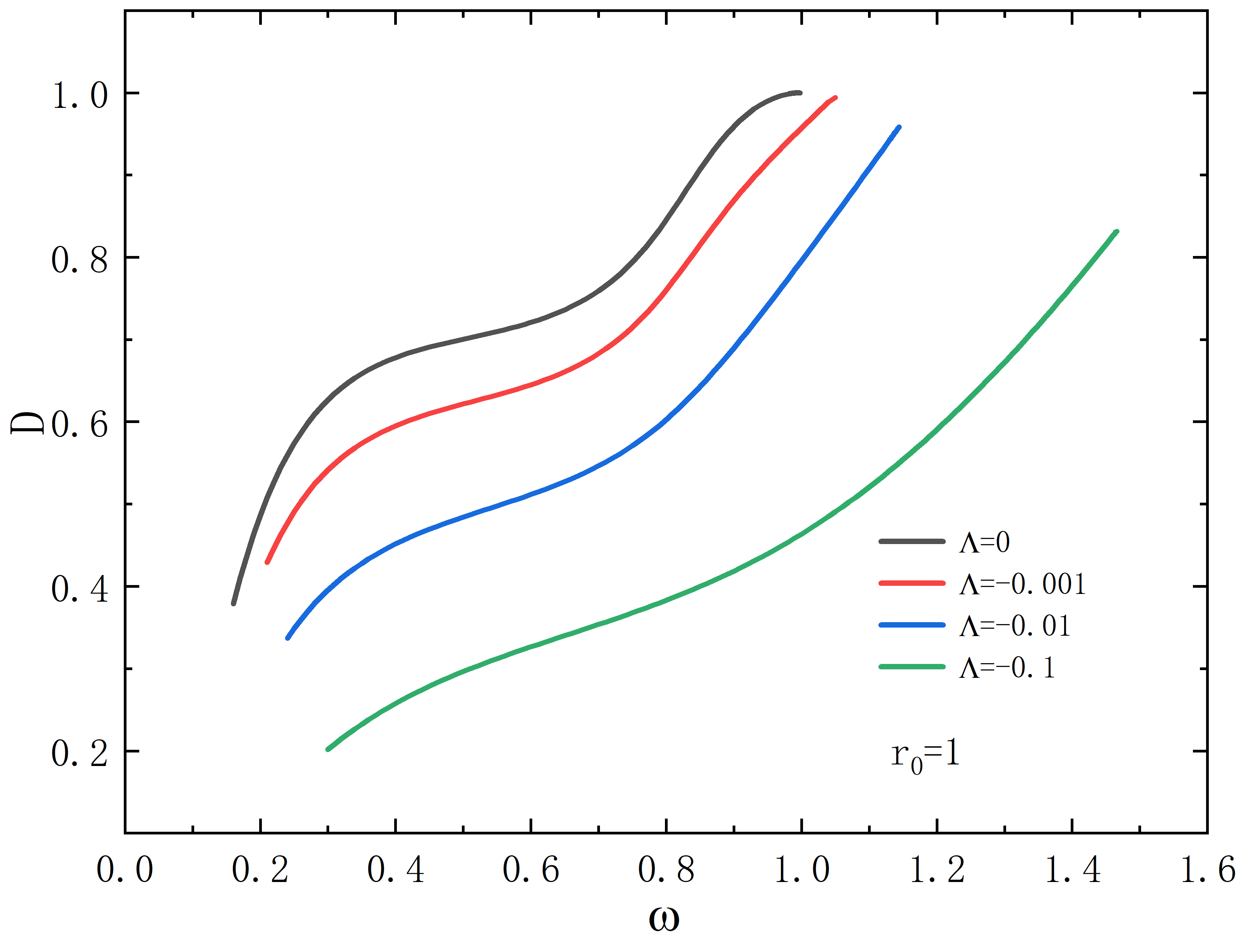}}
\subfigure{\includegraphics[width=0.45\textwidth]{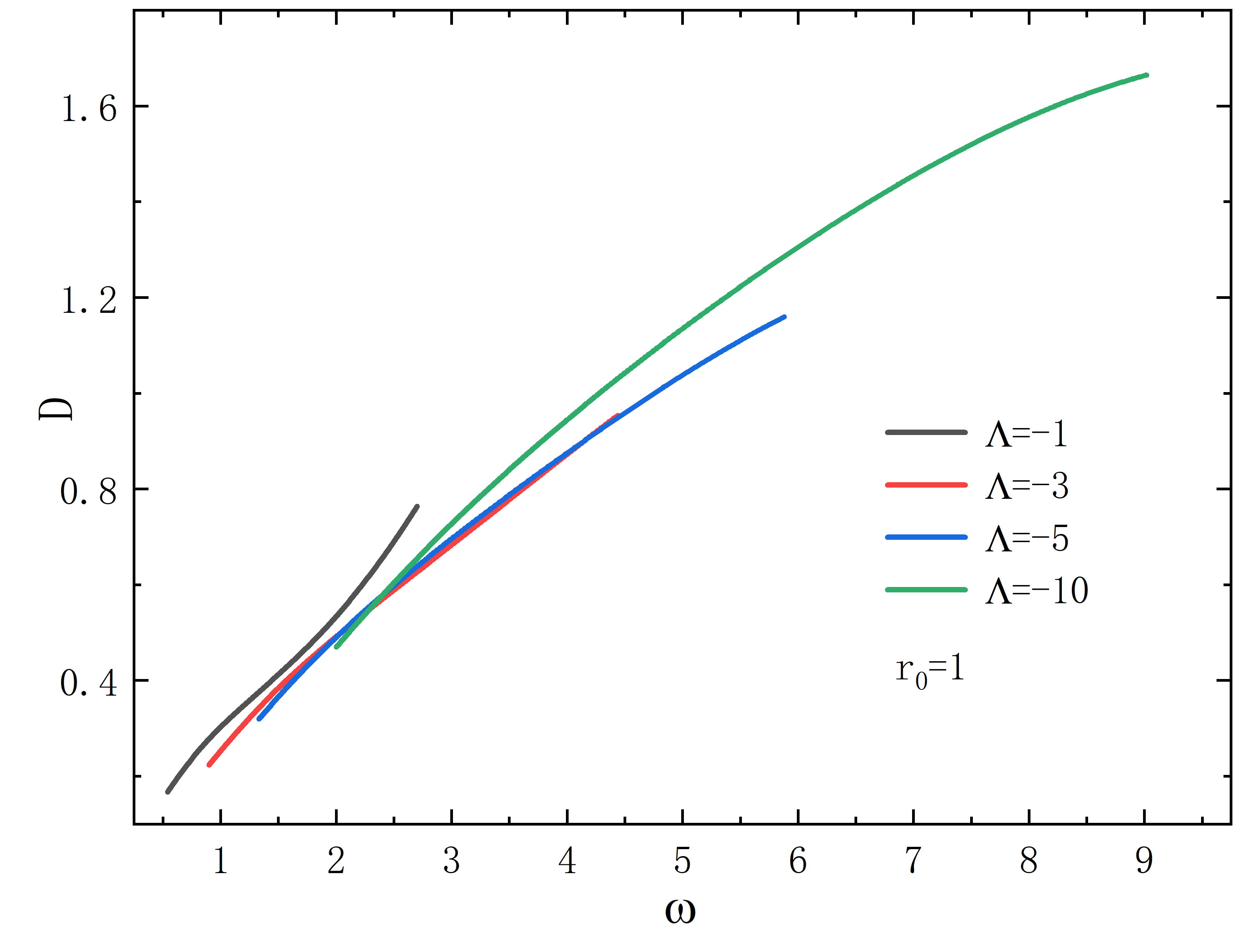}}
\caption{The scalar charge $\mathcal{D}$ of the phantom field as a function of frequency for different values of $\Lambda$. The throat radius is $r_0 = 1$. The upper panels correspond to symmetric case I, and the lower panels correspond to symmetric case II.} 
\label{phaseD}
\end{center}
\end{figure}

In the previous work, we defined the constant $\mathcal{D}$ to reflect the scalar charge of the phantom field and to test the accuracy of the numerical calculations. For a fixed throat radius $r_0 = 1$, its value as a function of frequency $\omega$ should remain consistent at different positions. From Figure \ref{phaseD}, we observe that the overall variation of $\mathcal{D}$ decreases initially to a minimum value as $\Lambda$ decreases, and then increases as $\Lambda$ decreases further. The result is consistent with the observations in the literature \cite{Blazquez-Salcedo:2020nsa}.

\subsection{The Geometry Properties}

To gain a deeper understanding of the geometric properties of the wormhole, we use the embedding diagram method to embed the two-dimensional equatorial plane geometry into three-dimensional cylindrical coordinates. This method visually presents the topological and geometric features of the wormhole.

When the time $t$ and the polar angle variable $\theta = \pi/2$ are fixed, the metric of the equatorial plane is given by:
\begin{equation}
ds^2 = \frac{p}{e^B N} dr^2 + \frac{p h}{e^B} d\varphi^2.
\end{equation}

By comparing this with the metric form of three-dimensional cylindrical coordinates:
\begin{equation}
ds^2 = d\rho^2 + dz^2 + \rho^2 d\varphi^2,
\end{equation}
we obtain the following embedding relations:
\begin{align}
\rho(r) &= \sqrt{\frac{p h}{e^B}}, \quad
z(r) = \pm \int \sqrt{\frac{p}{e^B N} - \left( \frac{d\rho}{dr} \right)^2}  dr.
\end{align}

Here, $\rho(r)$ is the circumference radius of the equatorial plane, and $z(r)$ describes the vertical coordinate of the embedding diagram. The throat corresponds to the minimum value of $\rho(r)$.

\begin{figure}[H]
\centering
\subfigure{\includegraphics[width=0.45\textwidth]{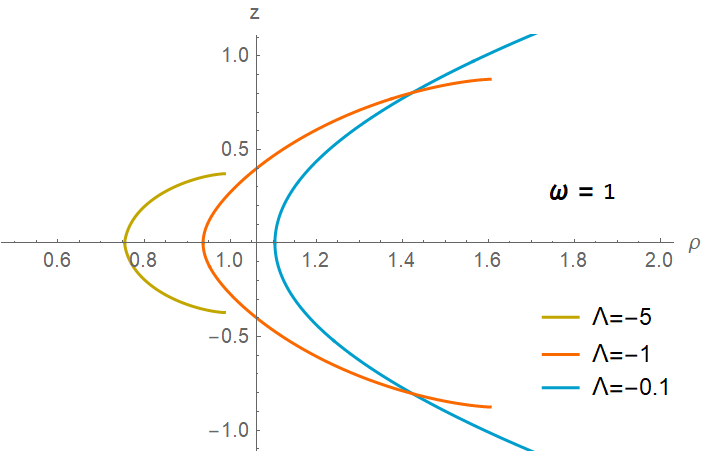}}
\subfigure{\includegraphics[width=0.45\textwidth]{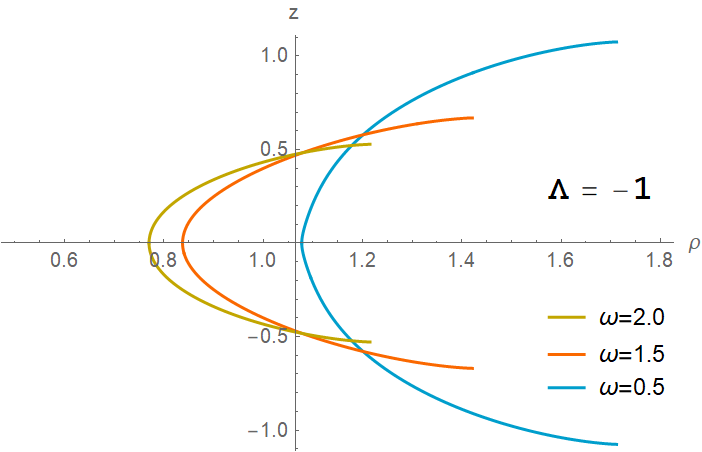}}
\subfigure{\includegraphics[width=0.45\textwidth]{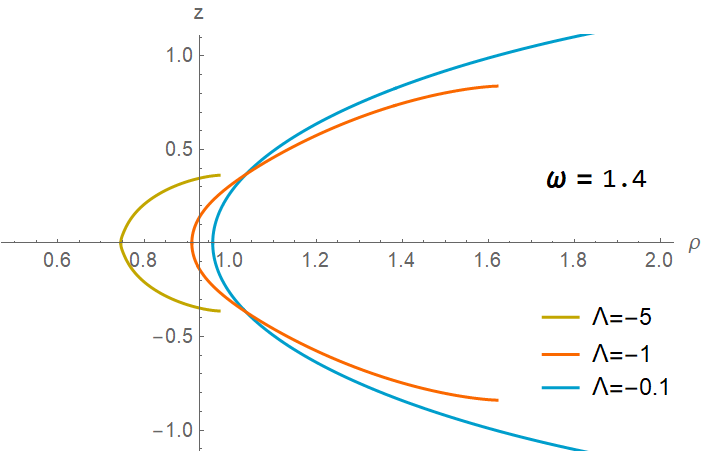}}
\subfigure{\includegraphics[width=0.45\textwidth]{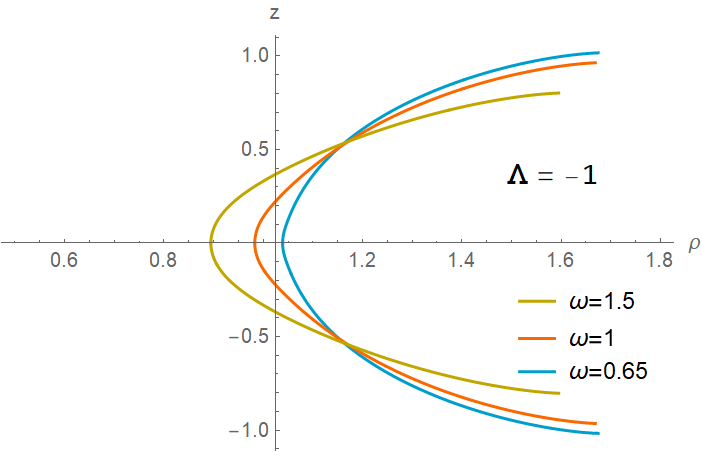}}
\caption{Two-dimensional wormhole embedding diagrams. The first row shows the embedding diagrams for symmetric case I with $\Lambda = -1$ and $\omega = 0.5, 1.5, 2$, and $\omega = 1$ with $\Lambda = -0.1, -1, -5$. The second row shows the embedding diagrams for symmetric case II with $\Lambda = -1$ and $\omega = 0.65, 1, 1.5$, and $\omega = 1.4$ with $\Lambda = -0.1, -1, -5$. The throat radius is $r_0 = 1$.}
\label{phaseWH}
\end{figure}

\begin{figure}[H]
\centering
\subfigure{\includegraphics[width=0.45\textwidth]{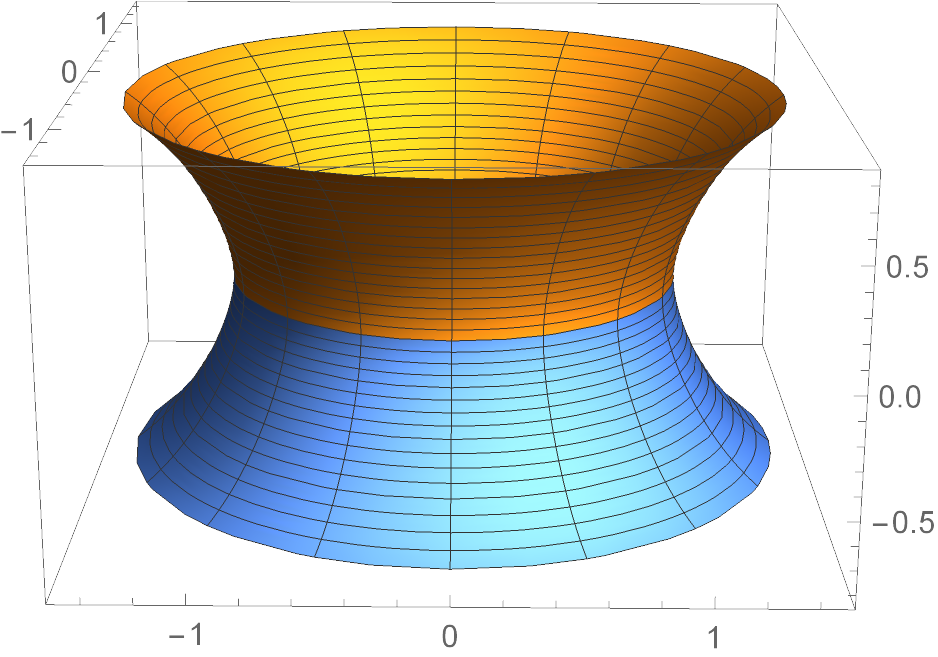}}
\subfigure{\includegraphics[width=0.45\textwidth]{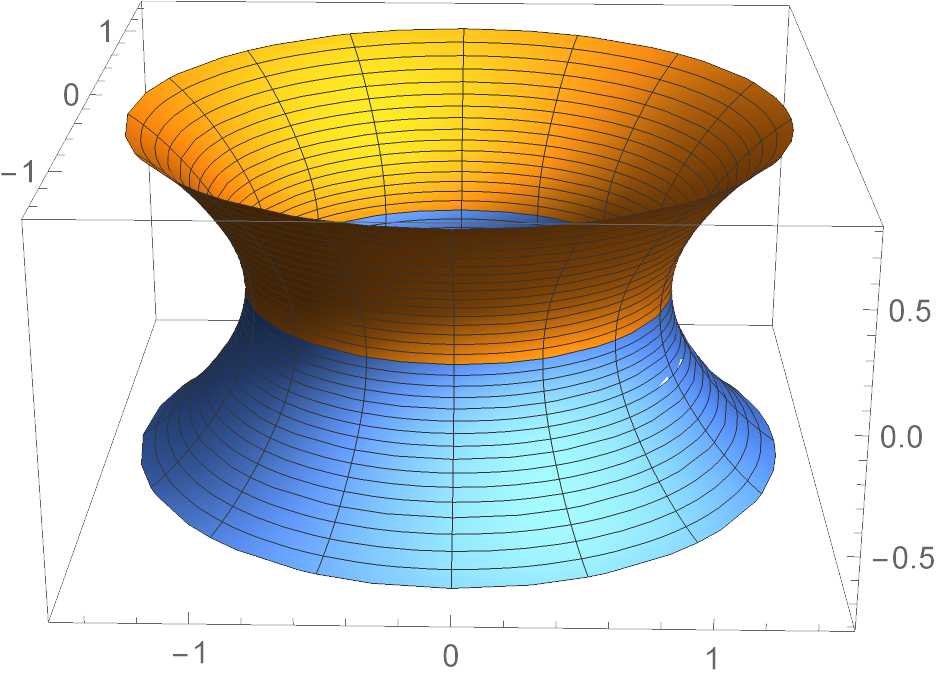}}
\caption{Three-dimensional wormhole embedding diagrams are shown below. The left panel displays the three-dimensional embedding diagram for the symmetric case 1, with $\Lambda = -1$ and $\omega = 1$. The right panel shows the three-dimensional embedding diagram for the symmetric case 2, with $\Lambda = -1$ and $\omega = 1.4$. In both cases, the throat radius is fixed at $r_0 = 1$.}
\label{phase3dWH}
\end{figure}

As shown in Figure \ref{phaseWH}, for both symmetric case I and symmetric case II, the effects of $\Lambda$ and $\omega$ on the geometric features are almost identical. For example, the right panel of the first row shows the effect of varying the frequency $\omega$ under the condition $\Lambda = -1$. The results indicate that the variation of frequency has a limited impact on the throat radius, but it significantly changes the depth and curvature of the throat. Similar effects are observed in symmetric case II. When $\omega$ is fixed, the influence of different cosmological constants $\Lambda$ on the embedding diagram can be observed. As $\Lambda$ increases, the wormhole throat radius gradually increases, and the geometric shape becomes more symmetric and smoother. Figure \ref{phase3dWH} shows the corresponding three-dimensional embedding diagram.

\section{CONCLUSION AND OUTLOOK}\label{sec5}

This paper presents a traversable wormhole model that explores the coupling properties of the phantom field and the Proca field within an asymptotically AdS Ellis wormhole spacetime. Two distinct symmetric solutions are derived. In AdS spacetime, the odd terms in the metric function vanish, leading to a zero total mass for the system. A detailed analysis of the Noether charge and the scalar field charge reveals key characteristics of this model. Additionally, the influence of the cosmological constant on the metric function and matter field is examined. Furthermore, embedding diagrams provide an intuitive way to visualize the wormhole geometry.

We computed the Noether charge for different throat radii, revealing the effect of the cosmological constant on the Noether charge. Unlike the case where $\Lambda = 0$, we found that when $\Lambda$ decreases, not only does the characteristic spiral structure expand, but the frequency solution region also increases and shifts to the right. By analyzing the metric functions, we found that when the cosmological constant $\Lambda$ and the frequency are sufficiently small, the minimum value of the Noether charge approaches zero. For Symmetric Solution I, we considered the formation of a horizon at the throat, while for Symmetric Solution II, two horizons appear on either side of the throat. This behavior is similar to the "black bounce" spacetime structure found in \cite{Simpson:2018tsi} . In contrast to the approach in the paper \cite{Su:2023zhh} , where the horizon is created by reducing the throat radius, we find that a reduction in the cosmological constant can also lead to the formation of the horizon. This further deepens our understanding of the horizon formation mechanism. When we couple the matter field, we find that the matter field distribution exhibits a concentration near the throat, specifically demonstrated by a significant peak of the Proca field function near $x = 0$, which is confirmed by the distribution of the Kretschmann scalar. Additionally, although the NEC condition is violated at the throat, the introduction of the Proca field allows certain regions on both sides of the throat to satisfy the NEC, depending on specific parameters. The study of the embedding geometry shows how variations in the cosmological constant $\Lambda$ and frequency $\omega$ affect the topology of the wormhole.

Future research directions include exploring whether asymmetric asymptotically AdS wormhole solutions exist when coupling matter fields. Moreover, the dynamical stability of the model remains an important open question. Further work will focus on the numerical evolution of the model to comprehensively assess its stability and explore its potential observational significance in astrophysics.

\section*{Acknowledgements}

This work is supported by the National Natural Science Foundation of China (Grant No. 12275110 and No. 12247101) and the National Key Research and Development Program of China (Grant No. 2022YFC2204101 and 2020YFC2201503).



\begin{thebibliography}{99}





\bibitem{Wheeler:1955zz}
J.~A.~Wheeler,
Geons,
Phys. Rev. \textbf{97} (1955), 511-536
doi:10.1103/PhysRev.97.511

\bibitem{Power:1957zz}
E.~A.~Power and J.~A.~Wheeler,
Thermal Geons,
Rev. Mod. Phys. \textbf{29} (1957), 480-495
doi:10.1103/RevModPhys.29.480

\bibitem{Kaup:1968zz}
D.~J.~Kaup,
Klein-Gordon Geon,
Phys. Rev. \textbf{172} (1968), 1331-1342
doi:10.1103/PhysRev.172.1331


\bibitem{Brito:2015pxa}
R.~Brito, V.~Cardoso, C.~A.~R.~Herdeiro and E.~Radu,
Proca stars: Gravitating Bose\textendash{}Einstein condensates of massive spin 1 particles,
Phys. Lett. B \textbf{752} (2016), 291-295
doi:10.1016/j.physletb.2015.11.051
[arXiv:1508.05395 [gr-qc]].


\bibitem{SalazarLandea:2016bys}
I.~Salazar Landea and F.~Garc\'\i{}a,
Charged Proca Stars,
Phys. Rev. D \textbf{94} (2016) no.10, 104006
doi:10.1103/PhysRevD.94.104006
[arXiv:1608.00011 [hep-th]].

\bibitem{Duarte:2016lig}
M.~Duarte and R.~Brito,
Asymptotically anti-de Sitter Proca Stars,
Phys. Rev. D \textbf{94} (2016) no.6, 064055
doi:10.1103/PhysRevD.94.064055
[arXiv:1609.01735 [gr-qc]].




\bibitem{Loginov:2015rya}
A.~Y.~Loginov,
Nontopological solitons in the model of the self-interacting complex vector field,
Phys. Rev. D \textbf{91} (2015) no.10, 105028
doi:10.1103/PhysRevD.91.105028

\bibitem{Brihaye:2017inn}
Y.~Brihaye, T.~Delplace and Y.~Verbin,
Proca Q Balls and their Coupling to Gravity,
Phys. Rev. D \textbf{96} (2017) no.2, 024057
doi:10.1103/PhysRevD.96.024057
[arXiv:1704.01648 [gr-qc]].

\bibitem{Heeck:2021bce}
J.~Heeck, A.~Rajaraman, R.~Riley and C.~B.~Verhaaren,
Proca Q-balls and Q-shells,
JHEP \textbf{10} (2021), 103
doi:10.1007/JHEP10(2021)103
[arXiv:2107.10280 [hep-th]].

\bibitem{Dzhunushaliev:2021tlm}
V.~Dzhunushaliev and V.~Folomeev,
Proca balls with angular momentum or flux of electric field,
Phys. Rev. D \textbf{105} (2022) no.1, 016022
doi:10.1103/PhysRevD.105.016022
[arXiv:2112.06227 [hep-th]].

\bibitem{Herdeiro:2023lze}
C.~Herdeiro, E.~Radu and E.~dos Santos Costa Filho,
Proca-Higgs balls and stars in a UV completion for Proca self-interactions,
JCAP \textbf{05} (2023), 022
doi:10.1088/1475-7516/2023/05/022
[arXiv:2301.04172 [gr-qc]].

\bibitem{Herdeiro:2019mbz}
C.~Herdeiro, I.~Perapechka, E.~Radu and Y.~Shnir,
Asymptotically flat spinning scalar, Dirac and Proca stars,
Phys. Lett. B \textbf{797} (2019), 134845
doi:10.1016/j.physletb.2019.134845
[arXiv:1906.05386 [gr-qc]].

\bibitem{Aoki:2022woy}
K.~Aoki and M.~Minamitsuji,
Resolving the pathologies of self-interacting Proca fields: A case study of Proca stars,
Phys. Rev. D \textbf{106} (2022) no.8, 084022
doi:10.1103/PhysRevD.106.084022
[arXiv:2206.14320 [gr-qc]].


\bibitem{Babichev:2017rti}
E.~Babichev, C.~Charmousis and M.~Hassaine,
Black holes and solitons in an extended Proca theory,
JHEP \textbf{05} (2017), 114
doi:10.1007/JHEP05(2017)114
[arXiv:1703.07676 [gr-qc]].

\bibitem{Aoki:2022mdn}
K.~Aoki and M.~Minamitsuji,
Highly compact Proca stars with quartic self-interactions,
Phys. Rev. D \textbf{107} (2023) no.4, 044045
doi:10.1103/PhysRevD.107.044045
[arXiv:2212.07659 [gr-qc]].

\bibitem{Minamitsuji:2018kof}
M.~Minamitsuji,
Vector boson star solutions with a quartic order self-interaction,
Phys. Rev. D \textbf{97} (2018) no.10, 104023
doi:10.1103/PhysRevD.97.104023
[arXiv:1805.09867 [gr-qc]].



\bibitem{Ma:2023bhb}
T.~X.~Ma, C.~Liang, J.~R.~Ren and Y.~Q.~Wang,
Dirac-Proca stars,
[arXiv:2309.11700 [gr-qc]].

\bibitem{Ma:2023bhb}
T.~X.~Ma, C.~Liang, J.~R.~Ren and Y.~Q.~Wang,
Dirac-Proca stars,
Phys. Rev. D \textbf{109} (2024) no.8, 084012
doi:10.1103/PhysRevD.109.084012
[arXiv:2309.11700 [gr-qc]].


\bibitem{Zhang:2023rwc}
R.~Zhang, S.~X.~Sun, L.~X.~Huang and Y.~Q.~Wang,
Rotating multistate Proca stars,
Phys. Rev. D \textbf{111} (2025) no.2, 024076
doi:10.1103/PhysRevD.111.024076
[arXiv:2312.15755 [gr-qc]].


\bibitem{Lazarte:2024jyr}
C.~Lazarte and M.~Alcubierre,
$\ell$-Proca stars,
Class. Quant. Grav. \textbf{41} (2024) no.13, 135003
doi:10.1088/1361-6382/ad4d4d
[arXiv:2401.16360 [gr-qc]].

\bibitem{Jockel:2023rrm}
C.~Jockel and L.~Sagunski,
Fermion Proca Stars: Vector Dark Matter Admixed Neutron Stars,
Particles \textbf{7} (2024) no.1, 52-79
doi:10.3390/particles7010004
[arXiv:2310.17291 [gr-qc]].

\bibitem{GRAVITY:2023azi}
A.~Foschi \textit{et al.} [GRAVITY],
Using the motion of S2 to constrain vector clouds around Sgr~A*,
Mon. Not. Roy. Astron. Soc. \textbf{530} (2024) no.4, 3740-3751
doi:10.1093/mnras/stae423
[arXiv:2312.02653 [astro-ph.GA]].

\bibitem{Romero-Shaw:2020thy}
I.~M.~Romero-Shaw, P.~D.~Lasky, E.~Thrane and J.~C.~Bustillo,
GW190521: orbital eccentricity and signatures of dynamical formation in a binary black hole merger signal,
Astrophys. J. Lett. \textbf{903} (2020) no.1, L5
doi:10.3847/2041-8213/abbe26
[arXiv:2009.04771 [astro-ph.HE]].

\bibitem{CalderonBustillo:2020fyi}
J.~Calder\'on Bustillo, N.~Sanchis-Gual, A.~Torres-Forn\'e, J.~A.~Font, A.~Vajpeyi, R.~Smith, C.~Herdeiro, E.~Radu and S.~H.~W.~Leong,
GW190521 as a Merger of Proca Stars: A Potential New Vector Boson of $8.7\times 10^{-13}$  eV,
Phys. Rev. Lett. \textbf{126} (2021) no.8, 081101
doi:10.1103/PhysRevLett.126.081101
[arXiv:2009.05376 [gr-qc]].


\bibitem{Morris:1988tu}
M.~S.~Morris, K.~S.~Thorne and U.~Yurtsever,
Wormholes, Time Machines, and the Weak Energy Condition,
Phys. Rev. Lett. \textbf{61} (1988), 1446-1449
doi:10.1103/PhysRevLett.61.1446



\bibitem{Einstein:1935tc}
A.~Einstein and N.~Rosen,
The Particle Problem in the General Theory of Relativity,
Phys. Rev. \textbf{48} (1935), 73-77
doi:10.1103/PhysRev.48.73


\bibitem{Misner:1957mt}
C.~W.~Misner and J.~A.~Wheeler,
Classical physics as geometry: Gravitation, electromagnetism, unquantized charge, and mass as properties of curved empty space,
Annals Phys. \textbf{2} (1957), 525-603
doi:10.1016/0003-4916(57)90049-0

\bibitem{Kruskal:1959vx}
M.~D.~Kruskal,
Maximal extension of Schwarzschild metric,
Phys. Rev. \textbf{119} (1960), 1743-1745
doi:10.1103/PhysRev.119.1743

\bibitem{Fuller:1962zza}
R.~W.~Fuller and J.~A.~Wheeler,
Causality and Multiply Connected Space-Time,
Phys. Rev. \textbf{128} (1962), 919-929
doi:10.1103/PhysRev.128.919




\bibitem{Ellis:1973yv}
H.~G.~Ellis,
Ether flow through a drainhole - a particle model in general relativity,
J. Math. Phys. \textbf{14} (1973), 104-118
doi:10.1063/1.1666161

\bibitem{Ellis:1979bh}
H.~G.~Ellis,
THE EVOLVING, FLOWLESS DRAIN HOLE: A NONGRAVITATING PARTICLE MODEL IN GENERAL RELATIVITY THEORY,
Gen. Rel. Grav. \textbf{10} (1979), 105-123
doi:10.1007/BF00756794

\bibitem{Bronnikov:1973fh}
K.~A.~Bronnikov,
Scalar-tensor theory and scalar charge,
Acta Phys. Polon. B \textbf{4} (1973), 251-266


\bibitem{Kodama:1978dw}
T.~Kodama,
General Relativistic Nonlinear Field: A Kink Solution in a Generalized Geometry,
Phys. Rev. D \textbf{18} (1978), 3529-3534
doi:10.1103/PhysRevD.18.3529

\bibitem{Morris:1988cz}
M.~S.~Morris and K.~S.~Thorne,
Wormholes in space-time and their use for interstellar travel: A tool for teaching general relativity,
Am. J. Phys. \textbf{56} (1988), 395-412
doi:10.1119/1.15620

\bibitem{Garattini:2019ivd}
R.~Garattini,
Casimir Wormholes,
Eur. Phys. J. C \textbf{79} (2019) no.11, 951
doi:10.1140/epjc/s10052-019-7468-y
[arXiv:1907.03623 [gr-qc]].

\bibitem{Mehdizadeh:2024oyd}
M.~R.~Mehdizadeh and A.~H.~Ziaie,
Novel Casimir wormholes in Einstein gravity,
Eur. Phys. J. Plus \textbf{139} (2024) no.11, 1001
doi:10.1140/epjp/s13360-024-05801-z
[arXiv:2406.03588 [gr-qc]].


\bibitem{Poisson:1995sv}
E.~Poisson and M.~Visser,
Thin shell wormholes: Linearization stability,
Phys. Rev. D \textbf{52} (1995), 7318-7321
doi:10.1103/PhysRevD.52.7318
[arXiv:gr-qc/9506083 [gr-qc]].



\bibitem{Blazquez-Salcedo:2020czn}
J.~L.~Bl\'azquez-Salcedo, C.~Knoll and E.~Radu,
Traversable wormholes in Einstein-Dirac-Maxwell theory,
Phys. Rev. Lett. \textbf{126} (2021) no.10, 101102
doi:10.1103/PhysRevLett.126.101102
[arXiv:2010.07317 [gr-qc]].


\bibitem{Lobo:2009ip}
F.~S.~N.~Lobo and M.~A.~Oliveira,
Wormhole geometries in f(R) modified theories of gravity,
Phys. Rev. D \textbf{80} (2009), 104012
doi:10.1103/PhysRevD.80.104012
[arXiv:0909.5539 [gr-qc]].

\bibitem{Banerjee:2021mqk}
A.~Banerjee, A.~Pradhan, T.~Tangphati and F.~Rahaman,
Wormhole geometries in $f(Q)$ gravity and the energy conditions,
Eur. Phys. J. C \textbf{81} (2021) no.11, 1031
doi:10.1140/epjc/s10052-021-09854-7
[arXiv:2109.15105 [gr-qc]].





\bibitem{Lobo:2005us}
F.~S.~N.~Lobo,
Phantom energy traversable wormholes,
Phys. Rev. D \textbf{71} (2005), 084011
doi:10.1103/PhysRevD.71.084011
[arXiv:gr-qc/0502099 [gr-qc]].

\bibitem{Sushkov:2005kj}
S.~V.~Sushkov,
Wormholes supported by a phantom energy,
Phys. Rev. D \textbf{71} (2005), 043520
doi:10.1103/PhysRevD.71.043520
[arXiv:gr-qc/0502084 [gr-qc]].

\bibitem{Lobo:2005yv}
F.~S.~N.~Lobo,
Stability of phantom wormholes,
Phys. Rev. D \textbf{71} (2005), 124022
doi:10.1103/PhysRevD.71.124022
[arXiv:gr-qc/0506001 [gr-qc]].

\bibitem{Bronnikov:2012ch}
K.~A.~Bronnikov, R.~A.~Konoplya and A.~Zhidenko,
Instabilities of wormholes and regular black holes supported by a phantom scalar field,
Phys. Rev. D \textbf{86} (2012), 024028
doi:10.1103/PhysRevD.86.024028
[arXiv:1205.2224 [gr-qc]].




\bibitem{Kleihaus:2014dla}
B.~Kleihaus and J.~Kunz,
Rotating Ellis Wormholes in Four Dimensions,
Phys. Rev. D \textbf{90} (2014), 121503
doi:10.1103/PhysRevD.90.121503
[arXiv:1409.1503 [gr-qc]].

\bibitem{Dzhunushaliev:2014bya}
V.~Dzhunushaliev, V.~Folomeev, C.~Hoffmann, B.~Kleihaus and J.~Kunz,
Boson Stars with Nontrivial Topology,
Phys. Rev. D \textbf{90} (2014) no.12, 124038
doi:10.1103/PhysRevD.90.124038
[arXiv:1409.6978 [gr-qc]].

\bibitem{Hoffmann:2017jfs}
C.~Hoffmann, T.~Ioannidou, S.~Kahlen, B.~Kleihaus and J.~Kunz,
Spontaneous symmetry breaking in wormholes spacetimes with matter,
Phys. Rev. D \textbf{95} (2017) no.8, 084010
doi:10.1103/PhysRevD.95.084010
[arXiv:1703.03344 [gr-qc]].

\bibitem{Yue:2023ela}
Y.~Yue, P.~B.~Ding and Y.~Q.~Wang,
Boson star with parity-odd symmetry in wormhole spacetime,
Eur. Phys. J. C \textbf{83} (2023) no.8, 732
doi:10.1140/epjc/s10052-023-11914-z
[arXiv:2305.04496 [gr-qc]].

\bibitem{Ding:2023syj}
P.~B.~Ding, T.~X.~Ma, T.~F.~Fang and Y.~Q.~Wang,
Study of boson stars with wormhole,
JHEP \textbf{04} (2024), 033
doi:10.1007/JHEP04(2024)033
[arXiv:2305.19819 [gr-qc]].

\bibitem{Hao:2023igi}
C.~H.~Hao, S.~X.~Sun, L.~X.~Huang, R.~Zhang, X.~Su and Y.~Q.~Wang,
Dirac stars in wormhole spacetime,
JCAP \textbf{04} (2024), 057
doi:10.1088/1475-7516/2024/04/057
[arXiv:2309.16379 [gr-qc]].

\bibitem{Su:2023zhh}
X.~Su, C.~H.~Hao, J.~R.~Ren and Y.~Q.~Wang,
Proca stars in wormhole spacetime,
JCAP \textbf{09} (2024), 010
doi:10.1088/1475-7516/2024/09/010
[arXiv:2311.17557 [gr-qc]].

\bibitem{Nozawa:2020gzz}
M.~Nozawa,
Static spacetimes haunted by a phantom scalar field III: asymptotically (A)dS solutions,
Phys. Rev. D \textbf{103} (2021) no.2, 024005
doi:10.1103/PhysRevD.103.024005
[arXiv:2010.07561 [gr-qc]].



\bibitem{Abe:2010ap}
F.~Abe,
Gravitational Microlensing by the Ellis Wormhole,
Astrophys. J. \textbf{725} (2010), 787-793
doi:10.1088/0004-637X/725/1/787
[arXiv:1009.6084 [astro-ph.CO]].

\bibitem{Ohgami:2015nra}
T.~Ohgami and N.~Sakai,
Wormhole shadows,
Phys. Rev. D \textbf{91} (2015) no.12, 124020
doi:10.1103/PhysRevD.91.124020
[arXiv:1704.07065 [gr-qc]].

\bibitem{Konoplya:2016hmd}
R.~A.~Konoplya and A.~Zhidenko,
Wormholes versus black holes: quasinormal ringing at early and late times,
JCAP \textbf{12} (2016), 043
doi:10.1088/1475-7516/2016/12/043
[arXiv:1606.00517 [gr-qc]].







\bibitem{Maldacena:1997re}
J.~M.~Maldacena,
The Large N limit of superconformal field theories and supergravity,
Adv. Theor. Math. Phys. \textbf{2} (1998), 231-252
doi:10.4310/ATMP.1998.v2.n2.a1
[arXiv:hep-th/9711200 [hep-th]].

\bibitem{Maldacena:2018lmt}
J.~Maldacena and X.~L.~Qi,
Eternal traversable wormhole,
[arXiv:1804.00491 [hep-th]].

\bibitem{Freivogel:2019lej}
B.~Freivogel, V.~Godet, E.~Morvan, J.~F.~Pedraza and A.~Rotundo,
Lessons on eternal traversable wormholes in AdS,
JHEP \textbf{07} (2019), 122
doi:10.1007/JHEP07(2019)122
[arXiv:1903.05732 [hep-th]].



\bibitem{Gao:2016bin}
P.~Gao, D.~L.~Jafferis and A.~C.~Wall,
Traversable Wormholes via a Double Trace Deformation,
JHEP \textbf{12} (2017), 151
doi:10.1007/JHEP12(2017)151
[arXiv:1608.05687 [hep-th]].




\bibitem{Maldacena:2013xja}
J.~Maldacena and L.~Susskind,
Cool horizons for entangled black holes,
Fortsch. Phys. \textbf{61} (2013), 781-811
doi:10.1002/prop.201300020
[arXiv:1306.0533 [hep-th]].




\bibitem{Maldacena:2017axo}
J.~Maldacena, D.~Stanford and Z.~Yang,
Diving into traversable wormholes,
Fortsch. Phys. \textbf{65} (2017) no.5, 1700034
doi:10.1002/prop.201700034
[arXiv:1704.05333 [hep-th]].



\bibitem{Dai:2020ffw}
D.~C.~Dai, D.~Minic, D.~Stojkovic and C.~Fu,
Testing the $\mathbf {ER=EPR}$ conjecture,
Phys. Rev. D \textbf{102} (2020) no.6, 066004
doi:10.1103/PhysRevD.102.066004
[arXiv:2002.08178 [hep-th]].


\bibitem{Kain:2023ore}
B.~Kain,
Probing the Connection between Entangled Particles and Wormholes in General Relativity,
Phys. Rev. Lett. \textbf{131} (2023) no.10, 101001
doi:10.1103/PhysRevLett.131.101001
[arXiv:2309.03314 [hep-th]].

\bibitem{Lemos:2003jb}
J.~P.~S.~Lemos, F.~S.~N.~Lobo and S.~Quinet de Oliveira,
Morris-Thorne wormholes with a cosmological constant,
Phys. Rev. D \textbf{68} (2003), 064004
doi:10.1103/PhysRevD.68.064004
[arXiv:gr-qc/0302049 [gr-qc]].

\bibitem{Lemos:2004vs}
J.~P.~S.~Lemos and F.~S.~N.~Lobo,
Plane symmetric traversable wormholes in an Anti-de Sitter background,
Phys. Rev. D \textbf{69} (2004), 104007
doi:10.1103/PhysRevD.69.104007
[arXiv:gr-qc/0402099 [gr-qc]].

\bibitem{Maeda:2008nz}
H.~Maeda and M.~Nozawa,
Static and symmetric wormholes respecting energy conditions in Einstein-Gauss-Bonnet gravity,
Phys. Rev. D \textbf{78} (2008), 024005
doi:10.1103/PhysRevD.78.024005
[arXiv:0803.1704 [gr-qc]].

\bibitem{Zhang:2024kqp}
S.~F.~Zhang and R.~H.~Lin,
Traversability of Schwarzschild-Anti-de Sitter Wormhole in f(T) gravity,
[arXiv:2410.09430 [gr-qc]].

\bibitem{Maeda:2012fr}
H.~Maeda,
Exact dynamical AdS black holes and wormholes with a Klein-Gordon field,
Phys. Rev. D \textbf{86} (2012), 044016
doi:10.1103/PhysRevD.86.044016
[arXiv:1204.4472 [gr-qc]].

\bibitem{Wu:2022gpm}
T.~Wu,
AdS wormholes from Ricci-flat/AdS correspondence,
Phys. Rev. D \textbf{108} (2023) no.4, 044001
doi:10.1103/PhysRevD.108.044001
[arXiv:2209.02278 [gr-qc]].




\bibitem{Anabalon:2018rzq}
A.~Anabal\'on and J.~Oliva,
Four-dimensional Traversable Wormholes and Bouncing Cosmologies in Vacuum,
JHEP \textbf{04} (2019), 106
doi:10.1007/JHEP04(2019)106
[arXiv:1811.03497 [hep-th]].


\bibitem{Lobo:2004uq}
F.~S.~N.~Lobo,
Thin shells around traversable wormholes,
[arXiv:gr-qc/0401083 [gr-qc]].



\bibitem{Blazquez-Salcedo:2020nsa}
J.~L.~Bl\'azquez-Salcedo, X.~Y.~Chew, J.~Kunz and D.~H.~Yeom,
Ellis wormholes in anti-de Sitter space,
Eur. Phys. J. C \textbf{81} (2021) no.9, 858
doi:10.1140/epjc/s10052-021-09645-0
[arXiv:2012.06213 [gr-qc]].



\bibitem{Hao:2024hba}
C.~H.~Hao, X.~Su and Y.~Q.~Wang,
AdS Ellis wormholes with scalar field,
[arXiv:2404.11002 [gr-qc]].




\bibitem{Simpson:2018tsi}
A.~Simpson and M.~Visser,
Black-bounce to traversable wormhole,
JCAP \textbf{02} (2019), 042
doi:10.1088/1475-7516/2019/02/042
[arXiv:1812.07114 [gr-qc]].

\end{thebibliography}
\end{document}